\documentclass[a4paper,fleqn,usenatbib]{mnras}

\usepackage{newtxtext,newtxmath}

\usepackage[T1]{fontenc}

\DeclareRobustCommand{\VAN}[3]{#2}
\let\VANthebibliography\thebibliography
\def\thebibliography{\DeclareRobustCommand{\VAN}[3]{##3}\VANthebibliography}

\usepackage{xr}
\usepackage{xcolor}
\usepackage{graphicx}	
\usepackage{amsmath}	

\usepackage{verbatim}
\usepackage{breqn}
\usepackage[export]{adjustbox}
\usepackage{booktabs}
\usepackage{pgfplotstable}
\usepackage{longtable}
 \usepackage{threeparttable}

\newcommand{\lya}{Ly$\alpha$}

\newcommand{\kms}{$km s^{-1}$}
\newcommand{\hi}{\mbox{\tiny H\,{\sc i}}}
\newcommand{\HI}{\mbox{H\,{\sc i}}}

\newcommand{\OIV}{\mbox{O\,{\sc iv}}}

\newcommand{\OVI}{\mbox{O\,{\sc vi}}}

\newcommand{\CIV}{\mbox{C\,{\sc iv}}}

\newcommand{\NeVIII}{\mbox{Ne\,{\sc viii}}}

\newcommand{\NHI} {$N_{\rm HI}$}
\newcommand{\Anand}[1]{{\color{red}[Anand: #1]}}

\title[Low-z metal absorbers]{Role of ionizing background on the statistics of metal absorbers in hydrodynamical simulations}
\author[Mallik et al.]{Sukanya Mallik$^{1}$\thanks{E-mail: sukanyam@iucaa.in},
	Raghunathan Srianand$^{1}$,
	Soumak Maitra$^{2}$, Prakash Gaikwad$^{3}$ \& \newauthor{Nishikanta Khandai$^{4, 5}$}
	\\
	\\
	$^{1}$ IUCAA, Postbag 4, Ganeshkhind, Pune - 411007, India\\
	$^{2}$ Istituto Nazionale di Astrofisica -Osservatorio Astronomico di Trieste, Via Tiepolo 11, Trieste, Italy\\
	$^{3}$ Max-Planck-Institut f\"ur Astronomie, K\"onigstuhl 17, D-69117 Heidelberg, Germany\\
	$^{4}$ School of Physical Sciences, National Institute of Science Education and Research, Jatni 752050, India\\
    $^{5}$ Homi Bhabha National Institute, Training School Complex, Anushaktinagar, Mumbai 400094, India}

\date{Accepted XXX. Received YYY; in original form ZZZ}

\pubyear{2022}

\begin{document}
\label{firstpage}
\pagerange{\pageref{firstpage}--\pageref{lastpage}}
\maketitle

\begin{abstract}

We study the statistical properties of \OVI, \CIV, and \NeVIII\ absorbers at low-$z$ (i.e., $z<0.5$) using  Sherwood simulations with "WIND" only and "WIND+AGN" feedback and Massive black-II simulation that incorporates both "WIND" i.e. outflows driven by stellar feedback and AGN feedbacks. For each simulation, by considering a wide range of metagalactic ionizing UV background (UVB), we show the statistical properties
%
such as 
distribution functions of column density ($N$), $b$-paramerer and velocity spread ($\Delta V_{90}$), the relationship between $N$ and $b$-parameter and the fraction of \lya\ absorbers showing detectable metal lines as a function of $N$(\HI) are influenced by the UVB used. This is because UVB changes the range in density, temperature, and metallicity of gas contributing to a given absorption line. For simulations considered here, we show the difference in some of the predicted distributions between different simulations is similar to the one obtained by varying the UVB for a given simulation.
Most of the observed properties of \OVI\ absorbers are roughly matched by Sherwood simulation with "WIND+AGN" feedback when using the UVB with a lower \OVI\ ionization rate. However, this simulation fails to produce observed distributions of \CIV\ and fraction of \HI\ absorbers with detectable metals.
%
  Therefore, in order to constrain different feedback processes and/or UVBs, using observed properties of \HI\ and metal ions,  it is important to perform simultaneous analysis of various observable parameters.

\end{abstract}

\begin{keywords}
		Cosmology: large-scale structure of Universe - Cosmology: diffuse radiation - Galaxies: intergalactic medium - Galaxies: quasars : absorption lines
\end{keywords}



\section{Introduction}
\label{Sec:introduction}


Metal absorption lines of ions such as \CIV\ and \OVI\ are frequently detected from low \HI\ column density (\NHI) absorbers  in the spectra of distant quasars \citep{Songaila1996}. These absorbers are believed to trace low-density regions \citep{bi1993}, such as circumgalactic medium \citep[CGM, see][for a review]{Tumlinson2017} or intergalactic medium \citep[IGM, see][for a review]{meiksin2009}, that can not sustain in situ star formation. In the standard Big Bang model, most of the elements heavier than Lithium (referred to as metals) are produced in stars. Therefore, analytical models and simulations  of structure formation strongly favor that the metals produced inside the stars are transported to in and around the galaxies (i.e., CGM) and into the IGM through large-scale winds driven by energy and momentum from SNe explosions, cosmic rays, and/or AGN activities in galaxies \citep[see for example,][]{Scannapieco2005,Samui2008,Samui2010}. 
On the other hand, observed statistical properties of galaxies demand sustained star formation, outflows, and feedback activities over a longer period of time \citep{croton2006,bower2006}. This requires a supply of cold gas to the star-forming regions through accretion. Therefore, the main challenge is to simulate the universe self-consistently, taking into account various feedback processes (such as outflows, infall, heating, and cooling of gas) and constrain them using the available observations.

For more than a decade, cosmological hydrodynamical simulations \citep[][]{ oppenheimer2009,tepper2011,oppenheimer2012,surech2015,rahmati2016,Nelson2018,bradley2022, Li2022} incorporating  a wide range of feedback processes are used to understand how feedbacks influence the observed column density (or equivalent width) distribution of different ions like \OVI, \CIV\ and \NeVIII\ \citep[for example, see figure 6 of][]{Nelson2018}.  Some simulations also explored the influence of non-equilibrium ionization, turbulence, and flickering AGNs \citep{tepper2011,oppenheimer2012,oppenheimer2013a, oppenheimer2013b, oppenheimer2016, oppenheimer2018}. All these studies have given some insights into the importance of various physical processes and sub-grid physics in cosmological simulations.
{The dependence of properties of \lya\ absorption on ionizing metagalactic UV-background(UVB) \citep[e.g.][]{kollmeier2014,shull2015,gaikwad2017a,khaire2019a,maitra2020,Tillman2022} is well explored using hydrodynamical simulations. }
%
It is interesting to know whether uncertainties in the UVB can lead to scatter in the measurable quantities of metal line absorbers similar to what one sees in simulations while varying the feedback parameters.

{
The detectability of absorption from a given metal ion depends on the physical state of the gas and the nature of the ionizing radiation field (global+ any local ionizing sources). It is a general procedure, while analysing the observed spectra, to assume the absorber to be a single or multi-phased region (slabs) in ionization equilibrium with the assumed UVB. The physical conditions in these phases are constrained by reproducing the observed column densities of different ions and their ratios consistently.
Such an approach has also been used to quantify how 
 various derived physical properties of the absorbers (i.e., metallicity, density, and size of the absorbing region) are affected by the uncertainties in the assumed radiation field \citep[see for example,][]{ simcoe2004, simcoe2011, howk2009, fechner2011, hussain2017,Haislmaier2021, Acharya2022}.
 By comparing the statistics of pixel optical depths using observed and simulated spectra, it has been found that the maximum uncertainty in the derived abundances of carbon, silicon, and oxygen arises due to variation in the spectral shape of UVB  \citep[e.g.,][]{ aguirre2004,aguirre2008, schaye2003}. 
Using different UVBs generated by \citet{haardt2001},
\citet{oppenheimer2009} have shown the statistics of \HI\ and \OVI\  absorption are affected by the choice of the UVB and the wind feedback model parameters.
Recently, \citet{appleby2021} using simulations of CGM, have shown 
that the observed properties of low and high ion absorption lines are sensitive to the assumed UVB by considering three different UVBs \citep[from][]{faucher2020,haardt2001,haardt2012}.
%
%
}

{ 
In cosmological simulations, an absorption line at a given wavelength can originate from a set of spatially distinct regions that happen to have required Doppler shifts to produce absorption at that wavelength \citep[]{peeples2019, Marra2021}. 
 It is possible that for a given line of sight, the detectable absorption lines can originate from slightly different regions and velocity fields when we consider different UVBs  due to changes in the ion fractions. Therefore, the UVB may influence not only the observed column density but also the profile shape of the absorption lines. Hence it is important to explore the influence of UVBs on the statistics of metal absorption that captures properties such as column density, b-parameter, number of required voigt profile components, velocity extent of absorption and fraction of \HI\ absorbers showing detectable metals. 
%
Also, how much influence a given UVB makes may depend on the assumed feedback processes in the simulation.
Therefore, one needs to perform a statistical analysis of a large number of simulated spectra for different UVBs and different simulations. In this regard, availability of 
the automated Voigt profile fitting code, VoIgt profile Parameter Estimation Routine \citep[VIPER;][]{gaikwad2017a} enables us to engage in such a statistically analysis to probe the effects of UVB. This forms the main motivation for this work.
}

In this work, we consider two sets of cosmological hydrodynamical simulations with different implementations of feedback prescriptions to explore the effect of ionizing background on the 
statistics of metal ions (in particular \CIV, \OVI\, and \NeVIII) detected in absorption in the quasar spectra. These ions are chosen as they tend to trace low-density regions where optically thin ionizing conditions assumed in this work may be more appropriate.
We use Sherwood simulations  \citep[]{Bolton2017} incorporating "WIND only"\footnote{This refers to the feedback introduced only by the stellar processes.} and "WIND+AGN" feedbacks  and Massive Black-II (hereafter MB-II) simulation \citep[]{khandai2015} that incorporates "WIND+AGN" feedback. All simulations considered here were performed using the parallel Tree-PM smoothed particle hydrodynamic (hereafter SPH) code P-GADGET-3, a modified version of the publicly available GADGET-2 code described in \citet[]{springel2005}.

The present manuscript is arranged as follows. In section~\ref{Sec:simlations}, we provide details of the simulations and UVB used in this study. We also discuss the distribution of SPH particles and different ions in the temperature-density plane for these simulations. We provide details of spectral generation and automatic Voigt profile fitting codes employed in our analysis. In section~\ref{Sec:results}, we present various statistical distributions of   \CIV, \OVI\ , and \NeVIII\  obtained from our simulations and show how they depend on the assumed UVB.
We discuss these results in Section~\ref{Sec:discussions} and provide a main summary of our study in Section~\ref{Sec:summary}.
  
\section{Details of simulations used and mock spectrum generation}
\label{Sec:simlations}

Sherwood simulations \citep[]{Bolton2017} used here have a box of size 80~h$^{-1}$cMpc containing 2$\times$512$^3$ dark matter and baryonic particles and use cosmological parameters \{$\Omega_m$, $\Omega_b$, $\Omega_\Lambda$, $\sigma_8$, $n_s$, $h$\} = \{0.308, 0.0482, 0.692, 0.829, 0.961, 0.678\} from \citet[]{planck2014}. The MB-II simulation has a box size of 100 h$^{-1}$cMpc and better resolution (2$\times$1792$^3$ dark matter and baryonic particles) and uses cosmological parameters  \{$\Omega_m$, $\Omega_b$, $\Omega_\Lambda$, $\sigma_8$, $n_s$, $h$\}=\{0.275, 0.046, 0.725, 0.816, 0.968, 0.701\} consistent with the Wilkinson Microwave Anisotropy Probe 7 cosmology \citep[]{komatsu2011}.
In all Sherwood simulations, initial conditions were generated at $z=99$ on a  regular grid using the N-\textsc{GEN}IC code \citep[]{springel2005} and transfer functions generated by CAMB \citep[]{lewis2020}. The initial condition for MB-II was generated with the CMBFAST transfer functions at $z=159$.
Initial particle mass in Sherwood simulation are $2.75 \times 10^8 h^{-1} M_{\odot}$(DM), $5.1 \times 10^7 h^{-1} M_{\odot}$(baryon) and for MB-II it is $1.1 \times 10^7 h^{-1} M_{\odot}$(DM), $2.2 \times 10^6 h^{-1} M_{\odot}$(baryon). The mass resolution and box size in both Sherwood and MB-II are comparable or better than the simulation boxes used in the past for IGM studies \citep[]{oppenheimer2009, oppenheimer2012, tepper2011, tepper2013}. \par


The radiative heating and cooling processes are incorporated in GADGET-3 by self-consistently solving the ionization equilibrium and  non-equilibrium thermal evolution for a given spatially uniform UVB. The photo-ionization and photo-heating  rates of the gas are calculated using spatially uniform \citet{haardt2012} UVB in Sherwood simulations and \citet{haardt1996} UVB in MB-II.  
Below we provide a detailed comparison of these two and other UVBs available in the literature.
%
The cooling in Sherwood simulation is implemented following \citet[]{sutherland1993} model, which considers collisional ionization equilibrium of a metal-enriched gas with a fixed relative abundance of metals. 
The radiative cooling in MB-II is implemented following \citet[]{katz1996} model, which considers radiative cooling of the gas with primordial composition. 
%
Both simulations follow the same star formation prescriptions developed in \citet[]{springel2003a}. The MB-II simulation uses Salpeter initial mass function (IMF), but Sherwood simulations use the Chabrier IMF. This increases the available supernovae feedback energy by a factor of $\sim$2 in the latter. 
Unlike in the case of Illustris and Illustris TNG where the abundances of individual elements are evolved, in the Sherwood and MB-II simulation, only the global metallicity of the gas particles are retained  \citep[as given by the equation (40) in][]{springel2003a}. For simplicity, we assume the relative abundance of different elements to follow the solar ratio.

%
The MB-II simulation follows the "WIND" prescription developed by \citet[]{springel2003a}. The "WIND" particles are selected stochastically from all the star-forming particles after a time delay ($\sim$ 30 Myr, corresponding to the maximum lifetime of stars that end up in core-collapse SNe) and given a constant velocity $v_w$ in a random direction.
The "WIND" particles remain hydrodynamically decoupled from the surrounding for the initial 50 Myr or until their density falls below 10 percent of the threshold density value used for star formation, which ensures that they escape the interstellar medium (ISM). The MB-II simulation uses energy-driven "WIND" with a constant "WIND" velocity ($v_w$ = 483.61 $km s^{-1}$ )  and mass loading factor ($\eta_w$ = 2). The Sherwood simulations follow the energy-driven outflow model of \citet[]{Puchwein2013} where the "WIND" velocity ($v_w$) scales with the escape velocity of the galaxy. 
The mass loading factor  is obtained from the star formation rate and the available energy, and it scales as $\eta_w\propto v_w^{-2}$.

The subgrid physics governing the formation and growth of black holes in both MB-II and Sherwood simulations is developed in \citet[]{springel2005a}. 
A seed black hole of a fixed mass ($m_{BH, seed}=5 \times 10^5 h^{-1} M_\odot$ for MB-II and $m_{BH, seed}=10^4 h^{-1} M_\odot$ in Sherwood simulations) is inserted into the halos having mass more than the predefined threshold mass  of $m_{halo}=5 \times 10^{10} h^{-1} M_\odot$ for both the simulations if the halo does not contain a black hole already. The black holes then grow by accreting gas from the surrounding regions with an accretion rate  given by the Bondi–Hoyle–Lyttleton prescription 
or by mergers with other black holes having relative velocity less than the local sound speed. Sherwood simulations  allow an accretion rate up
to the Eddington rate, and MB-II allows an accretion rate up to twice the Eddington rate. The black holes radiate with a luminosity that is proportional to the accretion rate, $L=\eta \dot{M}_{BH} c^2$ \citep[][where the standard value of 0.1 for radiative efficiency $\eta$ is used]{shakura1973}, and some fraction of this radiated energy couples with the neighboring particles in the form of feedback. The AGN feedback in the MB-II simulation is incorporated by assuming a constant heating efficiency of 0.5 percent of the rest mass energy of the accreted gas \citep[see][]{dimatteo2005}. 
As described in \citet[]{Puchwein2013},
Sherwood simulations consider two modes of AGN feedback depending on the accretion rate. The heating efficiency is assumed to be 0.5 percent of the rest mass energy of the accreted gas when the accretion rate is above 0.01 of the Eddington rate. For accretion rates below 0.01 of the Eddington rate, the Sherwood simulation assumes the `radio mode' AGN feedback, with a 2 percent mechanical feedback efficiency, which is in agreement with the X-ray observation of elliptical galaxies \citep[]{allen2006}.

\par 
\subsection{Different ionizing UVBs}
\label{sec:UVB}

\begin{figure}
 	\begin{center}
	\includegraphics[width=8.2cm, height=10cm]{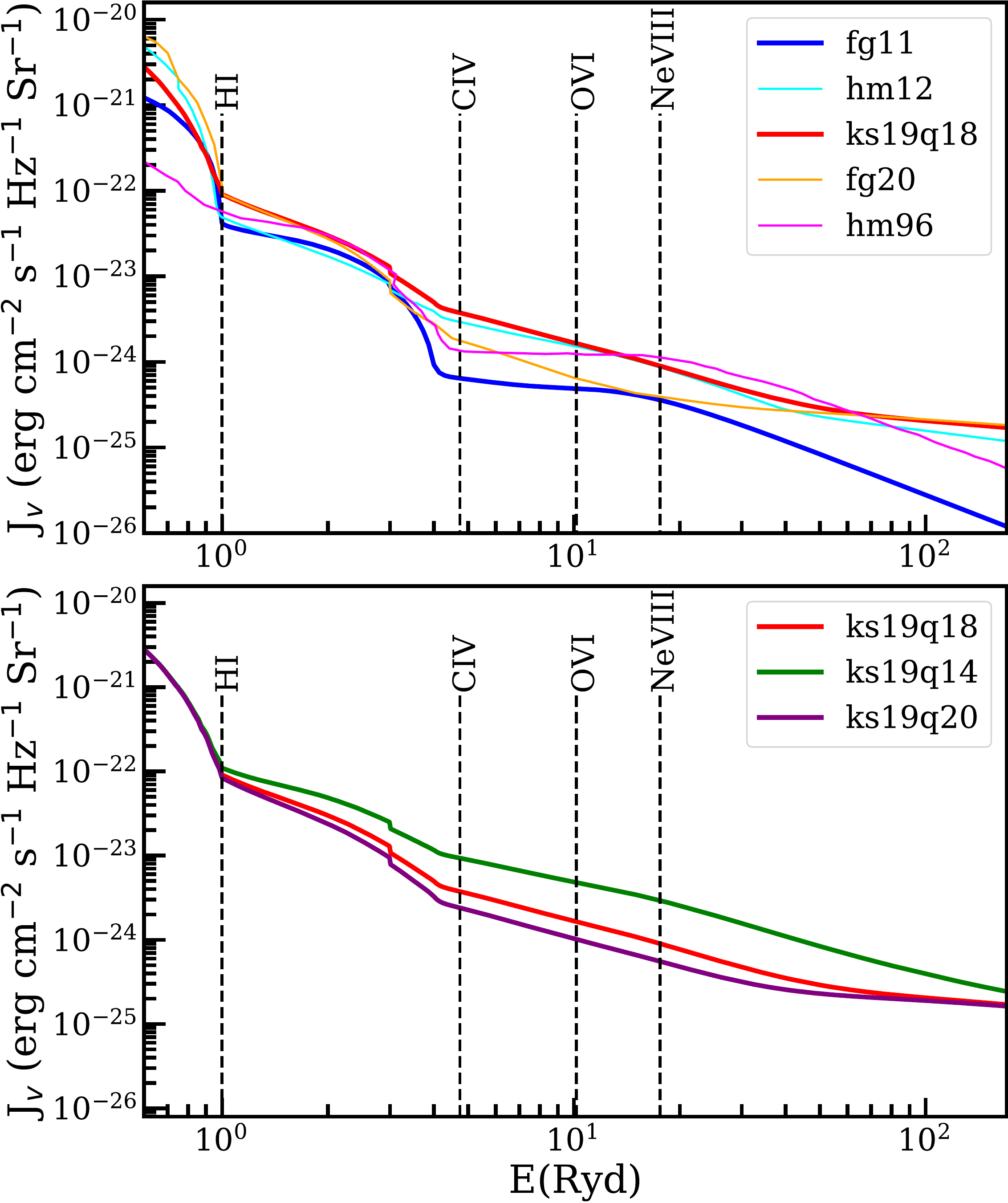}
		\end{center}
\caption{The upper panel shows various frequently used UVBs in the literature for z=0.5. These are from \citet{haardt1996} (denoted as "hm96"), updated version of \citet{faucher2009} UVB published in 2011 (denoted by "fg11"), \citet{haardt2012} (deonoted by "hm12"), \citet{khaire2019} with spectral index of $\alpha=-1.8$ (denoted by "ks19q18") and \citet{faucher2020} (denoted by "fg20"). The lower panel shows the variation of \citet{khaire2019} UVB for the allowed quasar UV to soft-X-ray spectral index range of $\alpha$=$-$1.4 (denoted by "ks19q14") to $\alpha$=$-$2.0 (denoted by "ks19q20"). The photoionization energy of different species are indicated by dashed lines and marked by the name of the species in both panels.
}
\label{fig:UVB}
\end{figure}  

The main aim of our present study is to explore the effect of the assumed UVB on the detectability of different high-ionization metal ions. For this, we consider different sets of UVB that are frequently used in the literature. The UVB affects both the temperature (through radiative heating) and the ionization state of the gas. In this exercise, we assume the influence of the UVB on the gas temperature is negligible (i.e., we do not change the gas temperature of the SPH particle) and focus mainly on  its effects in changing the ionization state of the gas. This assumption is reasonable as we consider gas at low redshifts (i.e., $z\leq0.5$) where the adiabatic expansion cooling dominates over the photo-heating.

In the bottom panel of Figure~\ref{fig:UVB}, we show a range of UVB computed by \cite{khaire2019} (hereafter "ks19") at $z\sim 0.5$ for a range of FUV to soft-X-ray spectral index ($-1.4\le \alpha\le-2.0$) of quasars. We consider UVB obtained for $\alpha=-1.8$ as our fiducial UVB model (denoted by "ks19q18") and also explore two other UVB models where we use UVB obtained for the extreme $\alpha$ values (denoted by "ks19q14" for index $\alpha=-1.4$ and "ks19q20" for  $\alpha=-2.0$). All three UVBs have similar \HI\ photoionization rate \citep[consistent within the range allowed by the low-$z$ \lya\ forest observations, see][]{gaikwad2017a,gaikwad2017b,khaire2019a} but differ appreciably in the high energy ranges.

As mentioned above \citet[]["hm12"]{haardt2012} UVB was used in the Sherwood simulations and \citet[]["hm96"]{haardt1996} UVB in MB-II.  In the top panel of Figure~\ref{fig:UVB} we plot these two
UVBs.
In this panel, along with our fiducial UVB, we also plot the 2011 December update of spectrum in \citet[]["fg11"]{faucher2009} and the UVB from \citet["fg20"]{faucher2020} for comparison. Note, the recent Illustris-TNG simulations use the "fg11" background \citep[see for example][]{Nelson2018}. 
As far as ionization energies relevant for \CIV, \OVI\ and \NeVIII\ (indicated by vertical dashed lines in Figure~\ref{fig:UVB}), the "ks19q18" and "fg11" UVBs roughly  cover the range spanned by the recent UVB models. So we mainly focus on these two models.

\begin{table}
    \centering
   \caption{Photoionization rate of metal ion species considered in this work for different UVB following \citet{shull2014}}
    \begin{tabular}{ ccccc } 
    \hline
    \hline
     & \multicolumn{4}{c}{Photoionization rate (s$^{-1}$) at z=0.5 for} \\
    ion & ks19q14 & ks19q18 & ks19q20 & fg11 \\
    \hline 
    H~{\sc i}   & 3.26E-13   & 2.46E-13 & 2.16E-13 & 1.28E-13  \\
    C~{\sc iv} & 4.31E-15 & 1.62E-15 & 1.02E-15 &  3.49E-16 \\
    O~{\sc vi} & 1.06E-15  & 3.43E-16  & 2.13E-16  &  1.21E-16 \\
    Ne~{\sc viii}& 3.51E-16 & 1.08E-16 & 6.99E-17 & 4.18E-17 \\
    \hline
    \end{tabular}
  
    \label{tab:Gamma_for_uvbs}
\end{table}

Following \citet[]{shull2014}, we calculate the photoionization rates for \HI\ and the  metal ions of our interest for different UVBs. Briefly, the photoionization rate ($\Gamma_{x}$) of a species $x$ is given by,
\begin{equation}
    \Gamma_{x}= 4 \pi \int_{\nu_{x}}^{\infty} \frac{J_{\nu}\sigma_{\nu}}{h {\nu}} \ d\nu ,
\end{equation}
where $I_{\nu}$ is the specific intensity of the radiation field, $\nu_{x}$ and $\sigma_{\nu}$ are threshold frequency for ionization and photoionization cross-section of species $x$ respectively. The photoionization cross sections for different species considered in this work were obtained from Table~1 of \citet[]{verner1996}. The photoionization rates for species considered in this work, viz. H~{\sc i}, C~{\sc iv}, O~{\sc vi} and Ne~{\sc viii} for \citet[]{khaire2019} UVB  with varying spectral index of $-$1.4 and $-$2.0 along with their fiducial model with $\alpha= -1.8$ and the "fg11" UVB model for z=0.5 are tabulated in Table \ref{tab:Gamma_for_uvbs}. We note that the photoionization rate for C~{\sc iv}, O~{\sc vi}, and Ne~{\sc viii} changes by 4.2, 4.9, and 5 times, respectively, between "ks19q20" and "ks19q14" and 4.64, 2.83 and 2.58 times between "ks19q18" and "fg11". As 
the optical depth from a species is inversely proportional to the photoionization rate, we expect stronger absorption signatures for simulations using UVB with the lower photoionization rate.
Therefore, naively we expect simulations using "fg11'' UVB to produce higher column densities of highly ionized ion species discussed above compared to the simulations using any of the "ks19" UVB.  However, the exact optical depth will also depend on the recombination rate that depends on density and temperature.

\subsection{Comparison of simulations in phase distribution}
\label{sec:comparison_among_sims}
\begin{figure*}
\centering
		\includegraphics[width=\textwidth]{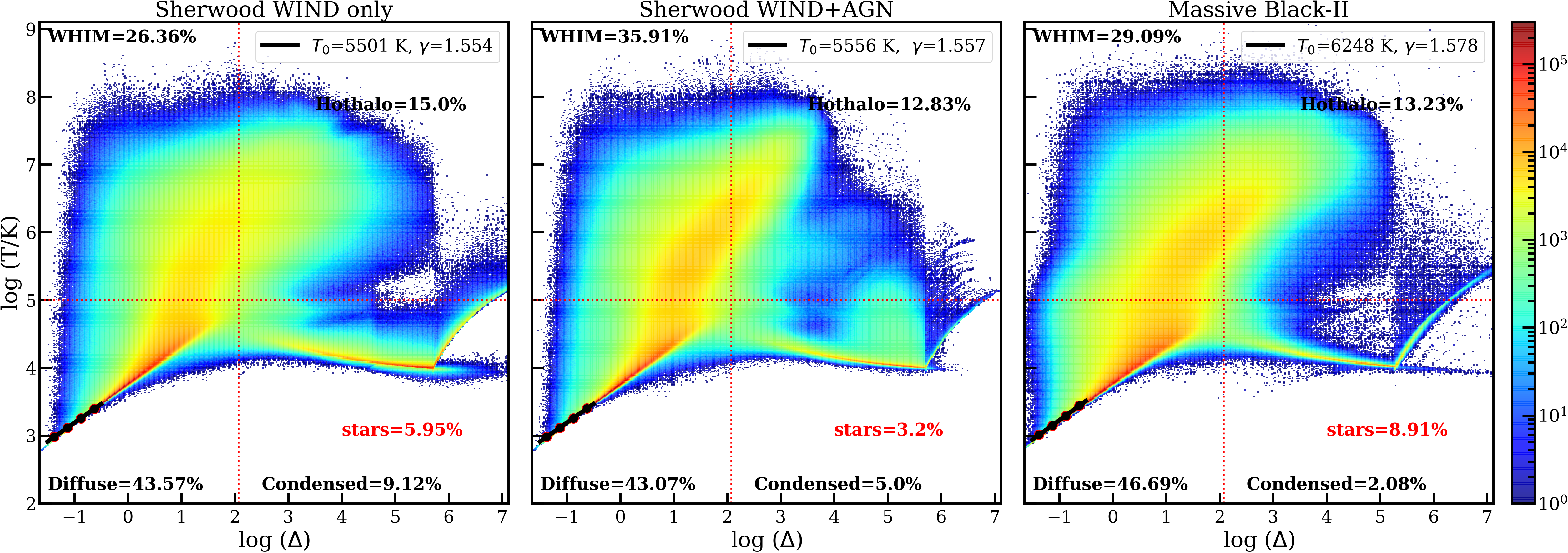}
		\includegraphics[width=\textwidth]{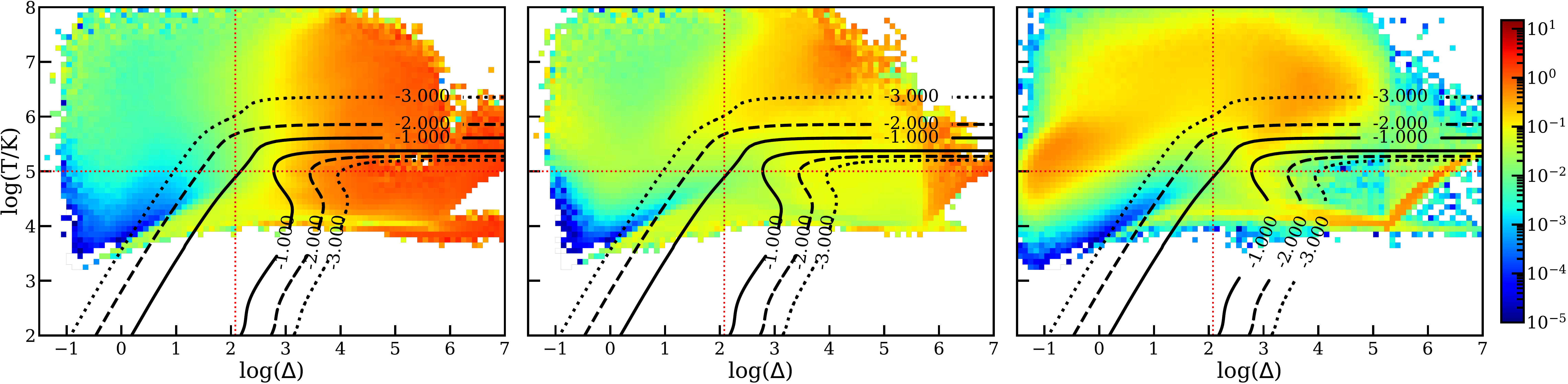}
  \caption{Distribution of the properties of SPH particles in the T-$\Delta$ plane for	three simulations used in our study: i.e., Sherwood "WIND" only simulation (left), Sherwood "WIND+AGN" simulation (middle), and MB-II simulation (right). {\it Upper panels}: The distribution of SPH particle density in $T-\Delta$ plane along with the fraction of particles in Stars, "Diffuse", "WHIM", "Hot-halo", and "Condensed" region as summarized in Table~\ref{table:phase_space}, are shown. The temperature-density relationship for the gas in the "diffuse" phase, as shown in black dots, is obtained by the median temperature within the $\Delta$ bin ranging from $-$1.5 to $-$0.5. The parameters $T_0$ and $\gamma$ of the temperature-density relationship are also provided in each panel. {\it Lower panels}: The mean metallicity of the SPH particles are plotted in $T-\Delta$ plane along with the contours for different O~{\sc vi} ionization fraction logarithmic scale, starting from $10^{-1}$ in solid line, $10^{-2}$ in dashed line and $10^{-3}$ in dotted line. These curves were obtained using the equilibrium photoionization code {\sc cloudy}. }
	\label{fig:phase_plots}
\end{figure*}

The distribution of SPH particles in different locations of temperature (T) - over-density ($\Delta$) plane is influenced by the effects of structure formation and different feedback processes. In the upper panels of Figure~\ref{fig:phase_plots}, we show the density of points (colour coded in logarithmic bins) of SPH particles in the $T-\Delta$ plane for all three simulations at $z=0.5$.
 Following the standard definition \citep[][]{dave2010, gaikwad2017a}, we divide this plot into four phases:
 "Diffuse" (T$< 10^5$K and $\Delta < 120$),  Warm Hot Ionized Medium ("WHIM": T$> 10^5$K and $\Delta <120$), "Hot halo" (T$> 10^5$K and $\Delta> 120$) and "Condensed" (T$< 10^5$K and $\Delta > 120$). The percentage of baryons present in different phases and locked in stars are summarized in Table~\ref{table:phase_space} and indicated in Figure~\ref{fig:phase_plots}. This table also gives the percentages in the case of Sherwood simulations without incorporating any feedback.

While comparing the three Sherwood simulations, we observe that the "WIND" only simulation has $\sim$3.5 times lesser baryon fraction locked in stars compared to the Sherwood Simulation without feedback. This fraction decreases even further (by another factor of 2) when the AGN feedback is included.
Although MB-II has both "WIND" and AGN feedback, the baryon fraction locked in stars is more than that of Sherwood "WIND+AGN" simulation due to differences in the star formation and feedback prescriptions. 
The reduction in the percentage of baryons locked in stars with the introduction of feedback is consistent with what has been found by \citet{dave2010}.

Interestingly the fraction of baryons in the "diffuse" phase (i.e., $\sim$43\%), which is probed by the \lya\ forest seen in the quasar spectra, depends weakly  on the feedback in the case of Sherwood simulations. These values are consistent with 41-43\% found by \citet{dave2010} for models run with a range of feedback prescriptions. \cite{Christiansen2020} reported $16.4-38.8\%$ 
for a range of AGN feedbacks considered in the SIMBA simulations.
They found that the inclusion of jet-feedback considerably reduces the fraction of gas in the "diffuse" phase while enhancing the same in the "WHIM" phase. 
The baryon fraction in the "diffuse" phase is close to 47\% in the case of MB-II, which is slightly higher than other reported values in the literature. 
%
It is well known that the gas in the "diffuse" phase follows temperature-density relation (TDR); $T= T_0 \Delta^{\gamma-1}$. We estimated the TDR of the SPH particles of different simulations from the straight line fit of the median temperature within logarithmic density bins of 0.25, as shown in Figure~\ref{fig:phase_plots} where black dots are the median temperature in corresponding density bins. 
It is evident from Table~\ref{table:phase_space} that the $T_0$ and $\gamma$ parameters do not change by more than 1.5\% between different Sherwood simulations. Thus, the influence of feedback (as implemented in the models considered here) on the $T-\Delta$ relations seems weak. The values of $T_0$ and $\gamma$ obtained in the case of MB-II are different from the Sherwood simulation with "WIND+AGN" feedback by only 12\% and 2.5\%, respectively.

The gas in the WHIM phase is expected to be probed by absorption produced by high ionization species.
It is also evident from Table~\ref{table:phase_space} and Figure~\ref{fig:phase_plots} that the baryon fraction in the "WHIM" phase is affected by different feedback processes. In particular, in the case of Sherwood simulations, the baryon fraction in this phase is increased by $\sim$13\% and $\sim$54\% for simulations incorporating "WIND" only and "WIND+AGN" feedback, respectively. The observed baryon fraction in WHIM found here is consistent with the range 24-33\% found by \citet{dave2010} in their simulations at $z=0$. 
%
\citet{Christiansen2020} reported a baryon fraction in "WHIM" phase spanning $29\%-71\%$ at $z=0$ for variation of AGN feedback in the SIMBA simulation. 
The higher fraction occurs in models with jet mode  AGN feedback included.

\citet{artale2022} reports the gas phase fractions in Illustris TNG simulations, in which the four phases are  defined as: "diffuse" (T$< 10^5$K and $\Delta < 1000$),  warm hot ionized medium (WHIM: $10^5$K$<$T$<10^7$K), hot-halo (T$> 10^7$K) and condensed (T$< 10^5$K and $\Delta > 1000$) respectively. For this definition, the percentage of baryons in the "WHIM" phase in our simulations varies from 38.3$\%$ to 45.6$\%$, which are consistent with  46.6$\%$-33.5$\%$ found by \cite{artale2022} for $z=0-1$. The percentage of baryons  in the "diffuse" phase in our simulations varies between 44$\%$ and 47$\%$  consistent with the range 37.3$\%$-56.4$\%$ found by \citet{artale2022}.

\par

\begin{table}
\centering
\caption{Results of phase space analysis of different simulations }
\setlength{\tabcolsep}{4pt}
\begin{tabular}{ccccc}
\hline

parameters & Sherwood  & Sherwood    & Sherwood    &    Massive Black \\
to         &   No      & "WIND"      & "WIND       &    "WIND      \\
           &           & only        &  +AGN"      &     +AGN"     \\
compare    & feedback  & feedback    & feedback    &    feedback      \\
\hline
\hline
$T_0$      &  5532     & 5501        & 5556        & 6248   \\
$\gamma$   &  1.555    & 1.554       & 1.557       & 1.578  \\
\hline
stars      & $21.50\%$  & $5.95\%$       & $3.20\%$     & $8.91\%$  \\
gas        & $78.50\%$  & $94.05\%$   & $96.80\%$    & $91.09\%$ \\
Diffuse    &$43.31\%$  & $43.57\%$   & $43.07\%$   & $46.69\%$ \\
WHIM       &$23.33\%$  & $26.36\%$   & $35.91\%$   & $29.09\%$ \\
Hothalo    &$11.08\%$  & $15.0\%$   & $12.83\%$   & $13.23\%$ \\ 
Condensed  &$0.79\%$   & $9.12\%$    & $5.00\%$    & $2.08\%$  \\
\hline
    \end{tabular}
    \label{table:phase_space}
\end{table}

\begin{table*}
    \centering
    \caption{Ion fraction and average metallicity in different phases}
     \begin{threeparttable}
   \begin{tabular}{c|c|cc|cc|cc}
     \hline
Ion &    \multicolumn{1}{c|}{Gas phases} & \multicolumn{2}{c|}{Sherwood WIND} & \multicolumn{2}{c|}{Sherwood WIND} & \multicolumn{2}{c}{MB-II WIND}\\
 Species &   \multicolumn{1}{c|}{to compare} & \multicolumn{2}{c|}{only feedback} & \multicolumn{2}{c|}{AGN feedback}     & \multicolumn{2}{c}{AGN feedback}\\
    \cline{3-8}
   &   & f1{\tnote{a}}   
      & $<Z/Z_\odot>${\tnote{b}}   & f1{\tnote{a}}   & $<Z/Z_\odot>${\tnote{b}}  & f1{\tnote{a}}  &   $<Z/Z_\odot>${\tnote{b}} \\
    \hline
\multicolumn{8}{c}{For "ks19q18" UVB}\\
 \OVI\   & Diffuse        & 4.83$\%$ & 0.0057 &  4.77$\%$ &  0.0096  &  5.46$\%$ &  0.0052   \\
    & WHIM           & 2.06$\%$ & 0.0127 &  2.19$\%$ & 0.0405  &  1.84$\%$ &  0.0324   \\
    & Hot halo       & 2.45$\%$ & 0.0825 &  1.23$\%$ &  0.0595  &  0.94$\%$ &  0.0818   \\ 
    & Condensed      & 1.96$\%$ & 0.1222 &  1.31$\%$ &  0.0322  &  0.93$\%$ &  0.0530   \\
 \\   
    \CIV\ & Diffuse        & 0.02$\%$  &   0.0659 &  0.02$\%$ &  0.0465  &  0.02$\%$ &  0.0459   \\
    &WHIM           & 0.00$\%$     &NA     &  0$\%$    &  NA      &  0$\%$  & NA   \\
    &Hot halo       & 0.16$\%$    &0.1465 &   0.17$\%$&  0.0617  &  0.03$\%$ &  0.1209   \\ 
    &Condensed      & 4.11$\%$    &0.1893 &  2.72$\%$ &  0.0379  & 1.43 $\%$ &  0.1396   \\
    
\\    
 \NeVIII &       Diffuse        & 18.78$\%$  &    0.0019&  18.55$\%$ &  0.0066  &  20.28$\%$ &  0.0025  \\
   & WHIM           & 19.70$\%$  &    0.0062&  23.59$\%$ &  0.0316  &   20.85$\%$ &  0.0265  \\
  &  Hot halo       & 5.93$\%$   &    0.1025&  3.45$\%$  &  0.0965  &   3.71$\%$ &   0.1387 \\ 
   & Condensed      & 0.76$\%$   &    0.0625&  0.63$\%$  &  0.0272  &   0.51$\%$ &   0.0324 \\

\multicolumn{8}{c}{For "fg11" UVB}\\
\OVI\    & Diffuse        & 18.32$\%$ & 0.0019  & 18.09$\%$ & 0.0065  & 19.74$\%$ & 0.0024  \\
    & WHIM           & 7.63$\%$  & 0.0087  & 10.59$\%$ & 0.0450  & 7.86$\%$  & 0.0441  \\
    & Hot halo       & 2.82$\%$  & 0.0748  & 1.55$\%$  & 0.0663  & 1.29$\%$  &0.0854  \\ 
    & Condensed      & 1.29$\%$  & 0.0884  & 0.96$\%$  & 0.0305  & 0.73$\%$  & 0.0411  \\
\\
 
 \CIV\   & Diffuse        & 0.88$\%$ &  0.0195  & 0.86$\%$  & 0.0173  & 0.95$\%$  & 0.0138  \\
    & WHIM           & 0.12$\%$ &  0.0163  & 0.11$\%$  & 0.0242  & 0.09$\%$  &  0.0198  \\
    & Hot halo       & 0.58$\%$ &  0.0587  & 0.42$\%$  & 0.0335  & 0.23$\%$  & 0.0485  \\ 
    &Condensed      & 3.41$\%$ &  0.1783  & 2.17$\%$  & 0.0348  & 1.32$\%$  & 0.0932  \\    
    \\
\NeVIII &       Diffuse        & 21.81$\%$ &   0.0018 & 21.95$\%$  & 0.0076 & 26.26$\%$  &  0.0046 \\
    & WHIM           & 43.20$\%$ &   0.0044 & 50.78$\%$  & 0.0268 & 47.31$\%$  &  0.0338 \\
    & Hot halo       & 5.90$\%$  &   0.1025 & 3.97$\%$  & 0.1052 & 4.26$\%$  &  0.1453 \\ 
    & Condensed      & 0.24$\%$  &   0.0397 & 0.22$\%$  & 0.0227 & 0.19$\%$  &  0.0234 \\
    \hline
    \end{tabular}
    \label{table:detectability}
\begin{tablenotes}
\item[a] Percentage of SPH particles having $f_{Ion}>0.01$ in different gas phases among all particles. 
\item[b] Average metallicity of the particles having $f_{Ion}>0.01$ in different gas phases.
\end{tablenotes}
     \end{threeparttable}
\end{table*}



\begin{figure}
	    \includegraphics[viewport=0 30 750 580,width=0.5\textwidth,clip=true]{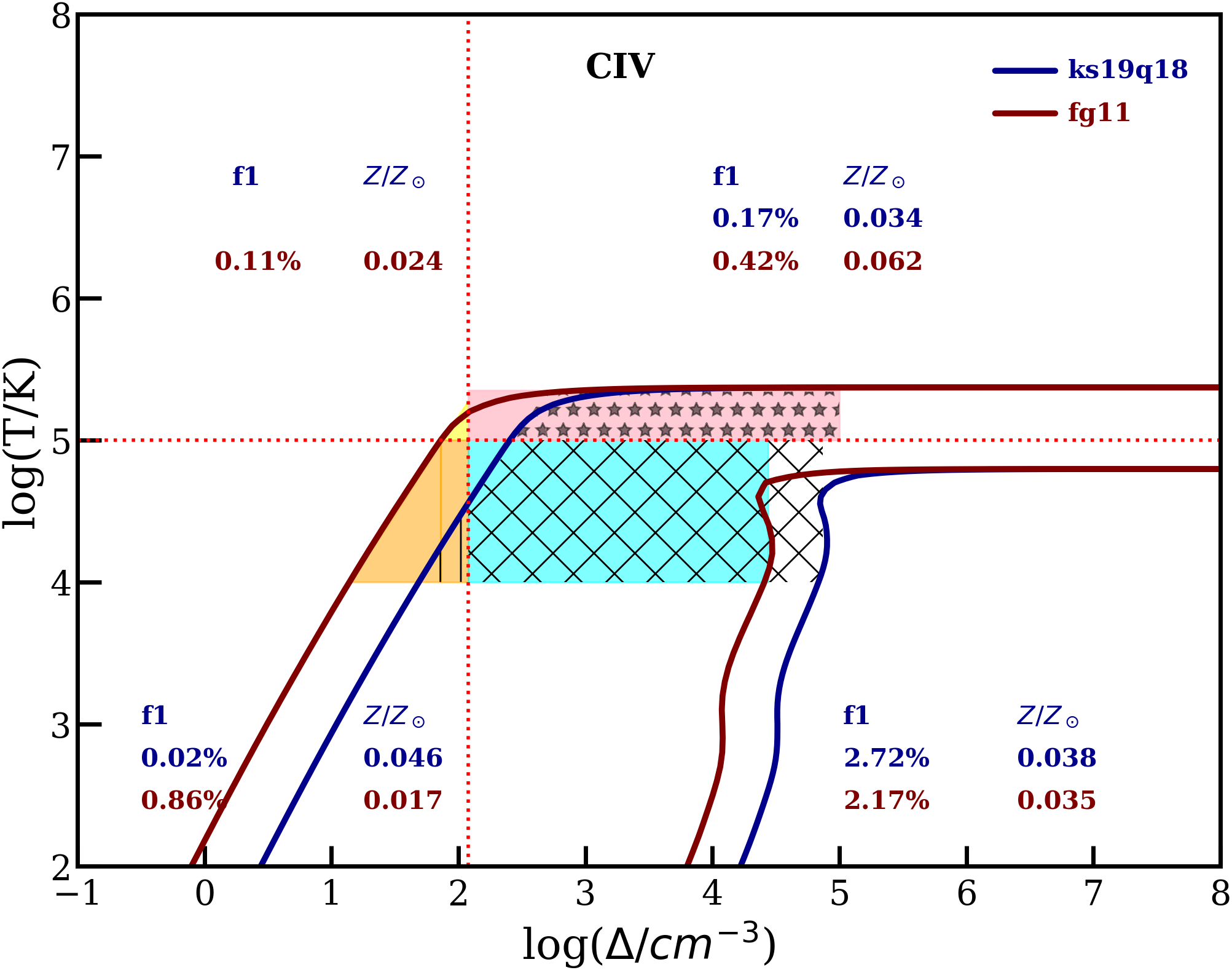}
	    \includegraphics[viewport=0 30 750 580,width=0.5\textwidth,clip=true]{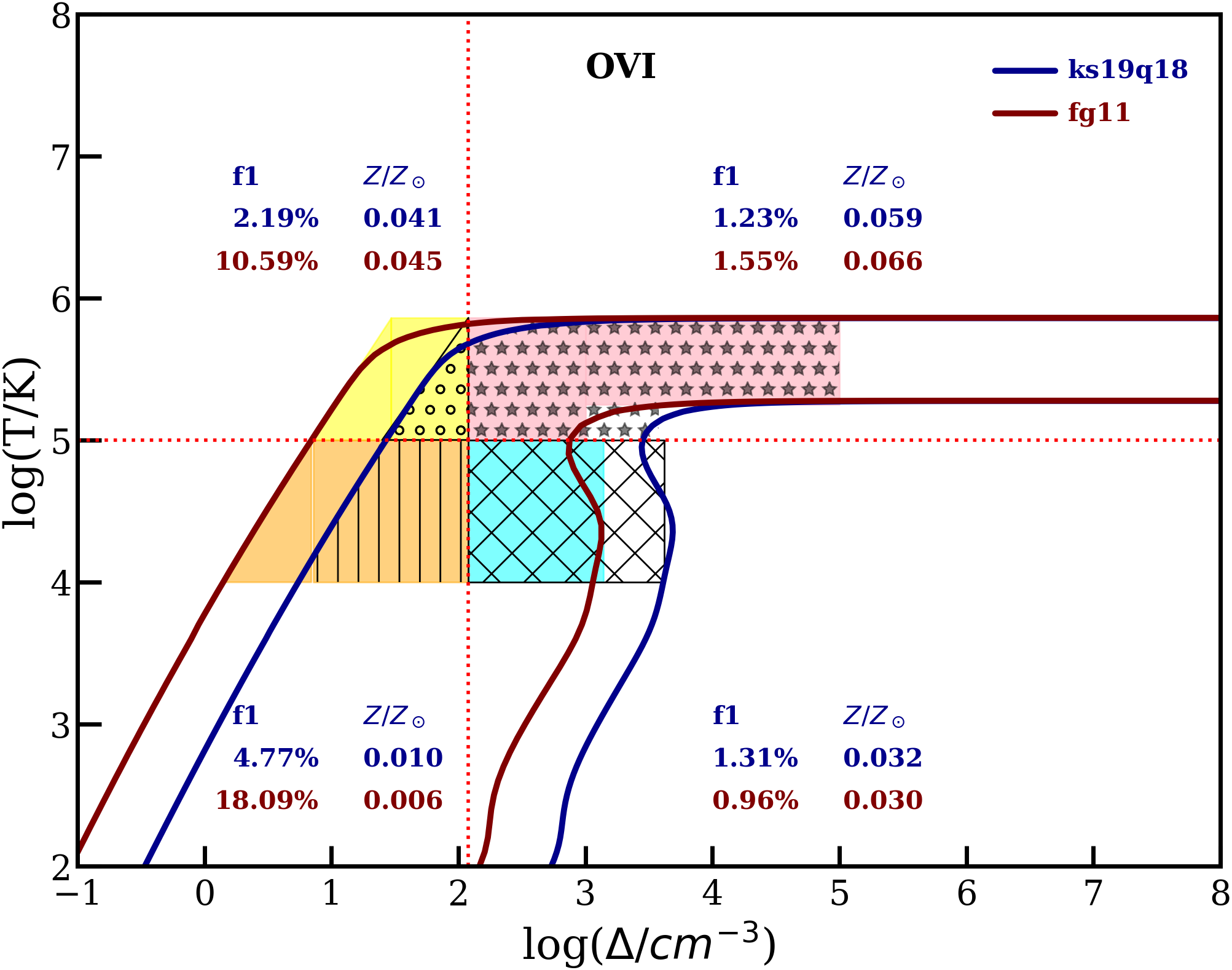}

	    \includegraphics[viewport=0 2 750 580,width=0.5\textwidth,clip=true]{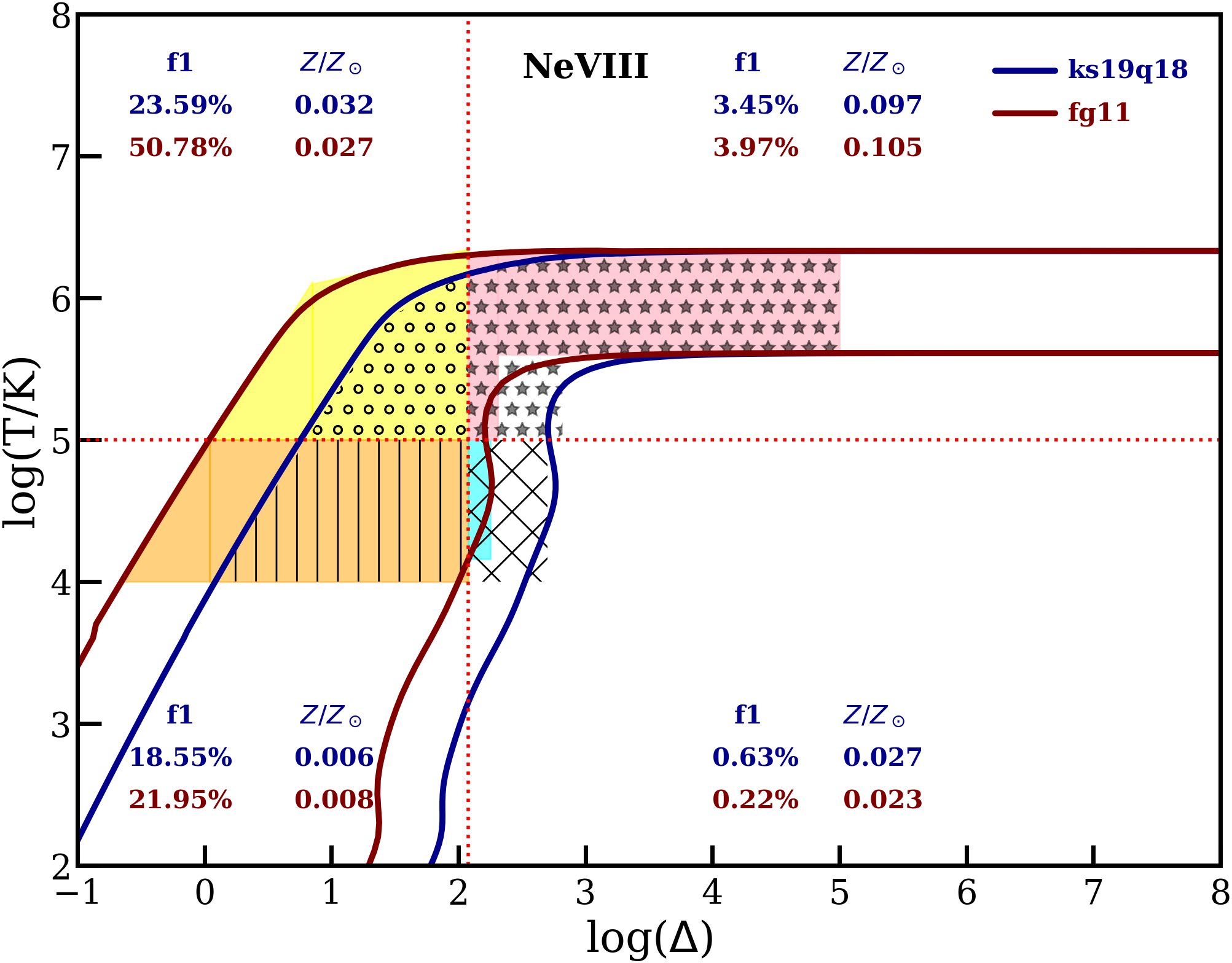}
\caption{{\it Top panel:}The contours for \CIV\ ionization fraction, $f_{\CIV}=0.01$ obtained using the photoionization code {\sc cloudy} are shown for "ks19q18" (blue) and "fg11" (dark red) UVBs.  
The regions having $f_{\CIV}>0.01$ in "WHIM", "diffuse", "condensed", and "hot halo" phases for the "ks19q18" UVB are marked with circles, vertical lines, hatches, and stars, respectively. The corresponding regions for the "fg11" UVB are colored yellow, orange, cyan, and pink, respectively. Percentage of particles ($f_1$) 
having  $f_{\CIV}>0.01$ and their average metallicity in these regions are mentioned for Sherwood "WIND+AGN" simulation, and the same quantities for all three simulations are summarized in Table~\ref{table:detectability}.
{\it Middle and bottom panel:} Same as top panel but for \OVI\ and \NeVIII\ . 
}
	\label{fig:detectability}
\end{figure}

The detectability of a given metal ion depends on the metallicity, fraction of volume covered by the metals, and ionization state of the gas. Feedback processes influence all these three factors. In the lower panel of Figure~\ref{fig:phase_plots} we plot the metallicity distribution in the T-$\Delta$ bins for all three simulations. The color scheme shows the metallicity in the units of solar metallicity on the logarithmic scale. In this figure, we also show the regions where the fraction of \OVI\  (i.e., $f_\OVI$ = $N$(\OVI)/N(O)) is more than 0.1(between the solid curves), 0.01(between dashed curves), and 0.001 (between the dotted curves) for our fiducial UVB "ks19q18". These curves were generated using the ion fraction in each T-$\Delta$ grid point computed using the photoionization code {\sc cloudy} \citep[version 17.02 of the code developed by][]{cloudy1998} assuming optically thin ionization conditions. It is evident  that the ion fraction of \OVI\ peaks in a restricted range in the $T-\Delta$ plane. To quantify the detectability of an ion in different simulations, we calculate the baryon fraction and average metallicity in different gas phases 
with ion fraction, $f_{Ion}>0.01$, for two different UVBs (i.e "ks19q18" and "fg11"). 
Results are summarized in Figure~\ref{fig:detectability}.
In Table~\ref{table:detectability}, for each simulation (and for two UVBs) we provide the percentage of SPH particles having $f_{Ion}>0.01$ in different phases (denoted as $f_1$), 
and average metallicity (Z/Z$_\odot$) of the particle having  $f_{Ion}>0.01$. 
We notice that the average metallicity is higher for gas with higher density. 

\vskip 0.1in
\noindent{\bf For \OVI:}
First we consider simulations using "ks19q18" UVB.  From Table~\ref{fig:detectability}, it is evident that only 11.3\%, 9.5\%, and 9.2\% of the total SPH particles  have  $f_{\OVI}>0.01$ in WIND, "WIND+AGN" Sherwood simulations, and MB-II simulation respectively. It is also evident that the percentage of SPH particles with  $f_{\OVI}>0.01$ in the "diffuse" ($\sim 4.8$\%) and "WHIM" ($\sim2.1$\%) phases are nearly the same in the "WIND" and "WIND+AGN" Sherwood simulations. 
However, we find the metallicity of particles with $f_{\OVI}>0.01$  in "diffuse" and "WHIM" phases are enhanced by a factor $\sim$1.7 and 3.2 in the "WIND+AGN" simulation compared to the "WIND" only simulation.  On the other hand, we find "WIND" only simulation has a higher fraction of baryons in the "hot halo" and "condensed" phases compared to the "WIND+AGN" Sherwood simulation. Also, the average metallicity of particles with $f_{\OVI}>0.01$  is high in the case of "WIND" only simulation in these two phases.  
\textit{Therefore, we expect detectable \OVI\ absorption  also from both diffuse and WHIM phases in the case "WIND+AGN" Sherwood simulations.  On the other hand, \OVI\ absorption will predominantly detected from the "hot halo" or "condensed" phase, in the "WIND" only simulation. }

While the MB-II simulation has similar (within 15\%) SPH fractions in the "diffuse" and "WHIM" phases compared to the "WIND+AGN" Sherwood simulation, they tend to have lower (i.e. $>26$\%) metallicity for particles with $f_{\OVI}>0.01$. On the other hand in, the "hot halo" and "condensed" phases of the MB-II simulation has higher metallicity (by $>$39\%) compared to those in  the Sherwood simulation incorporating "WIND+AGN" feedback.

From the middle panel in  Figure~\ref{fig:detectability} we notice that when we use "fg11" UVB instead of "ks19q18" UVB, the contours for a given $f_{\OVI}$ move towards lower density (as expected from the discussion presented in the previous section)
in all phases except in the "hot-halo" phase, where the collisional excitation is expected to dominate.  Therefore, we find the values of $f_1$ for the "diffuse" and "WHIM'" phases to be higher in the case of "fg11" UVB by a factor $\ge$3.6 in all three simulations.  The average metallicity of SPH particle with $f_{\OVI}>0.01$ in the "diffuse" phase is found to decrease by a factor 3, 1.5, and 1.16 in the case of "WIND" only, "WIND+AGN" Sherwood simulations and MB-II simulation respectively. However, in the case of "WHIM" the corresponding changes in the average metallicity of such SPH particles are by a factor of 1.45, 0.90, and 0.73 for the three simulations. As expected, $f_1$ in the "condensed" phase decreases when we use the "fg11" UVB. We also notice a reduction in the average metallicity (of SPH particles with $f_{\OVI}>0.01$)  in these cases.

\vskip 0.1in
\noindent{\bf For \CIV:}
Next, we explore the  detectability of \CIV\ ion in different models using plots shown in the top panel of Figure~\ref{fig:detectability} and results summarized in Table~\ref{table:detectability}. 
For simulations using "ks19q18" UVB, we find only 4.3$\%$, 2.9$\%$ and 1.5$\%$ of all SPH particles have C~{\sc iv} ionization fraction, $f_{\CIV}>0.01$, in WIND, "WIND+AGN" Sherwood simulations, and MB-II simulation respectively. The fraction $f_1$ in the "diffuse" phase is nearly 0.02$\%$ in all three simulations for "ks19q18" UVB. The mean metallicity for particles with $f_\CIV>0.01$ in the Sherwood "WIND" simulation is 42-44$\%$ higher than the Sherwood "WIND+AGN" and MB-II simulations. The \CIV\ ionization fraction is  found to be always less than 0.01 in the "WHIM" phase for "ks19q18" UVB.  Therefore, unlike \OVI,  the \CIV\ absorption in simulations considered here will mainly originate from high-density regions, predominantly in the "condensed" gas phase. The Sherwood "WIND" model has 1.5 times higher $f_1$  and 5 times more mean metallicity in the "condensed" phase compared to the Sherwood "WIND+AGN" simulation. The SPH fraction in the "hot halo" phase is similar in Sherwood "WIND" and "WIND+AGN" simulations, although the mean metallicity of particles with $f_\CIV>0.01$ is 2.4 times higher in the "WIND" simulation. The MB-II simulation has 2.9 times less $f_1$ and 35\% less mean metallicity for particles with $f_\CIV>0.01$ in the "condensed" phase 
compared to those of the Sherwood "WIND" simulation. Although the mean metallicity in "hot halo" is similar to that of Sherwood "WIND" only simulation, the SPH fraction is lesser by $>5$ times. Considering all these, we expect to detect stronger \CIV\ absorption in the Sherwood "WIND" simulation compared to that in "WIND+AGN" and MB-II simulations.


Using the "fg11" instead of "ks19q18" UVB increases the fraction $f_1$ in all phases except in the "condensed" phase. We also notice the average metallicity of particles with $f_\CIV>0.01$ to decrease in the case of "fg11" UVB (see Figure~\ref{fig:detectability}). The $f_1$ in the "diffuse" phase for all three simulations are very similar ($\leq$10$\%$). Sherwood "WIND" simulation has 13 and 41$\%$ higher metallicity  for particles with $f_\CIV>0.01$ in the "diffuse" phase compared to the same in Sherwood "WIND+AGN" and MB-II simulations, respectively. Unlike our fiducial UVB, a small fraction of particles in the "WHIM" phase also have   $f_\CIV>0.01$.  
This additional contribution from the "WHIM" phase in the case of the "fg11" UVB is expected to increase the detection of \CIV\ absorption in the low column densities. The fraction $f_1$ in the "hot halo" phase increases by 3.6, 2.5, and 7.7 times in the case of "fg11" UVB compared to the "ks19q18" UVB for Sherwood WIND, WIND+AGN, and MB-II simulations respectively. The average metallicity of the particle with $f_\CIV>0.01$ decreases by 2.5, 1.8, and 2.5 times for these simulations. The "condensed" phase has 21$\%$, 25$\%$, and 8$\%$ lesser $f_1$ fraction as well as 6-8$\%$ lesser mean metallicity (for particles with $f_\CIV>0.01$) for Sherwood WIND, WIND+AGN, and MB-II simulations. This could imply lesser C~{\sc iv} detectability in the "condensed" phase for "fg11" compared to "ks19q18" UVB. \textit{Therefore, we expect more \CIV\ detection in Sherwood "WIND" simulation compared to the other two simulations in all phases (except for the "WHIM" phase) with "fg11" UVB and enhanced (reduced) number of C~{\sc iv} absorbers in lower (higher) column densities for all three simulations.}

\vskip 0.1in
\noindent{\bf For \NeVIII:} In the bottom panel in Figure~
\ref{fig:detectability} we show the ion fraction contour for \NeVIII\ in the $T-\Delta$ plane.
Details of the fraction of SPH particles having  \NeVIII\ ion fraction ($f_\NeVIII$) more than 0.01 are also summarized in Table~\ref{table:detectability}.
When we consider the "ks19q18" UVB, we find roughly $\sim$45\% of the SPH particle have $f_\NeVIII >0.01$.
It is also clear that nearly 85\% of these particles are in the "diffuse" and "WHIM" phases. The average metallicity of such particles in the "diffuse" phase is found to be very low ($<$0.7\% of the solar value).  In the case of the WHIM phase, the "WIND+AGN" model tends to have higher values of $f_1$ and average metallicity. In the case of MB-II simulations, we find the "diffuse" phase to have a higher $f_1$ and the "WHIM" phase to have a slightly lower $f_1$ compared to the corresponding values for the "WIND+AGN" Sherwood simulation. The mean metallicity is found to be always higher in the "WIND+AGN" Sherwood simulation for the two phases discussed here.
We find the $f_1$ values to be 6\%, 3.5\%, and 3.7\%, respectively, for the WIND, "WIND+AGN" Sherwood simulations, and MB-II simulation, respectively for the "hot halo" phase. These particles also tend to have large metallicities (i.e., $\sim$0.1 solar values).  

When we consider simulations using "fg11" UVB, we find that $f_1$ values increase for the "diffuse" and "WHIM" phase without a steep increase in the average metallicity. However, in the "hot halo" phase, both $f_1$ (i.e., change is $\le 13$\%) and average metallicity of the gas with $f_\NeVIII>0.01$ do not depend strongly  on the UVB. \textit{This implies that if most of the high column density \NeVIII\ absorption originates from the "hot halo" phase, then the number of these systems and column density may not strongly depend on the UVB. On the other hand, if they originate from the "diffuse" or "WHIM" regions and the SNR of the spectra are good enough to detect low metallicity systems, the effect of UVB will be appreciable.}

\vskip 0.1in
All this suggests that in the simulations considered here, the effect of feedback (for a given assumed UVB) may be similar to the effect of the assumed radiation field (for a given simulation). It is also apparent that the effect of the radiation field is not as simple as one infers from single cloud photoionization models.
Therefore, to quantify the effect of changing UVB, it is important to compare various observables like distributions of column density and velocity width of absorbers. This is what we do in the following by analyzing the spectra generated from the simulations in the same way one analyses the observed spectrum.

\subsection{Simulated spectra and absorption line parameters}
\label{sec:sightline_generation}

\begin{figure*}
\centering
		\includegraphics[width=\textwidth]{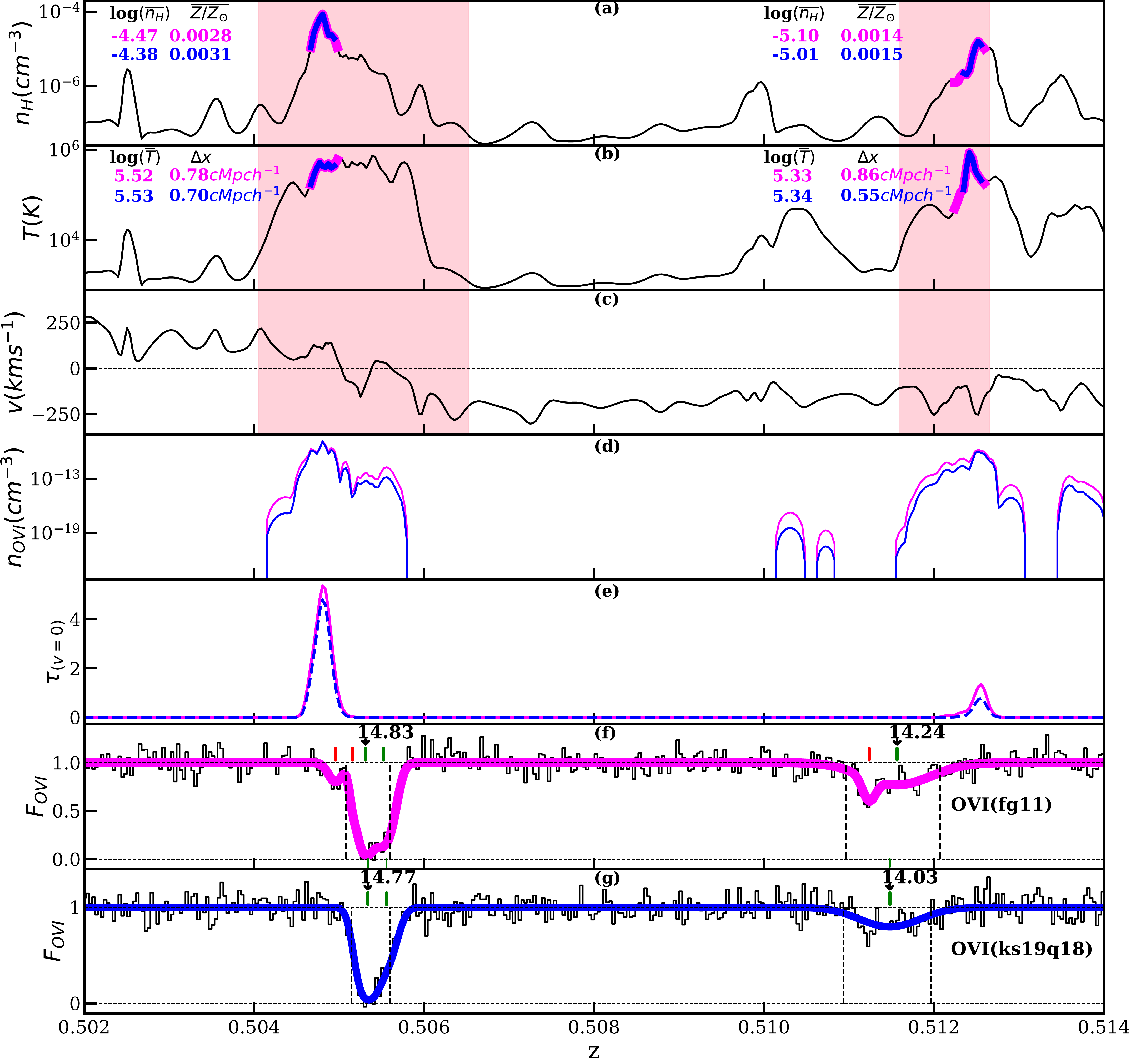}
		\caption{
		Panels (a), (b), (c), and (d) show hydrogen number density ($n_H$), temperature ($T$), peculiar velocity ($v$) field, and the \OVI\ number density field ($n_{OVI}$) along a sample line of sight from our "WIND+AGN" Sherwood simulations at $z\sim0.5$. Panel (e) shows the  optical depth profile generated for the  \OVI$\lambda$1031.92 transition assuming zero peculiar velocity and using ionization corrections obtained for the "fg11" (magenta)  and "ks19q18" (blue) UVBs. As expected the optical depth is more when we use "fg11" UVB. Panels "f" and "g" show the simulated normalised spectra (black) when we use "fg11" and "ks19q18" UVBs respectively, with the best-fitted Voigt profiles using {\sc viper} over-plotted. We also provide the total \OVI\ column density (in log units) measured for each system. Shaded regions in panels (a), (b) and (c) indicate grid points with correct peculiar velocity that can in principle contribute to the \OVI\ absorption seen in panel (f) and (g). 
  However, only a small fraction of them contribute substantial optical depth to the \OVI\ absorption. These are indicated with thick curves (magenta for "fg11"  UVB and blue for "ks19q18" UVB) in panels (a) and (b). Individual components are indicated using ticks (green for RSL$>4$ and red for RSL$<4$). These panels also provide the average density, temperature, metallicity, and comoving length ($\Delta x$) of the absorbing region. It is clear that these quantities are sensitive to the assumed UVB. The vertical dotted lines in the bottom two panels show the regions used to calculate $\Delta V_{90}$ and the arrows give the redshift of the system (assumed to be that of the strongest individual component).}
		\label{fig:demo_los}
\end{figure*}

\begin{figure*}
	\begin{minipage}{\textwidth}
		\includegraphics[width=\textwidth, trim={12cm 0 5cm 0},clip]{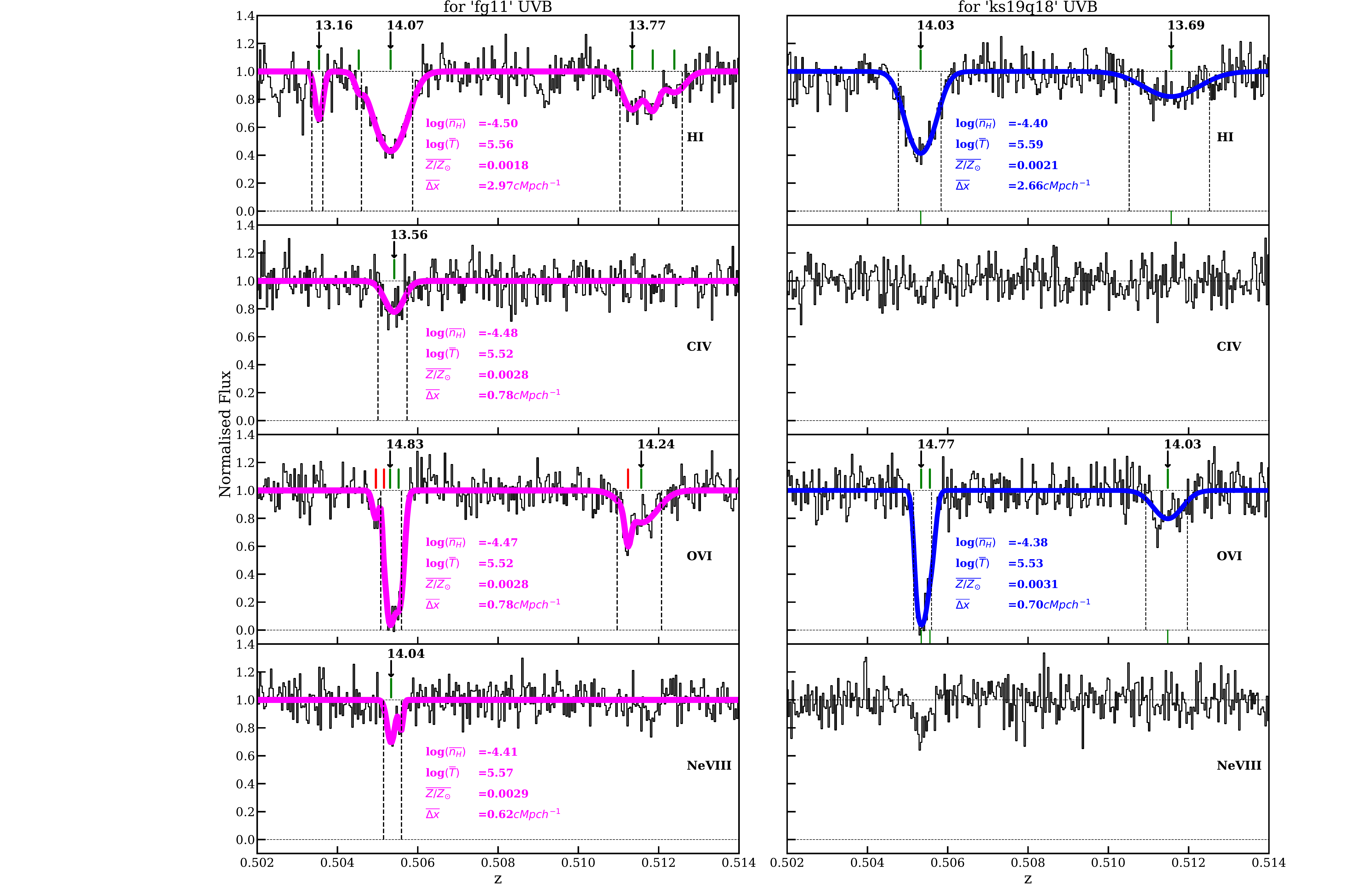}
		\caption{
  An example of a simulated absorption system showing all the ions of our interest in this work. 
  The spectra shown in the left and right columns were obtained using "fg11" and "ks19q18" UVBs respectively from the same line of sight.
        In each panel, we also summarize the mean density, temperature, and metallicity of the region of comoving length ($\Delta x$) contributing to the absorption. The vertical dashed lines mark the region used to measure $\Delta V_{90}$. It is clear that in this example even though the \lya\ and metal line absorption are aligned in velocity space they sample widely different regions. However, the average quantities are similar for the metal line transitions in this particular case. 
		}
		\label{fig:demo_los2}
	\end{minipage}%
\end{figure*}


The simulation outputs contain information such as mass, density, velocity, internal energy, metallicity, and electron abundance for all SPH particles. For calculating the hydrogen number density($n_H$), we account for the presence of helium at about 24$\%$ of the total baryonic mass of the Universe and ignore the metal contribution to the total mass. Assuming solar abundance ratio, we estimated the number density of different metal species for each SPH particle using its hydrogen number density and metallicity. We have considered \lya\ transition and only the strongest metal transitions of C~{\sc iv}$\lambda1548.20$, O~{\sc vi}$\lambda1031.92$ and Ne~{\sc viii}$\lambda770.41$ doublets in the present study. For simplicity, we do not simulate the doublets and mimic contamination from other absorbers along the line of sight in our simulated spectra. 
We obtained the number density of each ion species mentioned above for each SPH particle considering the temperature, density, metallicity and using 
{\sc cloudy} for a given UVB assuming the ionization equilibrium and optically thin conditions.

Following SPH formalism described in \citet[]{monaghan1992} and \cite[]{springeldematteo2005}, we assign density, temperature, velocity, and metal ion number density to each uniformly spaced grid point along the sightline drawn in any one direction of the simulation box. The estimated value of any quantity \textit{f} at any grid point \textit{i} in the SPH formalism is given by,
\begin{equation}
    f_i = \sum_j f_j \frac{m_j}{\rho_j} S_{ij}
\end{equation}
where the summation is performed over all particles ($j$). The quantities $m_j$, $\rho_j$, and $f_j$ are the mass, density, and value of the quantity $f$ of the $j^{th}$ particle respectively. The quantity $f$ could be overdensity($\Delta$), internal energy per unit mass ($u$), or any component of velocity (v). The smoothing kernel $S_{ij}$ is taken from \citet[]{monaghan1992}. 
The temperature is derived from the internal energy using,
\begin{equation}
    T= \frac{8m_H}{3k_B[(4+4N_e)-4YN_e-3Y]} \: u
\end{equation} 
where $m_H$ is the mass of proton, $k_B$ is Boltzmann constant, $Y$ is the mass fraction of helium abundance, $N_e$ is the electron abundance. 
We consider 1024 grid points in the case of Sherwood simulations and 2048 grid points in the case of MB-II simulation while generating the spectra.


The number density of neutral hydrogen and metal ions, temperature, velocity, and metallicity values at each grid point is used to calculate the optical depth profile along the line of sight as illustrated in Figure~\ref{fig:demo_los}. The hydrogen number density ($n_H$), temperature ($T$), and peculiar velocity field (v) along a sightline are shown in the upper three panels in Figure~\ref{fig:demo_los}. We use the $n_{\hi}$ obtained after appropriate ionization corrections, 
temperature (panel b), and velocity fields (panel c) to generate the absorption spectrum using equation 30 in \citet{choudhury2001}. In the case of metal ions, we use
the ion density field (see panel d for \OVI\ ion density field) together with the velocity and temperature field to generate the absorption spectrum of that ion. We convolve the spectra with a Gaussian profile of FWHM=17 $km s^{-1}$ (which is similar to the spectral resolution achieved with HST/COS) instead of HST/COS line spread function. Further, we add Gaussian noise corresponding to SNR=10 or 30 per pixel to obtain the spectra. 

In panel (e) of Figure~\ref{fig:demo_los}, we show the optical depth profile of \OVI$\lambda$1032 generated without applying the peculiar velocities for two different UVBs.
As expected, the optical depth when we use "fg11" UVB is found to be higher than the same for the "ks19q18" UVB.
The final simulated spectra (including the effects of peculiar velocity) for "fg11" and "ks19q18" UVBs are shown in panels "f" and "g" respectively. Clearly,  there are subtle changes in the absorption profiles when we use different UVB. To understand this, we use the
optical depth profile without including the effect of peculiar velocity (as shown in panel e of Figure~\ref{fig:demo_los}) to identify the grid points that contribute to a given absorption profile. As an example, in panels (a) and (b), we indicate the grid points that contribute to the \OVI\ absorption in the case of "fg11" and "ks19q18" UVBs with thick magenta and blue regions, respectively. {\it This clearly demonstrates that the effects of changing the UVB are equivalent to changing the density, velocity, temperature ranges,  and length scale of regions contributing to a given absorption system. }

\par

\begin{figure*}
	\begin{minipage}{0.3\textwidth}
		\includegraphics[width=5.3cm]{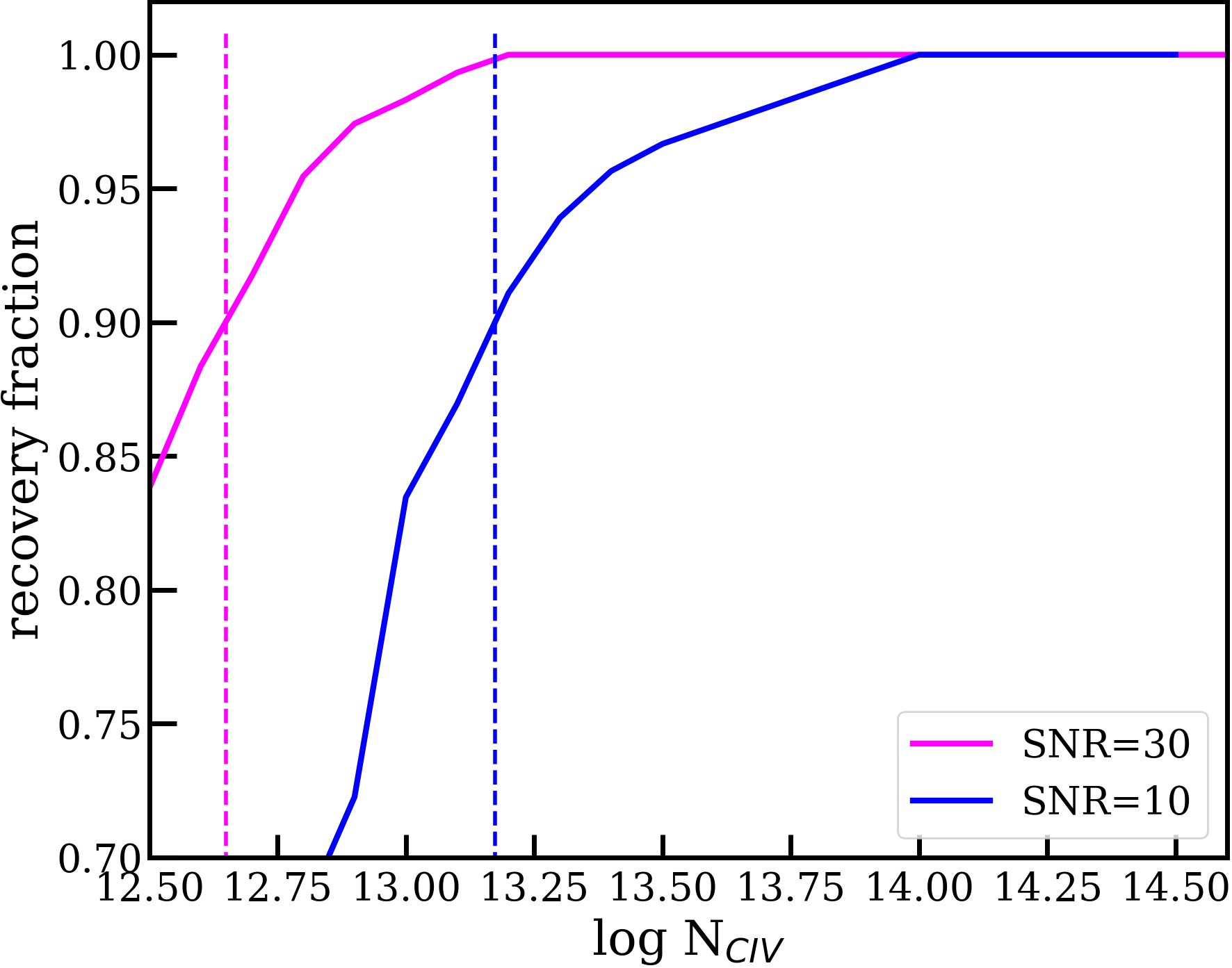}
	\end{minipage}%
	\begin{minipage}{0.3\textwidth}
	    \includegraphics[width=5.3cm]{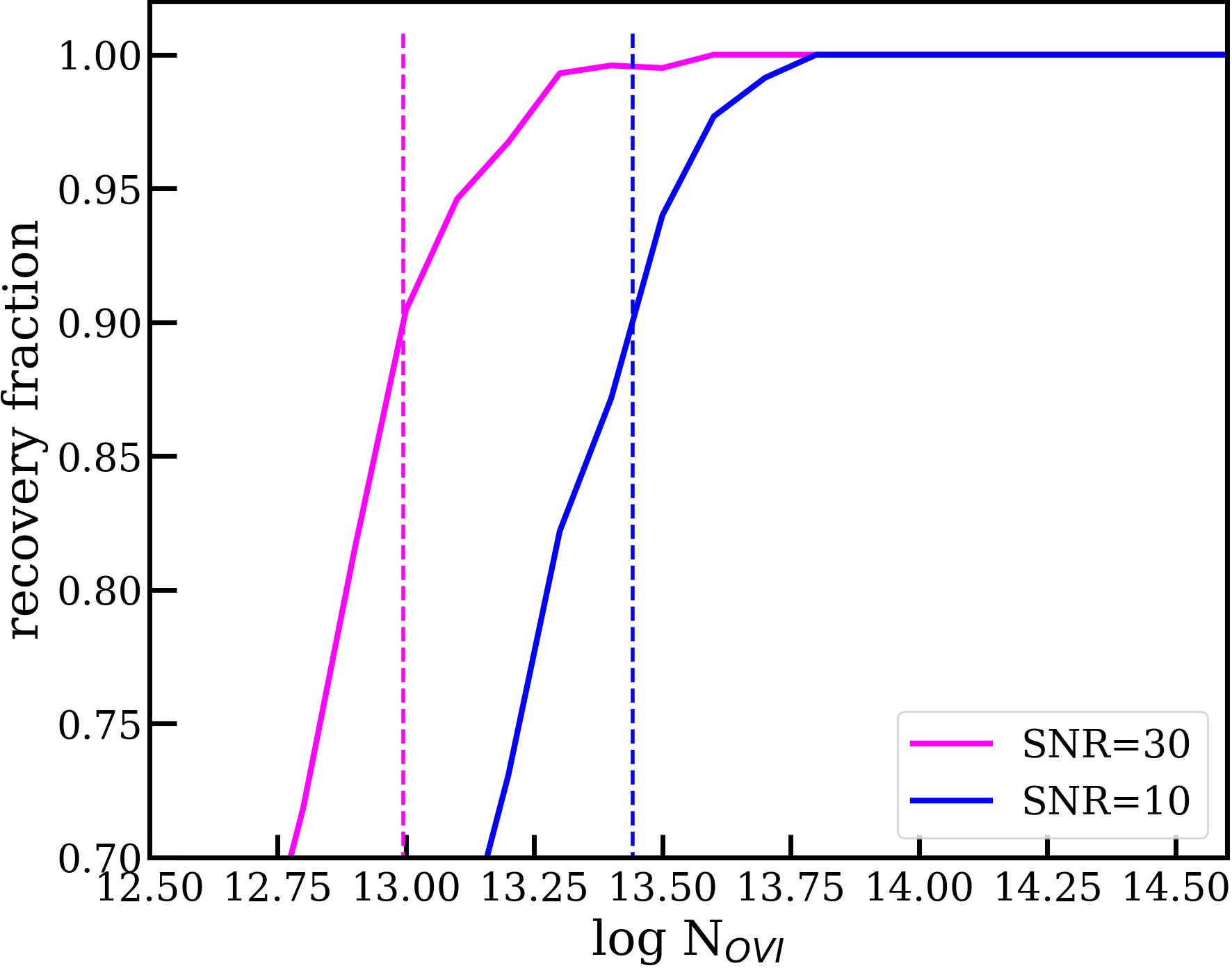}
	\end{minipage}
	\begin{minipage}{0.3\textwidth}
		\includegraphics[width=5.3cm]{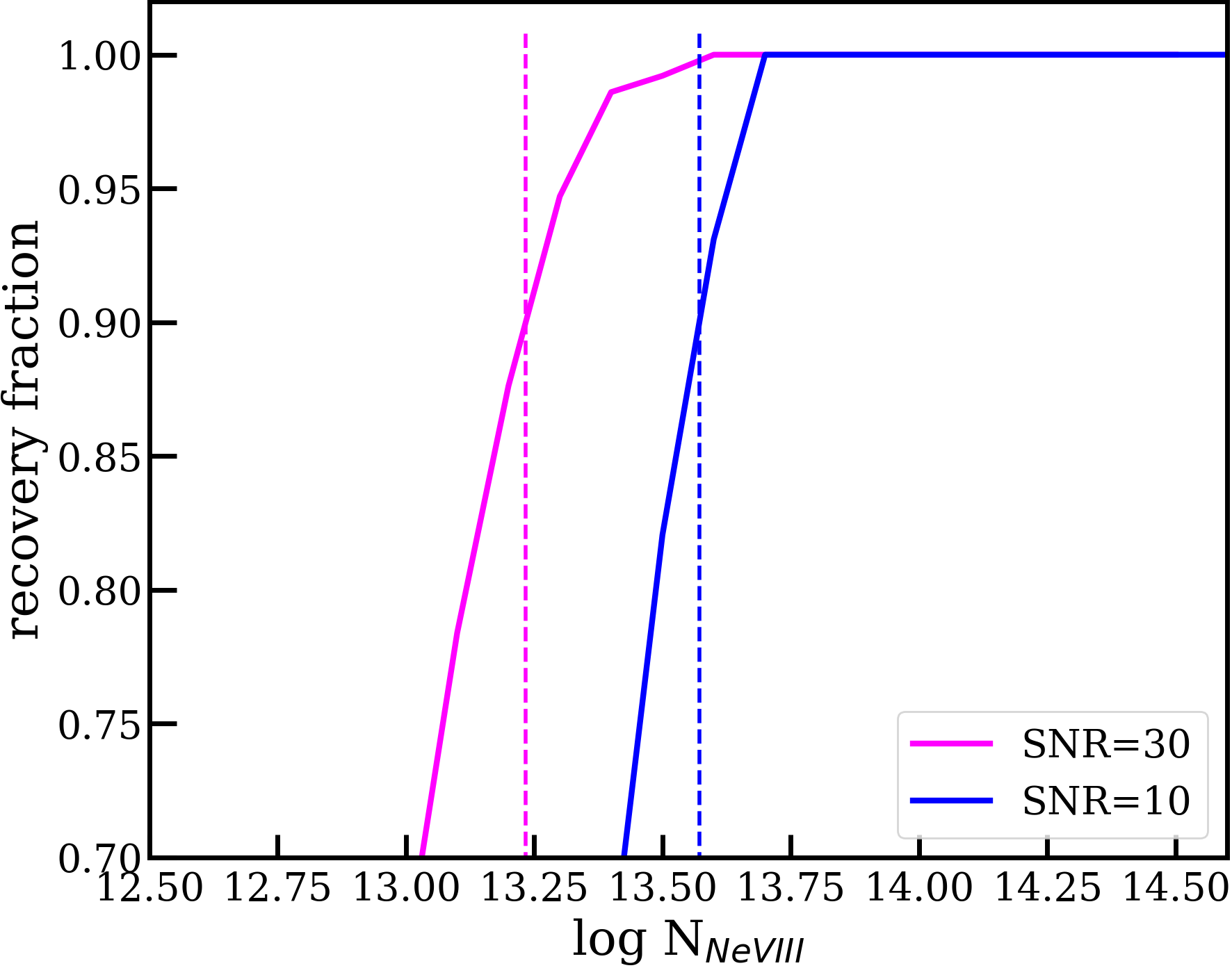}
	\end{minipage}%
	\caption{The recovery fraction of absorption lines (see end of section ~\ref{sec:sightline_generation} for details) detected with RSL$>$4 using {\sc viper} fits for \CIV, \OVI, and \NeVIII\ are shown in left, middle and right panel respectively. The column density threshold for detection depends on the SNR of the spectra. Here we considered simulated spectra with constant SNR of 30 (in magenta) and 10 (in blue). The $90\%$ completeness limit of recovery fraction(denoted by dashed line) for C~{\sc iv}, O~{\sc vi} and Ne~{\sc viii} in SNR=30 (SNR=10) spectra are $10^{12.65}$, $10^{13.03}$, $10^{ 13.24}$ cm$^{-2}$ ($10^{13.18}$, $10^{13.45}$, $10^{13.63}$ cm$^{-2}$).
 }
 
	\label{fig:recovery}
\end{figure*}

Decomposing absorption lines into Voigt profiles is important to compare the observed properties with the simulated ones. For this,
we use automated Voigt profile fitting parallel code "VoIgt profile Parameter Estimation Routine"  \citep[{\sc viper}; see][for details]{gaikwad2017b}. We use {\sc viper} to obtain the column density and Doppler parameter (b) for each of the metal absorption line components in the 10000 simulated spectra from each simulation suit.
Note till now {\sc viper} has been used to fit the \lya\ forest to study their statistical distribution as a function \HI\ column density
\citep[see for example,][]{maitra2020,maitra2020b,gaikwad2021}.
As of now {\sc viper} does not simultaneously fit multiple transitions from a given ion. Therefore, we use only the strongest metal line transitions for Voigt profile decomposition using {\sc viper}. This will mean we may underestimate the component structure when the transitions are highly saturated. The minimum number of components required to fit a given absorption profile with {\sc viper} is decided objectively using Akaike Information Criterion (AIC).
{\sc viper} assigns a rigorous significant level \citep[RSL; as described in][]{keeney2012} to each the fitted line. We have included the features having RSL$>$4 \citep[as was also used in][]{danforth2016} to avoid spurious detection. Sample {\sc viper} fits of the simulated \OVI\ absorption are also shown in panels f and g of Figure~\ref{fig:demo_los}. 

In Figure~\ref{fig:demo_los2}, we show another example of profiles of metal lines and \lya\ absorption along a simulated sight line for two UVBs. There are three \lya\ absorption systems identified when we use "fg11" UVB.
Two of them show associated \OVI\ absorption and only one system shows absorption from \CIV\ as well as \NeVIII.
In each panel, we also provide the average density, temperature, and metallicity of the gas contributing to the detected absorption feature. We also mention the line of sight comoving length ($\Delta x$) of the region contributing to the absorption. It is evident that while the absorption lines are coincident in the velocity space the \lya\ absorption is produced from a much larger region compared to the metal lines.  
In the right panels of Figure~\ref{fig:demo_los2} we plot the spectra from the same sightline but for the "ks19q18" UVB. It is clear that the \CIV\ and \NeVIII\ absorption are not statistically significant and \OVI\ absorption in the case of  "ks19q18" UVB is weaker than that when we use "fg11" UVB.
%
 
We use the column density, $b$-parameter, and redshifts for individual components obtained using {\sc viper} for our statistical analysis. In addition, we also consider the total column density of different ions for the system. For this purpose, we define a system as a simply connected region along the best-fitted spectrum with optical depth above a threshold optical depth. In our case, we assume the threshold optical depth for components within the connected region to be 0.05 \citep[see for example,][]{tepper2011}.  The column density for the systems is obtained by adding the column densities of individual components within the connected region. The redshift of the strongest component is considered as the redshift of the system. In addition to this, we also define the velocity width of absorption ($\Delta V_{90}$) as the velocity width of the region containing 5\% and 95\% of the total optical depth of a system (see Figure~\ref{fig:demo_los}).

We introduce "recovery fraction", defined as the fraction of true absorption in the noise-free spectrum at a given column density recovered by {\sc viper} when we use RSL$>$4.
%
The recovery fraction as a function of column density is shown in Figure~\ref{fig:recovery} for three ions of our interest for spectra simulated with an SNR of 10 and 30.
    The recovery fraction increases with column density and reaches 90$\%$ (marked with dashed line in Figure~\ref{fig:recovery}) at column densities $10^{12.65}$, $10^{13.03}$ and $10^{13.24}$ cm$^{-2}$ for C~{\sc iv}, O~{\sc vi} and Ne~{\sc viii} respectively. The same for SNR$=10$ spectra  are at column densities $10^{13.18}$, $10^{13.45}$ and $10^{13.63}$ cm$^{-2}$ for C~{\sc iv}, O~{\sc vi} and Ne~{\sc viii} respectively.

\section{Results}
\label{Sec:results}

In this section, we compare various observable parameters measured from our simulations with the available observations. In particular, we explore how they depend on the assumed UVB. As this is more exploratory in nature we do not attempt to produce a best fit to the observations rather use observed data as more of an indicator. We mostly use simulation results for $z=0.5$ in our discussions. However, in the case of \CIV\ , we also show simulation results for $z=0.1$ as most of the available observations are at $z<0.15$. 

\subsection{Column density distribution Function}
\begin{figure*}
	\begin{minipage}{\textwidth}
		\includegraphics[width=\textwidth]{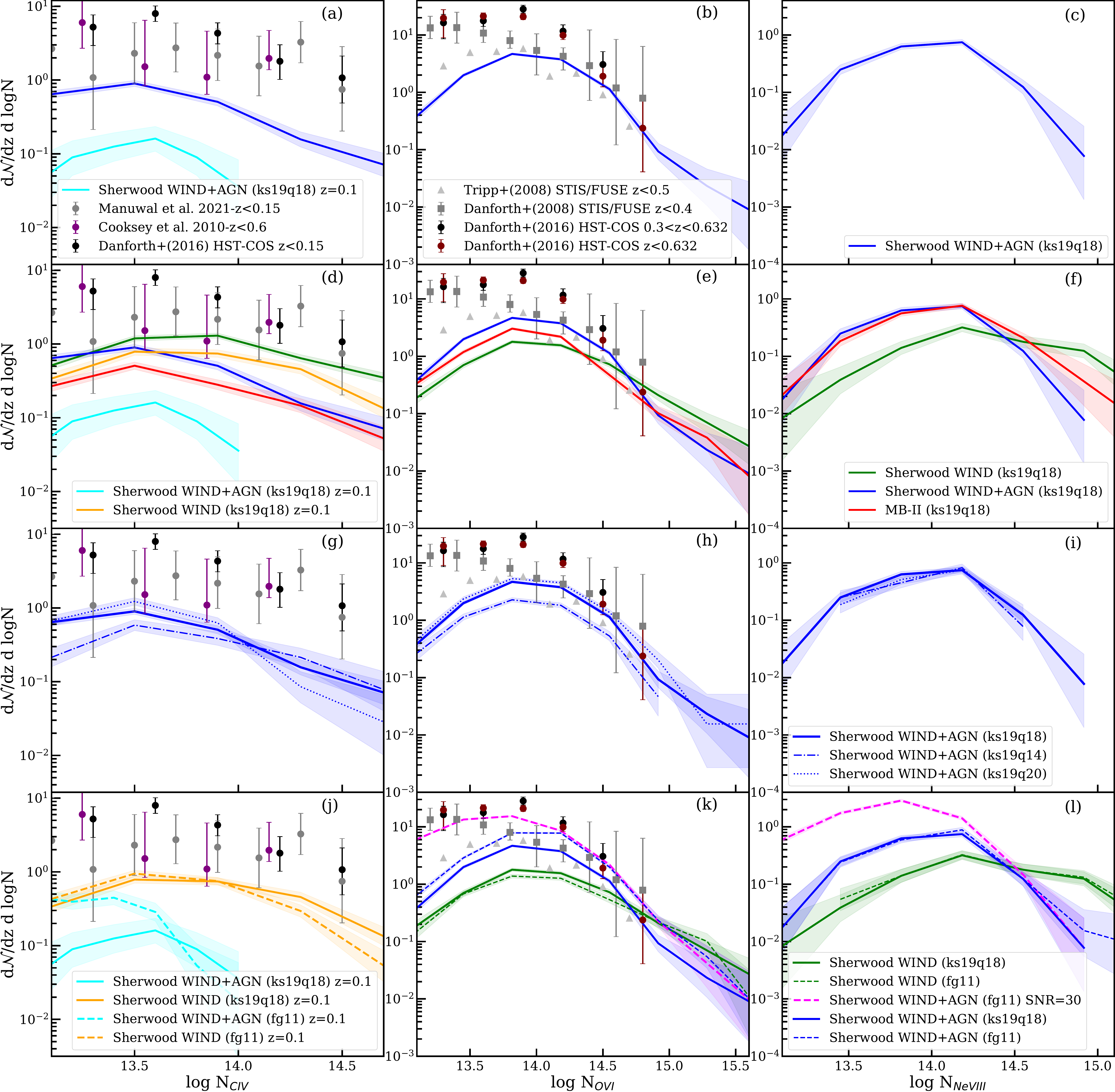}
		\caption{
  Comparison of CDDF predicted by our simulations (at $z=0.5$) with the available observations. 
  Left, middle and right panels show CDDF for C~{\sc iv}, O~{\sc vi} and Ne~{\sc viii} respectively. The top row shows the results for the "WIND+AGN" Sherwood simulation for "ks19q18" UVB. The second row shows the results for all our simulations for "ks19q18" UVB. The third row provides the effect of $\alpha$ in the "ks19" UVB on the predicted CDDF in the case of the "WIND+AGN" Sherwood simulation. 
  The last row provides results for the "fg11" UVB. In the case of \CIV, we also provide simulation results for $z=0.1$. Shaded regions provide Poisson errors in the CDDF.
  }
		\label{fig:cddf_all1}
	\end{minipage}%
\end{figure*}


The column density distribution function (hereafter CDDF) is defined as the number of absorbers, $\mathcal{N}$, of a species per logarithmic column density interval, $d logN$, and per redshift interval $dz$ sensitive for the detection of that species. The CDDF of H~{\sc i} has been used \citep{kollmeier2014, shull2015, gaikwad2017a} to constrain the \HI\ photoionization rate which can be used to constrain the nature of sources contributing to the hydrogen ionizing part of the UVB at low-$z$ \citep[see for example,][]{khaire2015b}. In addition, some previous works like \cite{oppenheimer2009}, \cite{tepper2011}, \cite{oppenheimer2012}, \cite{rahmati2016}, \cite{Nelson2018}, used the CDDF of metal ions to understand the effect of different feedback processes and their relative strengths on the predicted CDDF for a given UVB. Here, we mainly focus on the CDDF of the three identified metal ions and how they depend on the assumed UVB for the three simulations considered here. We use the column densities obtained from the Voigt profile decomposition using {\sc viper} and not a simple integration over a slice in the simulation \citep[as in][for example] {Nelson2018}.

\par 

\vskip 0.1in
\noindent{\bf Observed CDDF:}
 {The CDDF of \CIV\ (left panel), \OVI\ (middle panel), and \NeVIII\ (right panel) are shown in Figure~\ref{fig:cddf_all1} for the simulated spectra with SNR=10 using different UVBs for three simulations used here. In the case of \CIV\ we show the measurements  from \cite{cooksey2010} (for $z< 0.6$), \citet{danforth2016} (for $z<0.15$) and \cite{manuwal2021} (for $z< 0.15$). For \OVI\ we use  observational results from \cite{tripp2008} (for $z<0.5$), \cite{danforth2008} (for $z<0.4$) and \cite{danforth2016} (we obtained the CDDF for \OVI\ absorbers at $0.3<z<0.632$ from their table). In the case of \NeVIII\ until now we do not have a very sensitive measurement of CDDF. However, a wide range of values of $d\mathcal{N}/dz$ is quoted in the literature. For example, for rest equivalent width limit of 30 m\AA\ (i.e log~N(\NeVIII) = 13.7 assuming the linear part of the curve of growth) for the Ne~{\sc viii}$\lambda$770 line, reported $d\mathcal{N}/dz$ are typically in the range 2.7 to 14 \citep{narayanan2009,meiring2013,danforth2016}. However, based on non-detection of \NeVIII\ in their stacked spectrum and direct detections \citet{frank2018} have inferred a much lower $d\mathcal{N}/dz$ (i.e 1.38$^{+0.97}_{-0.82}$ for log $N$(\NeVIII)$\ge$13.7 cm$^{-2}$). Therefore, while we compare our model predicted CDDF with that of observations for \CIV\ and \OVI\ absorbers in the case of \NeVIII\ we compare the d$\mathcal{N}/dz$.
 
 \vskip 0.1in
\noindent{\bf \OVI\ absorption systems:}
In the top panels  of Figure~\ref{fig:cddf_all1}, we plot the CDDF of systems (panels "a", "b" and "c" for \CIV, \OVI, and \NeVIII\ respectively) from "WIND+AGN" Sherwood simulation (for a spectra SNR of 10 per pixel) assuming "ks19q18" UVB.  Recall, for the assumed SNR the limiting $N$(\OIV) is $10^{13.45}$ cm$^{-2}$. The CDDF of \OVI\ from the simulation roughly follows the observed distributions from \citet{tripp2008} and \citet{danforth2008} but under-predicts with respect to the HST-COS measurements of \citet{danforth2016}. In panel "e", we show the CDDF from all the three simulations considered here for the "ks19q18" UVB. More \OVI\ absorbers (i.e by a factor $\sim$1.6 for log~$N$(\OVI)$\ge$13.5) are seen in the "WIND+AGN" simulation compared to "WIND" only Sherwood simulation. We find this trend to be independent of the UVB used. 
We attribute the overall enhancement in the number of absorbers to the higher mean metallicity seen for the gas in the "WHIM" and "diffuse" phases in the case of "WIND+AGN" simulation (refer to Table~\ref{table:detectability}).
However, at the high column density end (i.e., log~$N$(\OVI)$>$14.7) we find the CDDF to be higher in the case of "WIND" only simulation compared to the "WIND+AGN" Sherwood simulation.
%
It is interesting to note that MB-II simulation produces less \OVI\ absorbers
compared to the "WIND+AGN" Sherwood simulation. However, in the case of "WIND" only Sherwood simulation we find the CDDF at high column density end (i.e log~$N$({\OVI})$>$ 14.30) to be higher than that from the MB-II. This enhancement in the CDDF in the case of "WIND" only simulation compared to the other two can be attributed to the high $N$(\OVI) systems predominantly originating from the "hot halo" and/or "condensed" regions where $f_1$ and average metallicity are higher in the case of "WIND" only Sherwood simulation (see Table~\ref{table:detectability}).

We show the effect of varying $\alpha$ used to generate the "ks19" UVBs on the CDDF of \OVI\ in panel "h" of Figure~\ref{fig:cddf_all1}. As expected, based on our discussions in section~\ref{sec:UVB}, the CDDF is found to be less (by a factor of $\sim$2.3) when we use $\alpha=1.4$ compared to that when
$\alpha=2$. It is interesting to note that while we use the "ks19" UVB for $\alpha=1.4$ in the "WIND+AGN" Sherwood simulation  the CCDF obtained is close to that of simulations with "WIND" only feedback when using the UVB computed with $\alpha=1.8$. Thus it appears that there could be degeneracy between the feedback parameters and the UVB (especially at the low $N$(\OVI) end). 

The observed CDDF by \citet{danforth2016} is not reproduced by any of our simulations when we use "ks19" UVB for any assumed value of $\alpha$ within the allowed range.
However, as can be seen from the panel "k" in Figure~\ref{fig:cddf_all1}, using the "fg11" UVB  increases the CDDF and brings it closer to the observations of \citet{danforth2016}. The increase in the number of \OVI\ absorbers is  by a factor of $\sim$1.8 compare to the simulation using "ks19q18" UVB. This can be  attributed to the increase in the number of SPH particles that can contribute to the \OVI\ absorption
when we use the "fg11" UVB. As an illustration, we show the CDDF for Sherwood "WIND+AGN" simulation for "fg11" UVB for SNR=30 per pixel. It is evident that the CDDF from the simulation passes through most of the observed points from \citet{danforth2016}.  Interestingly, in the case of "WIND" only simulation, the CDDF does not change appreciably when we use "fg11" UVB instead of "ks19q18" UVB(see panel "k" in Figure~\ref{fig:cddf_all1}).
This is interesting in view of the fact that $f_1$ values in the "WHIM" and "diffuse" regions have increased appreciably (see Table~\ref{table:detectability}). However, this has not translated into an increase in the CDDF mainly because of the low average metallicity of the SPH particle in these regions in the "WIND" only simulation compared to the "WIND+AGN" Sherwood simulation or MB-II.

Recently \citet{Nelson2018} and \citet{bradley2022}, have reproduced the observed \OVI\ CDDF from \citet{danforth2016} using their  IllustrisTNG and SIMBA simulations, respectively. It is interesting to note the former uses "fg11" UVB the latter simulations use the UVB given in \citet{faucher2020}.
As can be seen from Figure~\ref{fig:UVB}, these two UVBs have a much lower \OVI\ ionization rate (see also Table~\ref{tab:Gamma_for_uvbs}) compared to "ks19" or "HM12". Both these simulations also incorporate the AGN feedback that spreads metals over wide over-density ranges.
\citet{Nelson2018}, discuss the effect of various feedback prescriptions on the \OVI\ CDDF in detail. They also show the \OVI\ CDDF predicted by the earlier Illustris simulations \citep{suresh2017} under-produces the observed CDDF of \OVI. They attributed it to the low "WIND" velocities employed in Illustris simulations.
Similarly, \citet{bradley2022} have reported that the CDDF at high column densities is  sensitive to the jet feedback in their simulations.
%
In the past, several simulations have had problems reproducing the \OVI\ CDDF of \citet{danforth2016}. For example, the EAGLE simulations that use HM01 UVB underpredict the CDDF \citep[see][]{rahmati2016} for models with a wide range of feedback prescriptions. It is interesting to note that two of the simulations that reproduce the observed \OVI\ CDDF of \citet{danforth2016} use "fg11" background that tends to have a low \OVI\ ionization rate among the available UVBs in the literature. 

{\it In summary, to match the CDDF of \OVI\ observed by \citet{danforth2016} we favor the "WIND+AGN" Sherwood simulation that use "fg11" UVB among the simulations considered here. It is also evident that in order for the UVB to have significant effects on the CDDF the feedback employed should enrich the low-density regions (i.e "WHIM" and "diffuse" phases) with higher metallicity.}

\vskip 0.1in
\noindent{\bf \CIV\ absorption systems:} In panel "a" of Figure~\ref{fig:cddf_all1}, we compare the observed  CDDF \citep{cooksey2010, danforth2016,manuwal2021} for \CIV\ (at $z\le0.15$) with that predicted by our fiducial Sherwood "WIND+AGN" simulation using "ks19q18" UVB. The simulated curve for $z=0.5$ underpredicts the C~{\sc iv} CDDF compared to the observations. The discrepancy is even higher if we use the simulation results from $z=0.1$, that is appropriate as most of the \CIV\ observations are for $z<0.15$.

In panel "d" of Figure~\ref{fig:cddf_all1} we plot the CDDF for \CIV\ predicted by the three simulations for our fiducial UVB. Unlike for \OVI, the "WIND" only simulation produces more \CIV\ absorbers (3.3 times more d$\mathcal {N}$/dz for log~$N$(\CIV)$>$13.7) compared to "WIND+AGN" Sherwood simulation. The \CIV\  CDDF predicted by the "WIND" only Sherwood simulation broadly agrees with the observed distribution. Even at $z=0.1$, the "WIND" only simulation produces much more \CIV\ absorbers compared to the "WIND+AGN" Sherwood simulation.
%
 The CDDF of \CIV\ predicted in the MB-II simulation is even lower than the Sherwood "WIND+AGN" simulation, although the differences reduce at  higher  column densities (i.e., log~$N$(\CIV)$>$14.3).  As can be seen from Table~\ref{table:detectability},  \CIV\ ions mostly reside in the high-density regions (i.e "hot halo" and "condensed" regions). The value of $f_1$ in the "condensed" phase is higher in the case of the Sherwood simulation with "WIND" only feedback. Also, the metallicity of the gas in these two phases is also higher in the case of "WIND" only Sherwood simulation.  This to some extent explains the detection of more higher column density \CIV\ absorbers in the "WIND" only simulation compared to the other two simulations. 

 
Next, we study the effect of changing the UVB. In panel (g) of Figure~\ref{fig:cddf_all1} we show the predicted \CIV\ CDDF from the "WIND+AGN" simulations using "ks19" UVB generated using different values of $\alpha$.
It is evident that UVB generated with low values of $\alpha$ produce more \CIV\ absorbers at low \CIV\ column densities (i.e log $N$(\CIV)$<$ 14.0)  and less \CIV\ absorbers at high column densities.  In panel (j) of Figure~\ref{fig:cddf_all1} we compare the \CIV\ CDDF predicted by "WIND" and "WIND+AGN" simulations using "fg11" and "ks19q18" UVBs for $z=0.1$. For a given simulation when we use "fg11" UVB we get more low \CIV\ column density absorbers and less high \CIV\ column density absorbers compared to what we get while using "ks19q18" UVB.
The observed trends can be attributed to an increase in $f_1$ seen for the "hot halo" and a decrease in $f_1$ for the "Condensed" phases when we go from "ks19" to "fg11" UVB.

Our results are broadly consistent with earlier simulation results. \citet{rahmati2016} found the \CIV\ CDDF from their EAGLE simulation (that includes both stellar and AGN feedback and used HM01 UVB) to broadly agree with the observed CDDF of \citet{cooksey2010} but to underpredict the observed CDDF of \citet{danforth2016}. They also found the number of \CIV\ absorbers to reduce when AGN feedback is included along with the stellar feedback in a simulation box large enough to have an appreciable effect of AGN feedback. Confirming this trend, \citet{oppenheimer2012} have slightly overpredicted the \CIV\ CDDF compared to \citet{cooksey2010}, in their simulation that includes only stellar feedback.

{\it In summary, to match the observed CDDF of \CIV\ we favor the "WIND" only Sherwood simulation using "fg11" UVB among the simulations considered here. This is because, the \CIV\ absorption seems to originate mostly from the "condensed" phase followed by the "hot halo" phase. The inclusion of AGN feedback reduces $f_1$ and average metallicity in these phases. 
The discussion presented till now suggests that what feedback we will favor is dependent on which CDDF is considered. However, the overall observed \CIV\ and \OVI\ CDDFs seem to favor a softer ionizing background in the X-ray regime compared to the "ks19" UVBs.
}

\begin{figure*}
		\includegraphics[width=\textwidth]{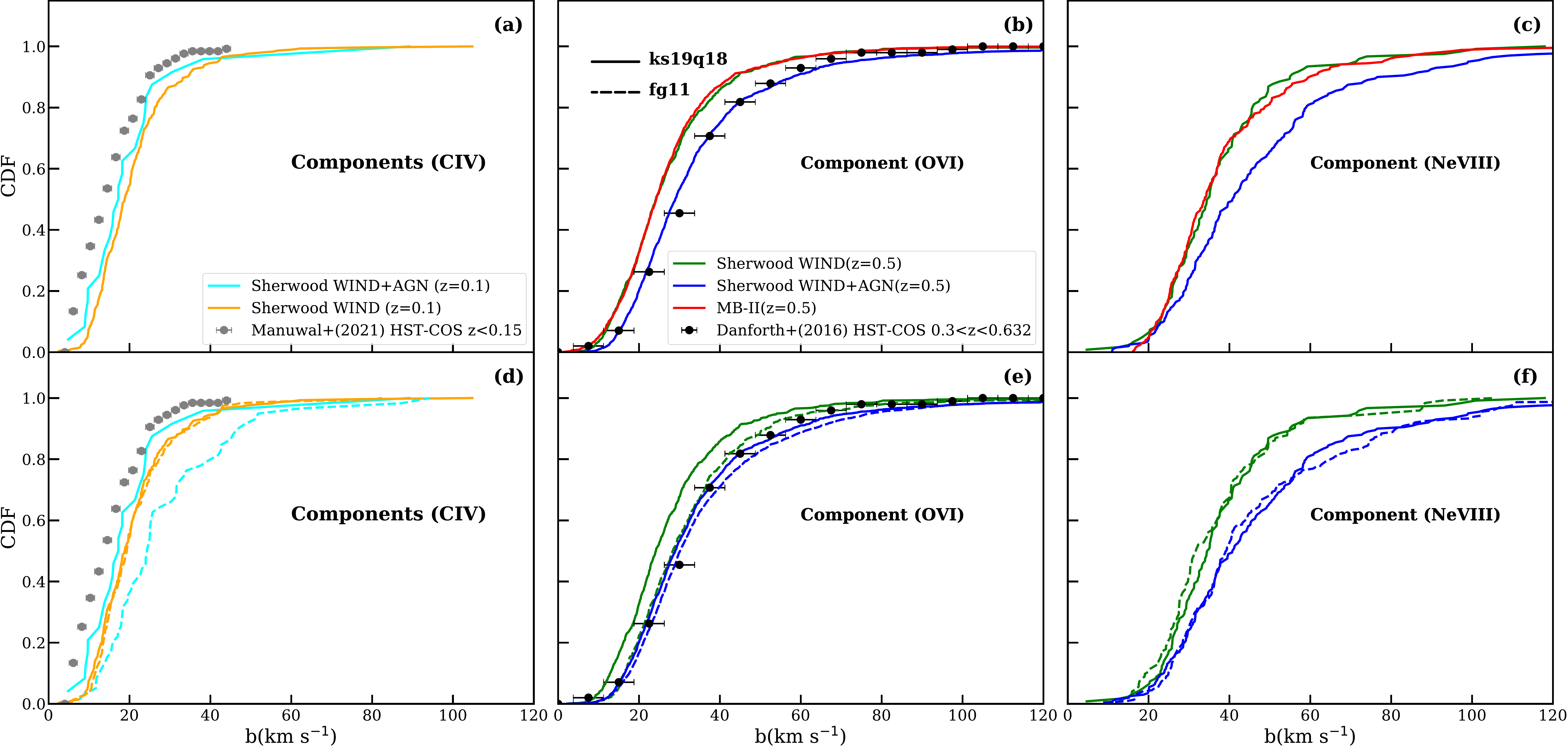}
		\caption{ The left, middle and right panels show cumulative distribution function of Doppler parameter (b-CDF) for C~{\sc iv}, O~{\sc vi} and Ne~{\sc viii} respectively, with simulated spectra using SNR=10. The upper panels show the comparison of b-CDF among different simulations for spectra obtained using UVB "ks19q18". The lower panels compare b-CDF of Sherwood "WIND+AGN" model for "ks19q18" UVB with "fg11" UVB. The observed b-CDF from \citet{danforth2016} for \OVI\ and \citet{manuwal2021} for \CIV\ are shown in black and grey circles, respectively and the bars along the abscissa indicate the bin width.
		}
		\label{fig:b_cdf}
\end{figure*}

\vskip 0.1in
\noindent{\bf \NeVIII\ absorption systems:}
The CDDF obtained for \NeVIII\ systems in our simulations are shown in the right panels of Figure~\ref{fig:cddf_all1}. For simulations using "ks19q18" UVB  we obtain $d\mathcal{N}/dz$ = 0.32, 0.60, and 0.59 for "WIND" only, "WIND+AGN" Sherwood simulations and MB-II simulation respectively.
Here, we consider a threshold column density of $N$(\NeVIII)=10$^{13.5}$ cm$^{-2}$. 
As noted before, typically quoted $d\mathcal{N}/dz$ values based on a few direct detections of \NeVIII\ systems are much higher than that from our simulations.  But simulation results are consistent with the values inferred by \citet{frank2018}.
We notice clear differences in the \NeVIII\ CDDF at low and high $N$(\NeVIII) ends between "WIND" only and "WIND+AGN" Sherwood simulations. This can be attributed to $f_1$ and average metallicity being high (low) in the "WIND" only (WIND+AGN) simulation with "condensed" and "hot halo"
("diffuse" and "WHIM") phases.
We also notice that changing the UVB (i.e change in $\alpha$ in the case of "ks19" or using "fg11") does not change the $d\mathcal{N}/dz$  values appreciably. 
Previous works like \citet{oppenheimer2012} 
(with a $d\mathcal{N}/dz$ $\sim$0.4 for rest equivalent width more than 30 m\AA)
and \citet{tepper2013} (with a $d\mathcal{N}/dz$ $\sim $ 0.1) also fail to produce $d\mathcal{N}/dz$ of \NeVIII\  as inferred from the direct detections. 
However, these measurements are consistent with the value inferred by \citet{frank2018} using the combination of direct detection and stacking method.

{\it Therefore, we conclude that for a range of UVB considered here all three simulations fail to produce high values of $d\mathcal{N}/dz$  inferred based on the direct detection of \NeVIII\ absorbers. However, all simulations produce values consistent with the $d\mathcal{N}/dz$ inferred by \citet{frank2018} based on stacking and direct detections.
Obtaining full CDDF for \NeVIII\ from  direct observations is important for  constraining the feedback parameters and UVB used in the simulations.}

}

\par

\subsection{Doppler parameter distribution}


Now we explore the effect of feedback and UVB on the distribution of $b$-parameter 
of the individual Voigt profile components and compare them with the observed distribution. It is usually assumed that the inferred $b$-parameter of an absorber obtained using Voigt profile fitting has contributions from both thermal and non-thermal motions.
As shown in Figure~\ref{fig:demo_los}, optical 
depth at a given wavelength is contributed by gas distributed over large spatial scales.
Therefore, the profile widths are influenced by the local gradients in the peculiar velocity and spread in the ion density and temperature over the region that contribute to the absorption. As seen before, the region contributing to an absorption line and average values of physical parameters in that region for a given simulation depends on the UVB used. This is the main motivation for the explorations presented in this section.

The cumulative distribution of Doppler parameters (hereafter "b-CDF") 
of C~{\sc iv}, O~{\sc vi} and Ne~{\sc viii} from our simulated spectra (for SNR=10 and "ks19q18" UVB) are shown in the top panels of Figure~\ref{fig:b_cdf} for all three simulations discussed in this work. 
We have considered only absorbers having column density above the recovery completeness limit (see section~\ref{sec:sightline_generation} and Figure~\ref{fig:recovery} for details). 

\vskip 0.1in
\noindent{\bf \OVI}: From panel (b) in Figure~\ref{fig:b_cdf}, we find b-CDF is distinctly different for "WIND" and "WIND+AGN" Sherwood simulations using "ks19q19" UVB. The "WIND+AGN" simulation systematically produces \OVI\ absorption components with large $b$ values (with a median $b$ of $\sim$28.9 \kms) compared to that (with a median $b$ of $\sim$24.2 \kms) of "WIND" only Sherwood simulation.  In the same panel, we plot the observed O~{\sc vi} b-CDF from \citet{danforth2016} for absorbers within the redshift range of $0.3<z<0.73$. For this, we used the Voigt profile fit results given by the authors.
The median $b$ value for the components in the observed sample is 30.9 \kms. This is closer to the value found for the "WIND+AGN" Sherwood simulation. Interestingly the predicted b-CDF from the "WIND+AGN" simulation closely follows the observed b-CDF, unlike that of the "WIND" only Sherwood simulation. In the same panel, we show the b-CDF for the MB-II simulation. The $b$-parameters are found to be narrower in this simulation compared to what we find for the "WIND+AGN" Sherwood simulation. This we attribute either to the higher spatial resolution achieved in this simulation or to the differences in the feedback processes between the two simulations. However, we confirm that the difference is not due to differences in the wavelength sampling we have adopted using the spectrum from MB-II simulations obtained with same wavelength sampling as we have used for the Sherwood simulations.

In panel  (e) of Figure~\ref{fig:b_cdf}, we compare the b-CDF of the Sherwood simulation using "ks19q18" and "fg11" UVBs. We notice a slight increase in the median $b$-value (i.e., 30.2 \kms\ from 28.9 \kms) when we use the "fg11" UVB in the case of the "WIND+AGN" simulation.
In the case of "WIND" only simulation
the increase in the $b$-values when we use "fg11" UVB is found to be higher (i.e 28.4 \kms\ from 24.2 \kms\ for the median $b$) than  that  obtained for the "WIND+AGN" Sherwood simulation.
We find that the b-CDF for the "WIND" only simulation that uses "fg11" UVB  is statistically indistinguishable (KS-test p-value of 0.28) from the b-CDF obtained from "WIND+AGN" simulation using "ks19q18" UVB. This demonstrates the effect of UVB  on the observed distribution of the $b$-parameter.
{ \it Overall, it appears that a good agreement to CDDF and b-CDF for the \OVI\ absorbers observed by \citet{danforth2016} can be reproduced with "WIND+AGN" models of Sherwood simulations using "fg11" UVB.}

\vskip 0.1in
\noindent{\bf \CIV:} In panel (a) of Figure~\ref{fig:b_cdf}, we compare the observed b-CDF of \CIV\ obtained by \citet{manuwal2021} with those obtained from our simulations at $z=0.1$. First of all, we notice that the median $b$ values for "WIND" ($\sim$18.8 \kms) and "WIND+AGN" ($\sim$17.2 \kms) are very similar (with the KS-test p-value of 0.62 between  the two distribution). 
However, the $b$-values seen in these simulations are higher than what is observed (where the median $b$-value is $\sim$13.8 \kms). 
In panel (d), we compare the \CIV\ b-CDF obtained for the "WIND+AGN" simulation that uses "fg11" UVB with those obtained using "ks19q18" UVB.  As in the case of \OVI, we notice that when we use "fg11," the $b$ values of individual components found (i.e., the median value of 24.4 \kms\ from 17.2 \kms) have slightly increased. The increase in the $b$-parameters is not appreciable (i.e., the median value of 19.3 \kms from 18.8 \kms) in the case of "WIND" only Sherwood simulation. This is confirmed by the large p-value (i.e., 0.81) returned by the KS-test.
The difference between the $b$-distribution of simulated \CIV\ absorbers with the observation increases if we use $z=0.5$ simulations. 

\vskip 0.1 in
\noindent{\bf \NeVIII:} In panel (c) of Figure~\ref{fig:b_cdf} we show the b-CDF obtained from our three simulations. The trend found is very similar to what we found for \OVI\ absorption.  That is, the median $b$ value found ($\sim$40.8 \kms) for the "WIND+AGN" simulation is higher than the median value found for the ($\sim$34.8 \kms) "WIND" only Sherwood simulation when using "ks19q18" UVB. 
The b-CDF (with a median $b$ value of 33.9 \kms) obtained for the MB-II simulation is closer to that of the "WIND" only simulation compared to the "WIND+AGN" Sherwood simulation. In panel (f) of Figure~\ref{fig:b_cdf}, we compare the $b$ distribution obtained for "fg11" UVB (with a median value of 39.3 \kms) with that obtained for "ks19q18" UVB for the "WIND+AGN" Sherwood simulations. The two b-CDF are found to be statistically indistinguishable (KS test p-value of 0.68).

{\it In summary, for a given UVB, our simulations produce larger $b$-parameters for ions with larger ionization energy as found in observations. In the case of \OVI\ and \NeVIII\ WIND+AGN Sherwood simulation produces larger $b$-parameters compared to "WIND" only Sherwood simulation.
All our simulations tend to produce larger $b$-values for \OVI\ when we  use the "fg11" UVB. However, no strong effect is seen in the case of \NeVIII. We also notice that the $b$-parameter for the \CIV\ components in our Sherwood simulations is higher than what has been observed.
This could possibly indicate that the \CIV\ absorption, in reality, may be originating from regions with much lower temperatures compared to what we find in our simulations. Also, as indicated by results from MB-II simulations, spatial resolution in the simulation could also have a role to play in producing large $b$-values.}


\subsection{Distribution of $\Delta V_{90}$}

\begin{figure*}
		\includegraphics[width=\textwidth]{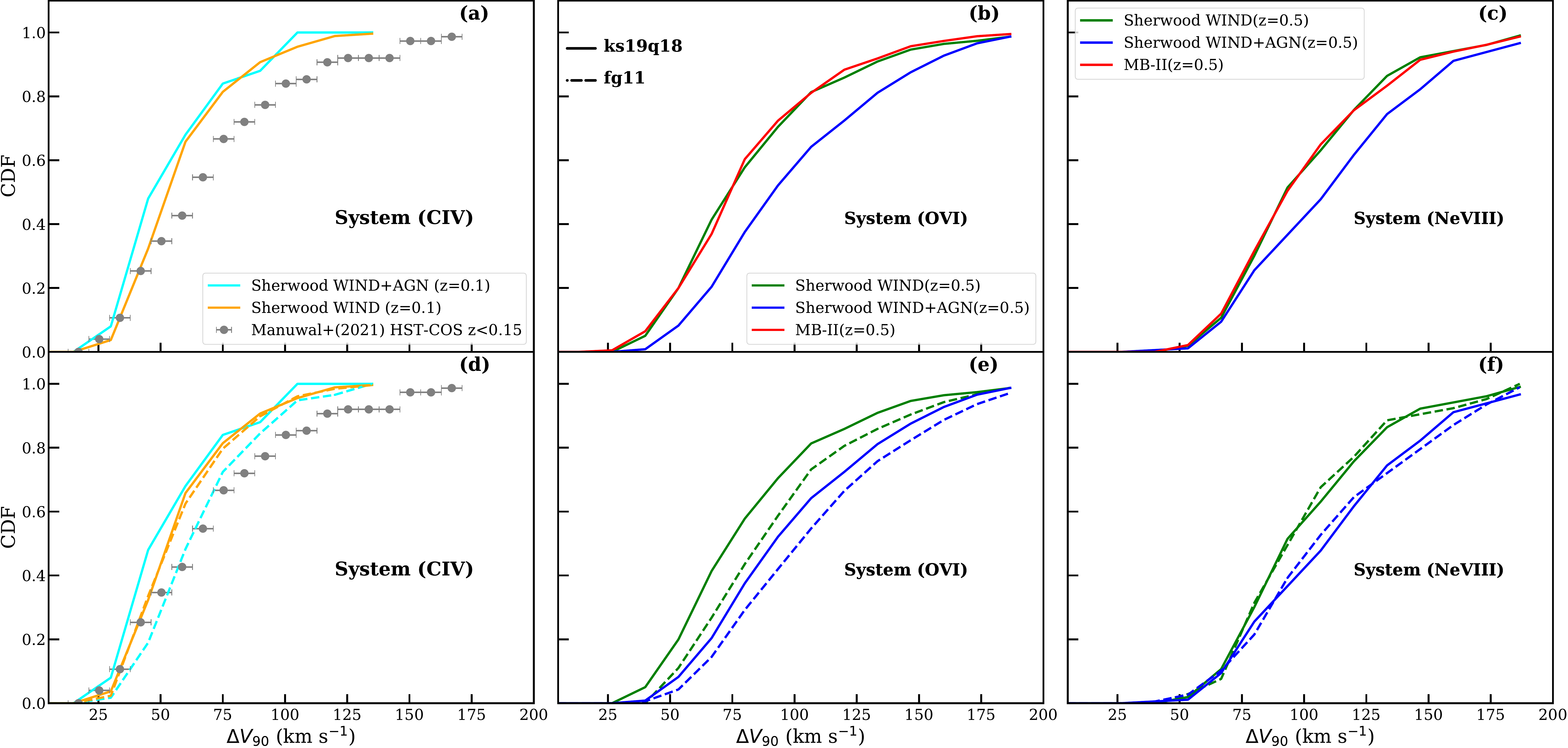}
		\caption{ 
  Same as that of Figure~\ref{fig:b_cdf}, but for  $\Delta V_{90}$. Observed data for \CIV\ are from \citet{manuwal2021}.
  %
		}
		\label{fig:d_cdf}
\end{figure*}

Next, we discuss the distribution of velocity width of 
absorption line systems quantified using 
the $\Delta V_{90}$.
The cumulative distribution of $\Delta V_{90}$ for C~{\sc iv}, O~{\sc vi} and Ne~{\sc viii} for different simulations using "ks19q18" UVB are shown in the upper panels of Figure~\ref{fig:d_cdf}. 

\vskip 0.1in
\par\noindent{\bf \OVI:}
It is clearly evident that $\Delta V_{90}$ measured for \OVI\ systems in the case of "WIND+AGN" Sherwood simulation (median $\Delta V_{90}$ = 95.9 \kms) are systematically larger than that for the "WIND" only (median $\Delta V_{90}$ = 75.3 \kms) simulation.
Interestingly, the cumulative distribution of $\Delta V_{90}$ obtained for the MB-II simulations  (with a median $\Delta V_{90}$ of 75.3\kms) are more closer to the "WIND" only simulation compared to "WIND+AGN" simulations. In panel (e) of Figure~\ref{fig:d_cdf}, we compare the cumulative distribution of $\Delta V_{90}$ predicted for simulations using "fg11" vs. those using "ks19q18" UVB.  These two distributions are statistically distinguishable indicating the importance of UVB for the $\Delta V_{90}$ distribution.
As expected for both "WIND" only and "WIND+AGN" Sherwood simulations, the absorption systems typically have larger $\Delta V_{90}$ when we use "fg11" UVB. This is
because a given absorption system gets contribution from a larger region when we use "fg11" UVB (see, Figure~\ref{fig:demo_los}).

\vskip 0.1in
\noindent{\bf \CIV:} In panel (a) of Figure~\ref{fig:d_cdf}, we plot the $\Delta V_{90}$ distribution for the \CIV\ absorption for different simulations (for "ks19q18") and compare them with the simulated distribution from \citet{manuwal2021}. First, we notice that typical  $\Delta V_{90}$ measured for the \CIV\ absorbers are less than that of  \OVI.
The cumulative distribution for "WIND" only and "WIND+AGN" Sherwood simulations are not similar (with a KS-test p-value of 0.23). The median $\Delta V_{90}$ for "WIND" only and "WIND+AGN" model are $\sim59.4$ \kms and $\sim55.6$ \kms, respectively.
None of the distributions from the simulations match with the observed distribution (with a median $\Delta V_{90} \sim 64$ \kms).
We show the results when we use "fg11" UVB in panel (d) of Figure~\ref{fig:d_cdf}. In the case of "WIND" only Sherwood simulation, we do not find any significant dependence of the cumulative distribution of $\Delta V_{90}$  on the assumed UVB. However, in the case of "WIND+AGN" Sherwood
simulations, we do see  $\Delta V_{90}$ increasing (with a median  $\Delta V_{90}\sim67.1$ \kms) when we use the "fg11" UVB at z=0.1. We noticed similar trend of having higher $\Delta V_{90}$ for "fg11" UVB in the case of "WIND+AGN" model also at z=0.5. 
\vskip 0.1in
\noindent{\bf \NeVIII:} In the panel (c) of Figure~\ref{fig:d_cdf} we plot the cumulative distribution of  $\Delta V_{90}$ for the \NeVIII\ absorption systems. We note that the $\Delta V_{90}$ measurements for \NeVIII\ are systematically larger than that of \CIV\ and \OVI.  Also "WIND" only simulation (with a median  $\Delta V_{90}\sim95.9$ \kms) produces lower  $\Delta V_{90}$ compared to "WIND+AGN" simulation (with a median  $\Delta V_{90}\sim109.7$ \kms). 
Like in other cases, the MB-II simulation produces  $\Delta V_{90}$ closer to that of "WIND" only simulations. In panel (f) of Figure~\ref{fig:d_cdf} we compare the cumulative distribution for two different UVB.
As seen before, in Sherwood simulations, properties of \NeVIII\ are independent of the choice of the UVB. We can attribute this to the \NeVIII\ being originating from regions (with high gas density) where collisional excitation dominate.

{\it In summary, discussions presented here clearly suggest that the distribution of  $\Delta V_{90}$ is another good probe of UVB for a given simulation.  The effect of UVB is found to be strong for species (like \OVI) that originate from regions spanning a wide range of density and temperature.}

\begin{figure*}
		\includegraphics[width=\textwidth]{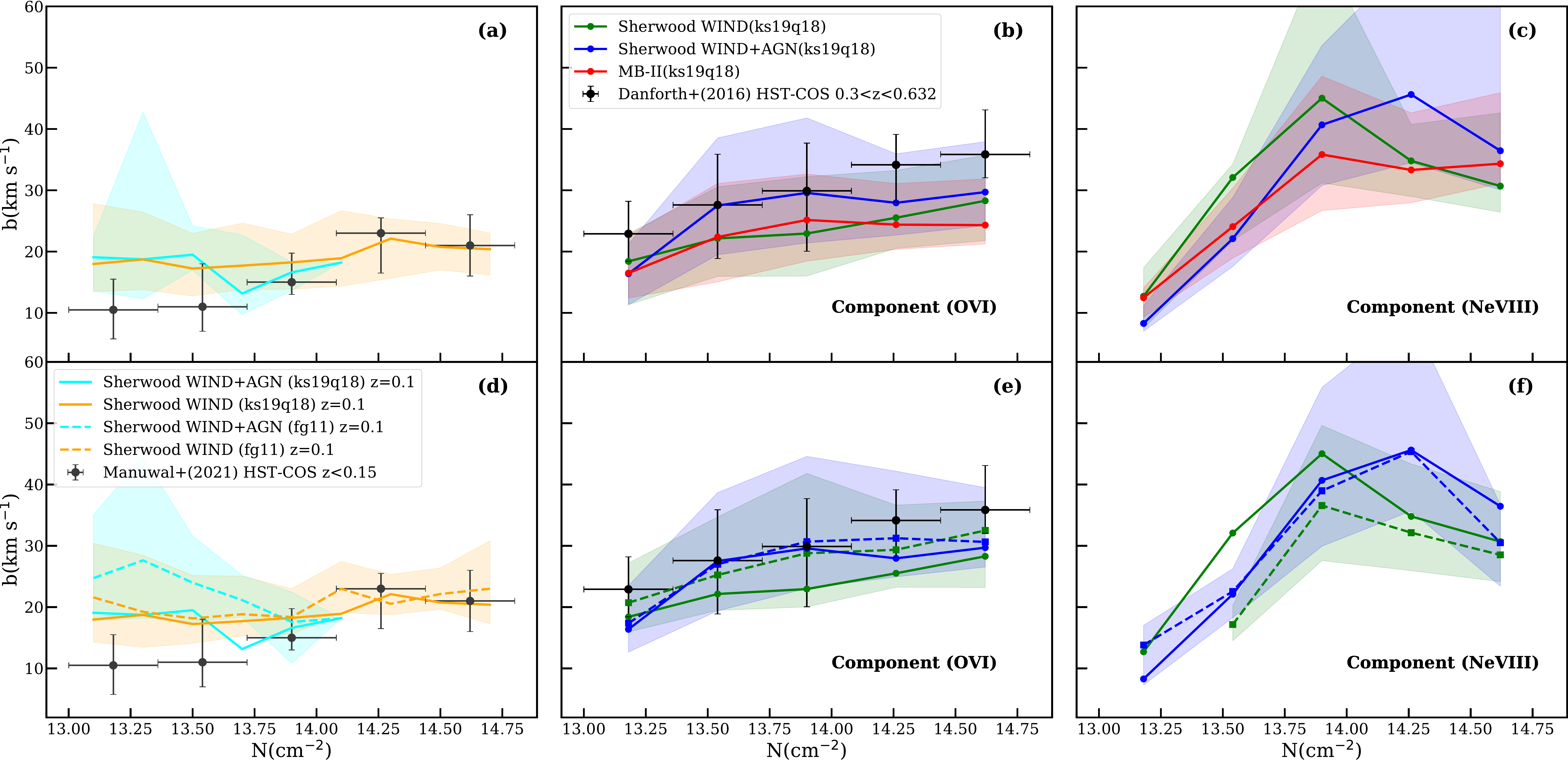}
		\caption{ The left, middle and right panels show the plot between  Doppler parameter and column density  (i.e., $b-N$) for C~{\sc iv}, O~{\sc vi} and Ne~{\sc viii} respectively. Simulated results shown are obtained for SNR$\sim$10 and the observations are from \citet{manuwal2021} for C~{\sc iv} and from \citet{danforth2016} for O~{\sc vi}. 
  The bars (shaded regions) on the observed (simulated) data along the abscissa and ordinate indicate the bin width and 25th-75th percentile range of the $b$-distribution respectively.  The upper panels show results from different simulations when "ks19q18" UVB is used. The lower panel compares the $b-N$ distribution of simulated spectra from the Sherwood "WIND+AGN" model for "ks19q18" UVB vs that obtained for "fg11" UVB. The Shaded region in the lower panel corresponds to simulated results obtained using the "fg11" UVB.
		}
		\label{fig:b_N}
\end{figure*}

\subsection{$N$ vs. $b$ correlations:}

Next, we study the possible correlations between the column density and $b$-parameter for the three ions in our simulations. A correlation between $N$ and $b$ is expected if there is a correlation between the density and temperature fields. Such a correlation is well established in the case of \HI\ from the \lya\ absorption and being regularly used to measure the temperature of the IGM at the mean over-density
\citep[see for example,][]{gaikwad2021}.

\vskip 0.1in
\par\noindent{\bf \OVI:}
 \cite{heckman2002} suggested a correlation between $b$ and $N$ for O~{\sc vi} absorbers can originate from radiatively cooling gas with temperature in the range of $10^5-10^6$K.  The existence of this correlation in the observed data is still debated.
 For example, 
 \cite{danforth2006} did not report any such correlation whereas \cite{lehner2006} found their data to follow the predicted correlation. \cite{tripp2008} also found a correlation between $b$ and $N$, but the significance level was not high enough to confirm the model predictions of \cite{heckman2002}. None of the studies with simulated \OVI\ absorbers fully confirm the existence of the $b-N$ correlation. For example, \cite{tepper2011} found a correlation in low column density absorbers (log $N$(\OVI)$<$13.5), but the Doppler parameter of high column density absorbers is much less than the trend seen in the observed data from \cite{tripp2008}. Similarly, \cite{oppenheimer2009} and \cite{oppenheimer2012} found any of the model variations like the velocity of wind, consideration of non-equilibrium ionization or constant metallicity does not confirm the observed trend, suggesting the need for inclusion of turbulent broadening.

In panel (b) in Figure~\ref{fig:b_N}, we show the relationship between $N$ and $b$ for \OVI\ absorbers in our simulations (for "ks19q18" UVB) and compare them with the observations from \citet{danforth2016}.
For observations, we have used all the \OVI\ absorbers in the redshift range $0.3<z<0.73$ and obtained the median (black circles) and error bars covering the 25th and 75th percentile. It is clear that for a given $N$(\OVI), the $b$-values predicted in the case of "WIND" only Sherwood simulations and MB-II simulations are lower than what has been observed. However, results from "WIND+AGN" simulations are roughly consistent with the observations. In panel (e) of Figure~\ref{fig:b_N}, we show the effect of using "fg11" UVB. As we have seen in Figure~\ref{fig:b_cdf}, for a given $N$(\OVI)
the $b$-values are higher when we use "fg11" UVB. Both "WIND" only and "WIND+AGN" simulations produce $N$-$b$ relationship for \OVI\ consistent with what has been observed when we use "fg11" UVB.

\vskip 0.1in
\noindent{\bf \CIV:} {In panel (a) of Figure~\ref{fig:b_N} we plot the results for the \CIV\ absorbers from our simulations for "ks19q18" UVB at $z=0.1$. We also plot the median and 25-75th percentile of the $b$ distribution at each $N$ bin of the observed data obtained from \citet{manuwal2021}. The observational data shows a mild correlation.  However, the simulated curves are nearly flat, with the median $b$-values at high column density broadly agreeing with the observation, but low column density end being more than what is observed. When we use the "fg11" background, both the Sherwood simulations show an increase in the $b$-parameter values. This increases the difference with respect to the observations at the low column density end. The $b-N$ distribution for simulated spectra at $z=0.5$ also does not have significant correlation.}

\vskip 0.1in
\noindent{\bf \NeVIII:} In panel (c) of Figure~\ref{fig:b_N} we plot the results for \NeVIII\ components in our simulations when using "ks19q18". We do see a correlation between $N$ and $b$ in the low column density end before flattening at the high column densities.
It is also interesting to note that in the low (high) column density end the $b$-values for the "WIND" only Sherwood simulations  are higher (lower) than that for the "WIND+AGN" Sherwood simulation.
We also find that when we use the "fg11" UVB
the b-values at a given $N$(\NeVIII) reduce in the case of "WIND" only
simulation while no strong effect is seen in the case of "WIND+AGN" simulation (see panel f of Figure~\ref{fig:b_N}).

{\it We find the mean relationship between $N$ and $b$ parameters are sensitive to the assumed UVB. The differences for a given simulation between "ks19q18" and "fg11" UVBs are as much as the difference between "WIND" only and "WIND+AGN" Sherwood simulations. It is well known that to some extent the correlation between $N$ and $b$ is influenced by the SNR at the low-$N$ end and Voigt profile decomposition at the high-$N$ end. Therefore, for realistic comparison with the data, one has to match the SNR distribution used in the simulations with the observations \citep[as in][]{maitra2020b}. }
\par
\subsection{Associated absorption of \lya\ and metal ions}
\begin{figure*}
	\begin{minipage}{\textwidth}
		\includegraphics[width=\textwidth]{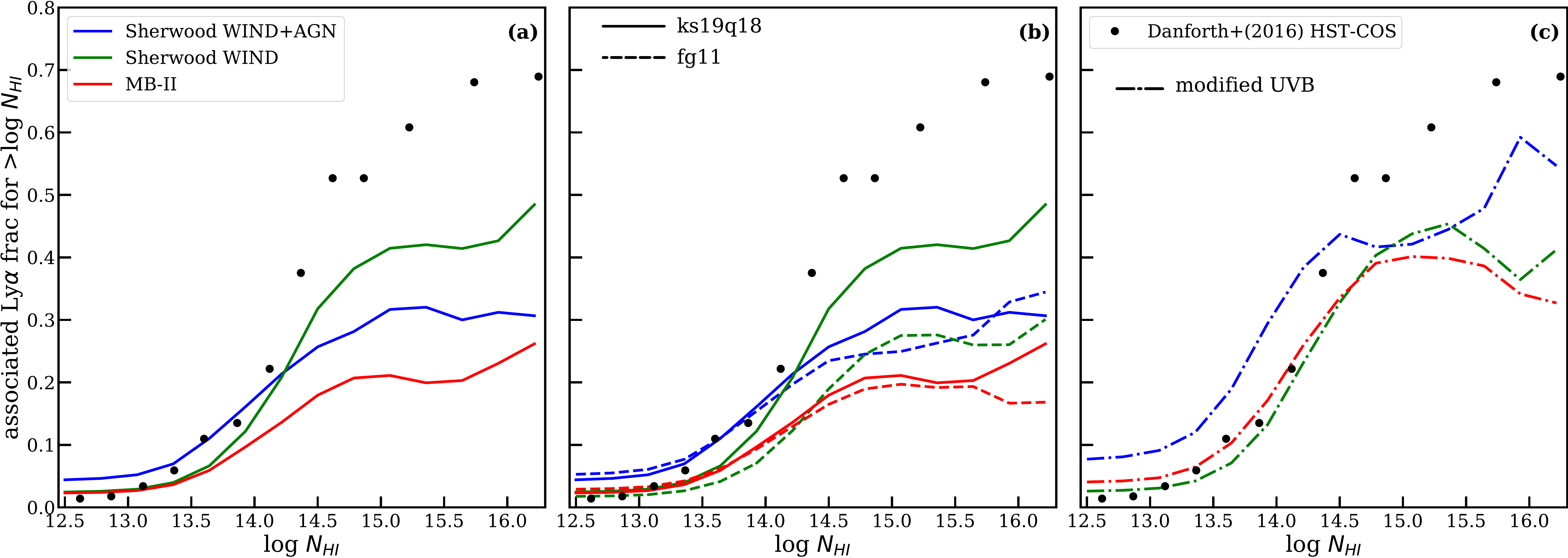}
		\caption{The fraction of \lya\ absorbers having associated absorption in high ionization species like \CIV\ , \OVI\ , \NeVIII\ as a function of $N$(\HI). 
  {\it Left panel:} Results from three simulations used here for the "ks19q18" UVB. {\it Middle panel:} The effect of changing UVB from "ks19q18" to "fg11" for different simulations. {\it Right panel:} Results for a modified UVB with the \HI\ photoionization rate is similar to "ks19q18" and the rest of the UVB is like that of "fg11". 
	}
		\label{fig:HI_align_all}
	\end{minipage}%
\end{figure*}

Another interesting observation by \citet{danforth2016} that is very useful for constraining the simulations is the fraction of \lya\ absorbers showing detectable metal ion absorption as a function of $N$(\HI). They found (see their figure 8) that more than 50\% of absorbers with log~$N$(\HI)$>$ 14.5 show detectable absorption from at least one of the high ion species (i.e., N~{\sc v}, \CIV, \OVI\ or \NeVIII).  In principle, this plot depends on the feedback processes ({which influences} how the metals are distributed over regions of different over-densities) and the
the shape of the UVB (i.e., the relative ratio of photoionization rate of \HI\ and other metal ions) assumed.

In Figure~\ref{fig:HI_align_all}, we compare this observation with results from our simulations for different UVBs. We plot the fraction of \lya\ absorbers having associated absorption by at least one of three high ionization metal species considered in this work as a function of \HI\ column density.
In panel (a) of Figure~\ref{fig:HI_align_all} we show the results for our three simulations when we use "ks19q18" UVB. It is clear that while the
metal absorption detection increases with $N$(\HI) in all simulations, none of the predicted curves are close to the observed data points. It is also interesting to note that the detectability of high ion absorption is higher in the "WIND" only simulation compare to the other two simulations at the high \HI\ column densities (i.e., for log~$N$(\HI)$>$14.25). As high column density absorbers mostly originate from regions having higher overdensities, this trend can be attributed to the higher mean metallicity and $f_1$ fraction in the "hot halo" and "condensed" regions in Sherwood "WIND" only model compared to the other two simulations.

In panel (b) of Figure~\ref{fig:HI_align_all}, we show the results when we use the "fg11" UVB. In all the discussions presented, "WIND+AGN" Sherwood simulation was producing CDDF, b-CDF, and $N$(\OVI) vs. $b$(\OVI) correlations reasonably well when we use the "fg11" UVB.  
It is interesting to note for all three simulations that for a given $N$(\HI), there is more metal absorption detected in the case of "ks19q18" UVB compared to "fg11" UVB. 
This is mainly because "fg11" UVB not only has low photoionization rate around the ionization energy of metal ions but also for \HI\ (see Table~\ref{tab:Gamma_for_uvbs}). Previous works like \citet{maitra2020b} report \HI\ CDDF with Sherwood simulations using "ks19q18" UVB matches well with the observations from \citet{danforth2016} at the low \HI\ column density end \citep[see figure 10 of][]{maitra2020b}. We confirm that as the \HI\ photoionization rate in "fg11" is half that  for the "ks19q18", Sherwood simulations using this UVB overproduce of \HI\ CDDF compared to the observed distribution. 

Next, we consider a modified UVB where we  enhanced the \HI\ photoionization rate in the case of "fg11" UVB to match that of "ks19q18" UVB. Results for this UVB are summarised in panel (c) of Figure~\ref{fig:HI_align_all}. As expected, this improves the fraction of \HI\ absorbers showing metal ion absorption in the range 14.5$<$log $N$(\HI)$<$16.0 for the case of "WIND+AGN" Sherwood simulation and MB-II simulation. However, the changes are marginal in the case of  "WIND" only Sherwood simulation. It is also clear from this figure that even after this ad hoc modification to the UVB, our simulations tend to produce fewer metal ion bearing high \HI\ column density absorbers.
Also, our "WIND+AGN" simulations tend to overproduce the fraction of metal-bearing absorbers at low $N$(\HI). 
 However, simulations used here may have some shortcomings, such as, (i) the high column density absorbers preferentially arising in the CGM, require higher mass and spacial resolution to capture the sub-grid physics correctly  \citep[e.g., previous works like][]{hafen2019,hafen2020, stern2021}, 
(ii) CGM may be influenced by the local radiation field from the host galaxy not included in the simulations used here, and (iii) the optically thin approximation used here is not adequate to capture the correct ion fraction. 

{\it The discussions presented here clearly demonstrate that the observed fraction of \lya\ absorbers with associated metals can provide a good constraint on different feedback processes in the simulations. For example, while the "WIND+AGN" simulations with "fg11" UVB reproduce CDDF, b-CDF, and $\Delta$V$_{90}$ distribution, it has problems reproducing the N(HI) vs. fraction.}


\section{Discussions}
\label{Sec:discussions}

It is now well documented in the literature that different feedback processes introduce large scatter in the observed properties of absorbers \citep[see for example,][]{oppenheimer2012,rahmati2016,Nelson2018,bradley2022}. Thus, one hopes to constrain different feedback processes using the observed statistical distributions.
As of now, simulations use one of the mean UVB computed by \citet{haardt2012,faucher2009,faucher2020}.
 Here we show, for a given simulation, a substantial spread in the statistical properties can be introduced by uncertainties in the UVB for species such as \OVI\ originating from regions spanning a wide range in density and temperature. 
Therefore, it is crucial to identify  observables that are sensitive to the changes in the UVB and the ones that depend mainly on feedback.

The differences in the computed UVBs, considered here originate from differences in the assumed comoving emissivities of sources, line of sight optical depth as a function of frequency, and the assumed spectral energy distribution of sources.
Interestingly the recently computed UVBs \citep{khaire2019, puchwein2019,faucher2020} using updated quasar luminosity function and IGM optical depths tend to have similar $\Gamma_{\hi}$ (at $z<0.5$) that are consistent with the measurements of \citet{gaikwad2017b}. However, they differ in the extreme UV to soft X-ray ranges due to the assumed SED. Most of the UVB calculations assume a single power law connecting 13.6 eV to soft X-ray regime ($\sim$2 KeV), while \citet{faucher2009} have assumed segmented power law with three components in this energy range \citep[see also][]{faucher2020}. This has resulted in lower $\Gamma$ values for high ions in their models. 
As noted above, UVB computed by \citet{faucher2009} also tends to produce lower $\Gamma_{\hi}$ due to the assumed quasar emissivity that got upgraded using recent measurements. On the other hand, non-LTE (Local Thermal Equilibrium) models of accretion disk emission suggest more complex spectra in this energy range that depends on both the mass of the Black Hole and accretion rate \citep[see, for example, figure 22 of][]{Hubeny2000}. Thus a realistic quasar SED could allow for even lower $\Gamma$ values of high ions compared to what has been found for \citet{faucher2009}.

All UVB models ignore  contributions from galaxies to the UV-Soft X-ray range. \citet{Upton2018} have investigated the contribution of X-ray binaries, hot ISM and CGM gas, virialized halo gas, and super-soft sources to the low-$z$ UVB.  They showed the contribution  from such sources in the 20-50 eV range could be comparable to the lower envelope of scatter in the quasar contribution.
Therefore, the contribution from galactic sources could be important when we consider softer QSO SEDs and sightlines closer to star-forming regions (i.e., high column density absorbers).

Note the uncertainties in the UVB translating into uncertainties in the measured metallicity, density, and size of the absorbing region are well documented \citep{Chen2017,hussain2017,Haislmaier2021,Acharya2022}. However, in these studies, absorbers are approximated to be a single phase or multiple phases in pressure equilibrium. Simulations clearly suggest given absorption line may originate from gas having a wide range of density, temperature, and physical locations. Here we show that the UVB influences not only the column density of a given species and column density ratios between different species, they also influence the $b$-parameter and $\Delta V_{90}$. The distributions of these parameters in simulations considered here (for a single UVB) are also found to depend on the feedback. 
There are indications that just changing the UVB alone will not be sufficient  to simultaneously match the properties of different species and the association of metals with \HI.
Most of the simulation studies listed above probe the effect feedback mainly on the CDDF of single ions. 

The main caveat of our work is that we explored the effect of UVB as a post-processing step assuming (i) optically thin ionization equilibrium conditions and (ii) ignoring the effect of UVB on the gas temperature. 
The \HI\ column density of most of the absorbers studied in our simulated spectra are much lower than the threshold used by \citet{rahmati2013} to introduce a correction for optically thick conditions. 
Nearly similar temperature density relationship derived for "diffuse" region in different cosmological simulations using different UVB suggests the photo-heating effects may not be that important at low-$z$ considered here. However, the temperature of the high-density SPH particle may be affected by the choice of the UVB and the inclusion of non-equilibrium cooling of the gas. This can have a secondary effect of changing the ion fraction. Therefore, it is good to check these effects by running the same simulation using two different UVBs. Alternatively, one could use post-processing approaches like `Code for Ionization and Temperature Evolution' \citep[CITE, as discussed in][]{gaikwad2019,gaikwad2021} to capture the changes in the photo heating, cooling, and non-equilibrium effects.

There are early indications in our ongoing study that statistics like (i) spatial correlation of metal line absorbers and (ii) cross-correlation of metal line absorbers with respect to the strong \lya\ absorbers may weakly depend on the UVB.  If confirmed, these may provide better tools for probing the feedback effects.

\par

\section{Summary}
\label{Sec:summary}

In this work,  we study the influence of the assumed UV ionizing background on the observable properties of high ionization metal line absorbers, using three cosmological hydrodynamic simulations (i.e., Sherwood simulations with "WIND" only and "WIND+AGN" feedback \citep{Bolton2017} and Massive Black-II simulations \citep{khandai2015}) at $z=0.5$. For this purpose, we use UVBs generated by \citet{khaire2019} for a range of quasar UV-to-soft X-ray spectral index ($\alpha$) and the one generated by \citet{faucher2009}. We compute the ion fractions for different elements using 
the photoionization code {\sc cloudy} \citep[version 17.02 of the code developed by][]{ferland1998} assuming optically thin ionization conditions. Here, we assume the influence of the UVB on the gas temperature to be negligible and probe only its effects in  changing the ionization state of the gas. Our main results  are:
\begin{enumerate}
    \item [1.]  For our fiducial UVB (i.e., "ks19q18" UVB), the "WIND+AGN" Sherwood simulation produces more (less) \OVI\ and \NeVIII\ absorbers at low (high) column densities compared to that of "WIND" only Sherwood simulation. This is because, in the "WIND" only simulation, the metallicity in the "hot halo" and "condensed" phases are higher than that in the "WIND+AGN" Sherwood simulation. The high column density absorbers tend to originate from these regions. On the other hand, metallicity in the "diffuse" and "WHIM" regions (where the low column density systems originate) are higher in the case of the "WIND+AGN" Sherwood simulation compared to that of "WIND" only simulation. Both Sherwood simulations underproduce the observed column density distributions of \OVI\ and \CIV. This may suggest that if the actual UVB is like that of \citet{khaire2019}, the feedback as implemented in  Sherwood simulations is not sufficient to explain the observations. The MB-II simulation that also incorporates "WIND+AGN" feedback produces less number of absorbers compared to the "WIND+AGN" Sherwood simulation. We attribute this to the reduction in the fraction of SPH particles having high ion fractions in the "WHIM", "hot halo", and "Condensed" regions in the MB-II simulation. The $d\mathcal{N}/dz$ predicted by the three simulations for \NeVIII\ are much lower than what has been found in direct measurements \citep[e.g.,][]{meiring2013} but consistent with what is inferred from stacking techniques \citep{frank2018}.

    \item [2.] In the case of the "WIND+AGN" Sherwood simulation, the predicted column density distribution of \OVI\ is found to be very sensitive to the assumed UVB. The CDDF is higher when the ionizing spectrum is softer (i.e. less photoionization rate for \OVI\ for a given \HI\ photoionization rate). For any choice of $\alpha$, over the range considered by \citet{khaire2019}, Sherwood "WIND+AGN" simulation under-predicts  CDDF of \OVI\ compared to the  observations by \citet{danforth2016}. On the other hand, for the "fg11", UVB of \citet{faucher2009}, the predicted CDDF is closer to the observed CDDF of  \citet{danforth2016}. However, we find that the observed CDDF of \HI\ in the case of "WIND+AGN" Sherwood simulation is not well reproduced by 
    "fg11" UVB while "ks19q18" provides a better match. Therefore, a UVB like "ks19q18" with the photoionization rate in the 50-200 eV range being $\sim$2.8 times lower will simultaneously produce the observed CDDFs of both \HI\ and \OVI.  Note the lack of direct observations of quasar SED in this regime allows such a possibility \citep[see for example,][for discussion on this]{Haislmaier2021}. However, when we consider "WIND" only Sherwood simulation, the predicted CDDF for \OVI\ is much lower than all the available observations. Also, CDDF is fairly insensitive to our choice of UVB for this simulation.  In the case of \CIV, we find an increase in the CDDF at the low column density end for WIND+AGNsimulation when we use the "fg11" background.  But this alone is not sufficient to cover the shortcoming of this simulation. However, in the case of "WIND" only simulations, we find the influence of the "fg11" UVB to be not very significant.

\item[3.]  We find "WIND+AGN" Sherwood simulation produces  higher values  of $b$-parameter and $\Delta V_{90}$  for \OVI\ and \NeVIII\ compared to those for the "WIND" only Sherwood simulation. However, these are not significantly different when we consider \CIV. In the case of MB-II simulations, the obtained value of $b$-parameter and  $\Delta V_{90}$ are lower than what we find for the "WIND+AGN" Sherwood simulation. We attribute this to the higher resolution of the simulation and higher sampling used for generating our spectra.
We show that both $b$ and $\Delta V_{90}$ distributions are affected by our choice of the UVB. This is because the region contributing to a given absorption line changes with the UVB. 
The general tendency is to produce larger values when we consider softer spectra. However, the magnitude of the enhancement varies from simulation to simulation for a given ion.
For example, in the case of \OVI\ we find the largest enhancement is seen in the case of "WIND" only simulation compared to "WIND+AGN" simulation. However, the trend is reversed in the case of \CIV. 

\item[4.] We also explore the correlation between $N$ and $b$ parameters. We do find a weak correlation between $b$(\OVI) and $N$(\OVI) for log $N$(\OVI)$\ge$13.0. The "WIND+AGN" Sherwood simulation produces a larger median $b$-value for a given \OVI\ column density bin compared to the "WIND" only Sherwood simulation. However, in the case of \CIV\ no clear correlation is evident between $b$(\CIV) and $N$(\CIV). Also, We do not find any significant difference between the results of the two Sherwood simulations.  We show the distribution can be influenced by the UVB. In particular, in the case of \OVI\ for a given $N$, the $b$-values show an enhancement if we use "fg11" UVB in the case of "WIND" only simulation. The enhancement is not that significant in the case of the "WIND+AGN" Sherwood simulation.

\item[5.] All our models show a trend of increase in the fraction of \lya\ absorbers showing metals with increasing $N$(\HI).
However, metal bearing \lya\ clouds at high \NHI\ are less than what has been found by \citet{danforth2016}. This seems to be the case for both our fiducial and "fg11" UVBs.  While "fg11" UVB increases the column density of metal ions, it has also increased $N${(\HI)} owing to the less $\Gamma_{\HI}$ in this UVB.  This leads to no substantial increase in the fraction of metal bearing \lya\ absorbers at a given $N$(\HI). This once again suggests that the feedback effects in the simulations considered here are not sufficient to reproduce the observations for the range UVBs considered here.
To explore this further, we also consider cases where $\Gamma_{\HI}$ is close to that of "ks19q18" UVB and the photoionization rate for metal ions close to that of "fg11" UVB. While such an UVB increases the fraction a little bit, we still find the observed distribution to be larger than that predicted by the simulation.   
\end{enumerate}

Results presented here confirm that the statistical distribution of absorbers can be significantly influenced by the assumed UVB. However, the magnitude of the effect varies between different simulations and
different species. The effect is higher for the ion that originates from a wide density and temperature range and produces detectable absorption. For example, in the case of "WIND+AGN" Sherwood simulation using "fg11" UVB, the predicted CDDF (panel k of Figure~\ref{fig:cddf_all1}), $b$-distribution (see Figure~\ref{fig:b_cdf}) and $N$ vs. $b$ correlations (see Figure~\ref{fig:b_N}) roughly follow the observed trend. While the same simulation underproduces the CDDF when we use "ks19q18" UVB. 
This clearly demonstrates that observed CDDF alone can not be a good probe of feedback processes.  Therefore, to reproduce the observed distributions, one should not only consider varying the feedback prescriptions but also consider the uncertainties in the UVB. Our exploration also indicates the importance of simultaneously comparing the properties of different ions. For example, "WIND+AGN" Sherwood simulation using "fg11" under predicts the CDDF of \CIV. We also notice that all models considered here fail to reproduce the observed relationship between $N$(\HI) and the fraction of absorbers having detectable metals.  Clearly, attempting to simultaneously reproduce various observational constraints is important for a clearer understanding of the role played by different feedback processes and the UVB.

\section*{Acknowledgments}
{
We acknowledge the use of High-performance computing facilities PERSEUS and PEGASUS at IUCAA. 
We are grateful to K. Subramanian, V. Khaire, and A. Mohapatra for insightful discussions regarding this work. We thank the anonymous referee and the scientific editor, Joop Schaye, for the valuable comments and suggestions. The Sherwood simulations were performed using the Curie supercomputer at the Tre Grand Centre de Calcul (TGCC), and the DiRAC Data Analytic system at the
University of Cambridge, operated by the University of Cambridge High
Performance Computing Service on behalf of the STFC DiRAC HPC Facility
(www.dirac.ac.uk) . This was funded by BIS National E-infrastructure
capital grant (ST/K001590/1), STFC capital grants ST/H008861/1 and
ST/H00887X/1, and STFC DiRAC Operations grant ST/K00333X/1. DiRAC is part of
the National E-Infrastructure. The MB-II simulation was run on the Cray XT5 supercomputer– Kraken – at the National Institute for Computational Sciences, supported by the National Science Foundation (NSF) PetaApps program, OCI-0749212. Soumak Maitra acknowledges support from PRIN INAF - NewIGM programme. Nishikanta Khandai acknowledges support from the IUCAA Associateship programme.
}

\section*{Data Availability}
The data underlying this article are available in the article and in its online supplementary material.




\bibliographystyle{mnras}
\bibliography{main} 

\begin{thebibliography}{}
\makeatletter
\relax
\def\mn@urlcharsother{\let\do\@makeother \do\$\do\&\do\#\do\^\do\_\do\%\do\~}
\def\mn@doi{\begingroup\mn@urlcharsother \@ifnextchar [ {\mn@doi@}
  {\mn@doi@[]}}
\def\mn@doi@[#1]#2{\def\@tempa{#1}\ifx\@tempa\@empty \href
  {http://dx.doi.org/#2} {doi:#2}\else \href {http://dx.doi.org/#2} {#1}\fi
  \endgroup}
\def\mn@eprint#1#2{\mn@eprint@#1:#2::\@nil}
\def\mn@eprint@arXiv#1{\href {http://arxiv.org/abs/#1} {{\tt arXiv:#1}}}
\def\mn@eprint@dblp#1{\href {http://dblp.uni-trier.de/rec/bibtex/#1.xml}
  {dblp:#1}}
\def\mn@eprint@#1:#2:#3:#4\@nil{\def\@tempa {#1}\def\@tempb {#2}\def\@tempc
  {#3}\ifx \@tempc \@empty \let \@tempc \@tempb \let \@tempb \@tempa \fi \ifx
  \@tempb \@empty \def\@tempb {arXiv}\fi \@ifundefined
  {mn@eprint@\@tempb}{\@tempb:\@tempc}{\expandafter \expandafter \csname
  mn@eprint@\@tempb\endcsname \expandafter{\@tempc}}}

\bibitem[\protect\citeauthoryear{{Acharya} \& {Khaire}}{{Acharya} \&
  {Khaire}}{2022}]{Acharya2022}
{Acharya} A.,  {Khaire} V.,  2022, \mn@doi [\mnras] {10.1093/mnras/stab3316},
  \href {https://ui.adsabs.harvard.edu/abs/2022MNRAS.509.5559A} {509, 5559}

\bibitem[\protect\citeauthoryear{{Aguirre}, {Schaye}, {Kim}, {Theuns}, {Rauch}
  \& {Sargent}}{{Aguirre} et~al.}{2004}]{aguirre2004}
{Aguirre} A.,  {Schaye} J.,  {Kim} T.-S.,  {Theuns} T.,  {Rauch} M.,
  {Sargent} W. L.~W.,  2004, \mn@doi [\apj] {10.1086/380961}, \href
  {https://ui.adsabs.harvard.edu/abs/2004ApJ...602...38A} {602, 38}

\bibitem[\protect\citeauthoryear{{Aguirre}, {Dow-Hygelund}, {Schaye}  \&
  {Theuns}}{{Aguirre} et~al.}{2008}]{aguirre2008}
{Aguirre} A.,  {Dow-Hygelund} C.,  {Schaye} J.,   {Theuns} T.,  2008, \mn@doi
  [\apj] {10.1086/592554}, \href
  {https://ui.adsabs.harvard.edu/abs/2008ApJ...689..851A} {689, 851}

\bibitem[\protect\citeauthoryear{Allen, Dunn, Fabian, Taylor  \&
  Reynolds}{Allen et~al.}{2006}]{allen2006}
Allen S.~W.,  Dunn R.,  Fabian A.,  Taylor G.,   Reynolds C.,  2006, Monthly
  Notices of the Royal Astronomical Society, 372, 21

\bibitem[\protect\citeauthoryear{{Appleby}, {Dav{\'e}}, {Sorini},
  {Storey-Fisher}  \& {Smith}}{{Appleby} et~al.}{2021}]{appleby2021}
{Appleby} S.,  {Dav{\'e}} R.,  {Sorini} D.,  {Storey-Fisher} K.,   {Smith} B.,
  2021, \mn@doi [\mnras] {10.1093/mnras/stab2310}, \href
  {https://ui.adsabs.harvard.edu/abs/2021MNRAS.507.2383A} {507, 2383}

\bibitem[\protect\citeauthoryear{{Artale} et~al.,}{{Artale}
  et~al.}{2022}]{artale2022}
{Artale} M.~C.,  et~al., 2022, \mn@doi [\mnras] {10.1093/mnras/stab3281}, \href
  {https://ui.adsabs.harvard.edu/abs/2022MNRAS.510..399A} {510, 399}

\bibitem[\protect\citeauthoryear{{Bi}}{{Bi}}{1993}]{bi1993}
{Bi} H.,  1993, \mn@doi [\apj] {10.1086/172380}, \href
  {http://adsabs.harvard.edu/abs/1993ApJ...405..479B} {405, 479}

\bibitem[\protect\citeauthoryear{{Bolton}, {Puchwein}, {Sijacki}, {Haehnelt},
  {Kim}, {Meiksin}, {Regan}  \& {Viel}}{{Bolton} et~al.}{2017}]{Bolton2017}
{Bolton} J.~S.,  {Puchwein} E.,  {Sijacki} D.,  {Haehnelt} M.~G.,  {Kim} T.-S.,
   {Meiksin} A.,  {Regan} J.~A.,   {Viel} M.,  2017, \mn@doi [\mnras]
  {10.1093/mnras/stw2397}, \href
  {https://ui.adsabs.harvard.edu/abs/2017MNRAS.464..897B} {464, 897}

\bibitem[\protect\citeauthoryear{{Bower}, {Benson}, {Malbon}, {Helly}, {Frenk},
  {Baugh}, {Cole}  \& {Lacey}}{{Bower} et~al.}{2006}]{bower2006}
{Bower} R.~G.,  {Benson} A.~J.,  {Malbon} R.,  {Helly} J.~C.,  {Frenk} C.~S.,
  {Baugh} C.~M.,  {Cole} S.,   {Lacey} C.~G.,  2006, \mn@doi [\mnras]
  {10.1111/j.1365-2966.2006.10519.x}, \href
  {https://ui.adsabs.harvard.edu/abs/2006MNRAS.370..645B} {370, 645}

\bibitem[\protect\citeauthoryear{{Bradley}, {Dav{\'e}}, {Cui}, {Smith}  \&
  {Sorini}}{{Bradley} et~al.}{2022}]{bradley2022}
{Bradley} L.,  {Dav{\'e}} R.,  {Cui} W.,  {Smith} B.,   {Sorini} D.,  2022,
  arXiv e-prints, \href {https://ui.adsabs.harvard.edu/abs/2022arXiv220315055B}
  {p. arXiv:2203.15055}

\bibitem[\protect\citeauthoryear{{Chen}, {Johnson}, {Zahedy}, {Rauch}  \&
  {Mulchaey}}{{Chen} et~al.}{2017}]{Chen2017}
{Chen} H.-W.,  {Johnson} S.~D.,  {Zahedy} F.~S.,  {Rauch} M.,   {Mulchaey}
  J.~S.,  2017, \mn@doi [\apjl] {10.3847/2041-8213/aa762d}, \href
  {https://ui.adsabs.harvard.edu/abs/2017ApJ...842L..19C} {842, L19}

\bibitem[\protect\citeauthoryear{{Choudhury}, {Srianand}  \&
  {Padmanabhan}}{{Choudhury} et~al.}{2001}]{choudhury2001}
{Choudhury} T.~R.,  {Srianand} R.,   {Padmanabhan} T.,  2001, \mn@doi [\apj]
  {10.1086/322327}, \href {http://cdsads.u-strasbg.fr/abs/2001ApJ...559...29C}
  {559, 29}

\bibitem[\protect\citeauthoryear{{Christiansen}, {Dav{\'e}}, {Sorini}  \&
  {Angl{\'e}s-Alc{\'a}zar}}{{Christiansen} et~al.}{2020}]{Christiansen2020}
{Christiansen} J.~F.,  {Dav{\'e}} R.,  {Sorini} D.,   {Angl{\'e}s-Alc{\'a}zar}
  D.,  2020, \mn@doi [\mnras] {10.1093/mnras/staa3007}, \href
  {https://ui.adsabs.harvard.edu/abs/2020MNRAS.499.2617C} {499, 2617}

\bibitem[\protect\citeauthoryear{{Cooksey}, {Thom}, {Prochaska}  \&
  {Chen}}{{Cooksey} et~al.}{2010}]{cooksey2010}
{Cooksey} K.~L.,  {Thom} C.,  {Prochaska} J.~X.,   {Chen} H.-W.,  2010, \mn@doi
  [\apj] {10.1088/0004-637X/708/1/868}, \href
  {https://ui.adsabs.harvard.edu/abs/2010ApJ...708..868C} {708, 868}

\bibitem[\protect\citeauthoryear{{Croton} et~al.,}{{Croton}
  et~al.}{2006}]{croton2006}
{Croton} D.~J.,  et~al., 2006, \mn@doi [\mnras]
  {10.1111/j.1365-2966.2005.09675.x}, \href
  {https://ui.adsabs.harvard.edu/abs/2006MNRAS.365...11C} {365, 11}

\bibitem[\protect\citeauthoryear{{Danforth} \& {Shull}}{{Danforth} \&
  {Shull}}{2008}]{danforth2008}
{Danforth} C.~W.,  {Shull} J.~M.,  2008, \mn@doi [\apj] {10.1086/587127}, \href
  {http://adsabs.harvard.edu/abs/2008ApJ...679..194D} {679, 194}

\bibitem[\protect\citeauthoryear{{Danforth}, {Shull}, {Rosenberg}  \&
  {Stocke}}{{Danforth} et~al.}{2006}]{danforth2006}
{Danforth} C.~W.,  {Shull} J.~M.,  {Rosenberg} J.~L.,   {Stocke} J.~T.,  2006,
  \mn@doi [\apj] {10.1086/500191}, \href
  {https://ui.adsabs.harvard.edu/abs/2006ApJ...640..716D} {640, 716}

\bibitem[\protect\citeauthoryear{{Danforth} et~al.,}{{Danforth}
  et~al.}{2016}]{danforth2016}
{Danforth} C.~W.,  et~al., 2016, \mn@doi [\apj] {10.3847/0004-637X/817/2/111},
  \href {http://adsabs.harvard.edu/abs/2016ApJ...817..111D} {817, 111}

\bibitem[\protect\citeauthoryear{{Dav{\'e}}, {Oppenheimer}, {Katz}, {Kollmeier}
   \& {Weinberg}}{{Dav{\'e}} et~al.}{2010}]{dave2010}
{Dav{\'e}} R.,  {Oppenheimer} B.~D.,  {Katz} N.,  {Kollmeier} J.~A.,
  {Weinberg} D.~H.,  2010, \mn@doi [\mnras] {10.1111/j.1365-2966.2010.17279.x},
  \href {http://adsabs.harvard.edu/abs/2010MNRAS.408.2051D} {408, 2051}

\bibitem[\protect\citeauthoryear{{Di Matteo}, {Springel}  \& {Hernquist}}{{Di
  Matteo} et~al.}{2005}]{dimatteo2005}
{Di Matteo} T.,  {Springel} V.,   {Hernquist} L.,  2005, \mn@doi [\nat]
  {10.1038/nature03335}, \href
  {https://ui.adsabs.harvard.edu/abs/2005Natur.433..604D} {433, 604}

\bibitem[\protect\citeauthoryear{{Faucher-Gigu{\`e}re}}{{Faucher-Gigu{\`e}re}}{2020}]{faucher2020}
{Faucher-Gigu{\`e}re} C.-A.,  2020, \mn@doi [\mnras] {10.1093/mnras/staa302},
  \href {https://ui.adsabs.harvard.edu/abs/2020MNRAS.493.1614F} {493, 1614}

\bibitem[\protect\citeauthoryear{{Faucher-Gigu{\`e}re}, {Lidz}, {Zaldarriaga}
  \& {Hernquist}}{{Faucher-Gigu{\`e}re} et~al.}{2009}]{faucher2009}
{Faucher-Gigu{\`e}re} C.-A.,  {Lidz} A.,  {Zaldarriaga} M.,   {Hernquist} L.,
  2009, \mn@doi [\apj] {10.1088/0004-637X/703/2/1416}, \href
  {http://adsabs.harvard.edu/abs/2009ApJ...703.1416F} {703, 1416}

\bibitem[\protect\citeauthoryear{{Fechner}}{{Fechner}}{2011}]{fechner2011}
{Fechner} C.,  2011, \mn@doi [\aap] {10.1051/0004-6361/201117080}, \href
  {https://ui.adsabs.harvard.edu/abs/2011A&A...532A..62F} {532, A62}

\bibitem[\protect\citeauthoryear{{Ferland}, {Korista}, {Verner}, {Ferguson},
  {Kingdon}  \& {Verner}}{{Ferland} et~al.}{1998a}]{ferland1998}
{Ferland} G.~J.,  {Korista} K.~T.,  {Verner} D.~A.,  {Ferguson} J.~W.,
  {Kingdon} J.~B.,   {Verner} E.~M.,  1998a, \mn@doi [\pasp] {10.1086/316190},
  \href {http://adsabs.harvard.edu/abs/1998PASP..110..761F} {110, 761}

\bibitem[\protect\citeauthoryear{{Ferland}, {Korista}, {Verner}, {Ferguson},
  {Kingdon}  \& {Verner}}{{Ferland} et~al.}{1998b}]{cloudy1998}
{Ferland} G.~J.,  {Korista} K.~T.,  {Verner} D.~A.,  {Ferguson} J.~W.,
  {Kingdon} J.~B.,   {Verner} E.~M.,  1998b, \mn@doi [\pasp] {10.1086/316190},
  \href {https://ui.adsabs.harvard.edu/abs/1998PASP..110..761F} {110, 761}

\bibitem[\protect\citeauthoryear{{Frank}, {Pieri}, {Mathur}, {Danforth}  \&
  {Shull}}{{Frank} et~al.}{2018}]{frank2018}
{Frank} S.,  {Pieri} M.~M.,  {Mathur} S.,  {Danforth} C.~W.,   {Shull} J.~M.,
  2018, \mn@doi [\mnras] {10.1093/mnras/sty294}, \href
  {https://ui.adsabs.harvard.edu/abs/2018MNRAS.476.1356F} {476, 1356}

\bibitem[\protect\citeauthoryear{{Gaikwad}, {Khaire}, {Choudhury}  \&
  {Srianand}}{{Gaikwad} et~al.}{2017a}]{gaikwad2017a}
{Gaikwad} P.,  {Khaire} V.,  {Choudhury} T.~R.,   {Srianand} R.,  2017a,
  \mn@doi [\mnras] {10.1093/mnras/stw3086}, \href
  {http://adsabs.harvard.edu/abs/2017MNRAS.466..838G} {466, 838}

\bibitem[\protect\citeauthoryear{{Gaikwad}, {Srianand}, {Choudhury}  \&
  {Khaire}}{{Gaikwad} et~al.}{2017b}]{gaikwad2017b}
{Gaikwad} P.,  {Srianand} R.,  {Choudhury} T.~R.,   {Khaire} V.,  2017b,
  \mn@doi [\mnras] {10.1093/mnras/stx248}, \href
  {http://adsabs.harvard.edu/abs/2017MNRAS.467.3172G} {467, 3172}

\bibitem[\protect\citeauthoryear{{Gaikwad}, {Srianand}, {Khaire}  \&
  {Choudhury}}{{Gaikwad} et~al.}{2019}]{gaikwad2019}
{Gaikwad} P.,  {Srianand} R.,  {Khaire} V.,   {Choudhury} T.~R.,  2019, \mn@doi
  [\mnras] {10.1093/mnras/stz2692}, \href
  {https://ui.adsabs.harvard.edu/abs/2019MNRAS.490.1588G} {490, 1588}

\bibitem[\protect\citeauthoryear{{Gaikwad}, {Srianand}, {Haehnelt}  \&
  {Choudhury}}{{Gaikwad} et~al.}{2021}]{gaikwad2021}
{Gaikwad} P.,  {Srianand} R.,  {Haehnelt} M.~G.,   {Choudhury} T.~R.,  2021,
  \mn@doi [\mnras] {10.1093/mnras/stab2017}, \href
  {https://ui.adsabs.harvard.edu/abs/2021MNRAS.506.4389G} {506, 4389}

\bibitem[\protect\citeauthoryear{{Haardt} \& {Madau}}{{Haardt} \&
  {Madau}}{1996}]{haardt1996}
{Haardt} F.,  {Madau} P.,  1996, \mn@doi [\apj] {10.1086/177035}, \href
  {http://adsabs.harvard.edu/abs/1996ApJ...461...20H} {461, 20}

\bibitem[\protect\citeauthoryear{{Haardt} \& {Madau}}{{Haardt} \&
  {Madau}}{2001}]{haardt2001}
{Haardt} F.,  {Madau} P.,  2001, in {Neumann} D.~M.,  {Tran} J.~T.~V.,  eds,
  Clusters of Galaxies and the High Redshift Universe Observed in X-rays. p.~64
  (\mn@eprint {arXiv} {astro-ph/0106018})

\bibitem[\protect\citeauthoryear{{Haardt} \& {Madau}}{{Haardt} \&
  {Madau}}{2012}]{haardt2012}
{Haardt} F.,  {Madau} P.,  2012, \mn@doi [\apj] {10.1088/0004-637X/746/2/125},
  \href {http://adsabs.harvard.edu/abs/2012ApJ...746..125H} {746, 125}

\bibitem[\protect\citeauthoryear{{Hafen} et~al.,}{{Hafen}
  et~al.}{2019}]{hafen2019}
{Hafen} Z.,  et~al., 2019, \mn@doi [\mnras] {10.1093/mnras/stz1773}, \href
  {https://ui.adsabs.harvard.edu/abs/2019MNRAS.488.1248H} {488, 1248}

\bibitem[\protect\citeauthoryear{{Hafen} et~al.,}{{Hafen}
  et~al.}{2020}]{hafen2020}
{Hafen} Z.,  et~al., 2020, \mn@doi [\mnras] {10.1093/mnras/staa902}, \href
  {https://ui.adsabs.harvard.edu/abs/2020MNRAS.494.3581H} {494, 3581}

\bibitem[\protect\citeauthoryear{{Haislmaier}, {Tripp}, {Katz}, {Prochaska},
  {Burchett}, {O'Meara}  \& {Werk}}{{Haislmaier} et~al.}{2021}]{Haislmaier2021}
{Haislmaier} K.~J.,  {Tripp} T.~M.,  {Katz} N.,  {Prochaska} J.~X.,  {Burchett}
  J.~N.,  {O'Meara} J.~M.,   {Werk} J.~K.,  2021, \mn@doi [\mnras]
  {10.1093/mnras/staa3544}, \href
  {https://ui.adsabs.harvard.edu/abs/2021MNRAS.502.4993H} {502, 4993}

\bibitem[\protect\citeauthoryear{{Heckman}, {Norman}, {Strickland}  \&
  {Sembach}}{{Heckman} et~al.}{2002}]{heckman2002}
{Heckman} T.~M.,  {Norman} C.~A.,  {Strickland} D.~K.,   {Sembach} K.~R.,
  2002, \mn@doi [\apj] {10.1086/342232}, \href
  {https://ui.adsabs.harvard.edu/abs/2002ApJ...577..691H} {577, 691}

\bibitem[\protect\citeauthoryear{{Howk}, {Ribaudo}, {Lehner}, {Prochaska}  \&
  {Chen}}{{Howk} et~al.}{2009}]{howk2009}
{Howk} J.~C.,  {Ribaudo} J.~S.,  {Lehner} N.,  {Prochaska} J.~X.,   {Chen}
  H.-W.,  2009, \mn@doi [\mnras] {10.1111/j.1365-2966.2009.14805.x}, \href
  {https://ui.adsabs.harvard.edu/abs/2009MNRAS.396.1875H} {396, 1875}

\bibitem[\protect\citeauthoryear{{Hubeny}, {Agol}, {Blaes}  \&
  {Krolik}}{{Hubeny} et~al.}{2000}]{Hubeny2000}
{Hubeny} I.,  {Agol} E.,  {Blaes} O.,   {Krolik} J.~H.,  2000, \mn@doi [\apj]
  {10.1086/308708}, \href
  {https://ui.adsabs.harvard.edu/abs/2000ApJ...533..710H} {533, 710}

\bibitem[\protect\citeauthoryear{{Hussain}, {Khaire}, {Srianand}, {Muzahid}  \&
  {Pathak}}{{Hussain} et~al.}{2017}]{hussain2017}
{Hussain} T.,  {Khaire} V.,  {Srianand} R.,  {Muzahid} S.,   {Pathak} A.,
  2017, \mn@doi [\mnras] {10.1093/mnras/stw3265}, \href
  {https://ui.adsabs.harvard.edu/abs/2017MNRAS.466.3133H} {466, 3133}

\bibitem[\protect\citeauthoryear{{Katz}, {Weinberg}  \& {Hernquist}}{{Katz}
  et~al.}{1996}]{katz1996}
{Katz} N.,  {Weinberg} D.~H.,   {Hernquist} L.,  1996, \mn@doi [\apjs]
  {10.1086/192305}, \href {http://adsabs.harvard.edu/abs/1996ApJS..105...19K}
  {105, 19}

\bibitem[\protect\citeauthoryear{{Keeney}, {Danforth}, {Stocke}, {France}  \&
  {Green}}{{Keeney} et~al.}{2012}]{keeney2012}
{Keeney} B.~A.,  {Danforth} C.~W.,  {Stocke} J.~T.,  {France} K.,   {Green}
  J.~C.,  2012, \mn@doi [\pasp] {10.1086/667392}, \href
  {http://adsabs.harvard.edu/abs/2012PASP..124..830K} {124, 830}

\bibitem[\protect\citeauthoryear{{Khaire} \& {Srianand}}{{Khaire} \&
  {Srianand}}{2015}]{khaire2015b}
{Khaire} V.,  {Srianand} R.,  2015, \mn@doi [\mnras] {10.1093/mnrasl/slv060},
  \href {http://adsabs.harvard.edu/abs/2015MNRAS.451L..30K} {451, L30}

\bibitem[\protect\citeauthoryear{{Khaire} \& {Srianand}}{{Khaire} \&
  {Srianand}}{2019}]{khaire2019}
{Khaire} V.,  {Srianand} R.,  2019, \mn@doi [\mnras] {10.1093/mnras/stz174},
  \href {http://adsabs.harvard.edu/abs/2019MNRAS.484.4174K} {484, 4174}

\bibitem[\protect\citeauthoryear{{Khaire} et~al.,}{{Khaire}
  et~al.}{2019}]{khaire2019a}
{Khaire} V.,  et~al., 2019, \mn@doi [\mnras] {10.1093/mnras/stz344}, \href
  {https://ui.adsabs.harvard.edu/abs/2019MNRAS.486..769K} {486, 769}

\bibitem[\protect\citeauthoryear{{Khandai}, {Di Matteo}, {Croft}, {Wilkins},
  {Feng}, {Tucker}, {DeGraf}  \& {Liu}}{{Khandai} et~al.}{2015}]{khandai2015}
{Khandai} N.,  {Di Matteo} T.,  {Croft} R.,  {Wilkins} S.,  {Feng} Y.,
  {Tucker} E.,  {DeGraf} C.,   {Liu} M.-S.,  2015, \mn@doi [\mnras]
  {10.1093/mnras/stv627}, \href
  {https://ui.adsabs.harvard.edu/abs/2015MNRAS.450.1349K} {450, 1349}

\bibitem[\protect\citeauthoryear{{Kollmeier} et~al.,}{{Kollmeier}
  et~al.}{2014}]{kollmeier2014}
{Kollmeier} J.~A.,  et~al., 2014, \mn@doi [\apjl]
  {10.1088/2041-8205/789/2/L32}, \href
  {http://adsabs.harvard.edu/abs/2014ApJ...789L..32K} {789, L32}

\bibitem[\protect\citeauthoryear{{Komatsu} et~al.,}{{Komatsu}
  et~al.}{2011}]{komatsu2011}
{Komatsu} E.,  et~al., 2011, \mn@doi [\apjs] {10.1088/0067-0049/192/2/18},
  \href {https://ui.adsabs.harvard.edu/abs/2011ApJS..192...18K} {192, 18}

\bibitem[\protect\citeauthoryear{{Lehner}, {Savage}, {Wakker}, {Sembach}  \&
  {Tripp}}{{Lehner} et~al.}{2006}]{lehner2006}
{Lehner} N.,  {Savage} B.~D.,  {Wakker} B.~P.,  {Sembach} K.~R.,   {Tripp}
  T.~M.,  2006, \mn@doi [\apjs] {10.1086/500932}, \href
  {https://ui.adsabs.harvard.edu/abs/2006ApJS..164....1L} {164, 1}

\bibitem[\protect\citeauthoryear{{Lewis}, {Challinor}  \& {Lasenby}}{{Lewis}
  et~al.}{2000}]{lewis2020}
{Lewis} A.,  {Challinor} A.,   {Lasenby} A.,  2000, \mn@doi [\apj]
  {10.1086/309179}, \href
  {https://ui.adsabs.harvard.edu/abs/2000ApJ...538..473L} {538, 473}

\bibitem[\protect\citeauthoryear{{Li} et~al.,}{{Li} et~al.}{2022}]{Li2022}
{Li} R.,  et~al., 2022, \mn@doi [\apj] {10.3847/1538-4357/ac8359}, \href
  {https://ui.adsabs.harvard.edu/abs/2022ApJ...936...11L} {936, 11}

\bibitem[\protect\citeauthoryear{{Maitra}, {Srianand}, {Gaikwad}  \&
  {Khandai}}{{Maitra} et~al.}{2020a}]{maitra2020b}
{Maitra} S.,  {Srianand} R.,  {Gaikwad} P.,   {Khandai} N.,  2020a, arXiv
  e-prints, \href {https://ui.adsabs.harvard.edu/abs/2020arXiv201205926M} {p.
  arXiv:2012.05926}

\bibitem[\protect\citeauthoryear{{Maitra}, {Srianand}, {Gaikwad}, {Choudhury},
  {Paranjape}  \& {Petitjean}}{{Maitra} et~al.}{2020b}]{maitra2020}
{Maitra} S.,  {Srianand} R.,  {Gaikwad} P.,  {Choudhury} T.~R.,  {Paranjape}
  A.,   {Petitjean} P.,  2020b, \mn@doi [\mnras] {10.1093/mnras/staa2847},
  \href {https://ui.adsabs.harvard.edu/abs/2020MNRAS.498.6100M} {498, 6100}

\bibitem[\protect\citeauthoryear{{Manuwal}, {Narayanan}, {Udhwani}, {Srianand},
  {Savage}, {Charlton}  \& {Misawa}}{{Manuwal} et~al.}{2021}]{manuwal2021}
{Manuwal} A.,  {Narayanan} A.,  {Udhwani} P.,  {Srianand} R.,  {Savage} B.~D.,
  {Charlton} J.~C.,   {Misawa} T.,  2021, \mn@doi [\mnras]
  {10.1093/mnras/stab1556}, \href
  {https://ui.adsabs.harvard.edu/abs/2021MNRAS.505.3635M} {505, 3635}

\bibitem[\protect\citeauthoryear{{Marra} et~al.,}{{Marra}
  et~al.}{2021}]{Marra2021}
{Marra} R.,  et~al., 2021, \mn@doi [\mnras] {10.1093/mnras/stab2896}, \href
  {https://ui.adsabs.harvard.edu/abs/2021MNRAS.508.4938M} {508, 4938}

\bibitem[\protect\citeauthoryear{{Meiksin}}{{Meiksin}}{2009}]{meiksin2009}
{Meiksin} A.~A.,  2009, \mn@doi [Reviews of Modern Physics]
  {10.1103/RevModPhys.81.1405}, \href
  {http://adsabs.harvard.edu/abs/2009RvMP...81.1405M} {81, 1405}

\bibitem[\protect\citeauthoryear{{Meiring}, {Tripp}, {Werk}, {Howk}, {Jenkins},
  {Prochaska}, {Lehner}  \& {Sembach}}{{Meiring} et~al.}{2013}]{meiring2013}
{Meiring} J.~D.,  {Tripp} T.~M.,  {Werk} J.~K.,  {Howk} J.~C.,  {Jenkins}
  E.~B.,  {Prochaska} J.~X.,  {Lehner} N.,   {Sembach} K.~R.,  2013, \mn@doi
  [\apj] {10.1088/0004-637X/767/1/49}, \href
  {https://ui.adsabs.harvard.edu/abs/2013ApJ...767...49M} {767, 49}

\bibitem[\protect\citeauthoryear{{Monaghan}}{{Monaghan}}{1992}]{monaghan1992}
{Monaghan} J.~J.,  1992, \mn@doi [\araa] {10.1146/annurev.aa.30.090192.002551},
  \href {http://adsabs.harvard.edu/abs/1992ARA%26A..30..543M} {30, 543}

\bibitem[\protect\citeauthoryear{{Narayanan}, {Wakker}  \&
  {Savage}}{{Narayanan} et~al.}{2009}]{narayanan2009}
{Narayanan} A.,  {Wakker} B.~P.,   {Savage} B.~D.,  2009, \mn@doi [\apj]
  {10.1088/0004-637X/703/1/74}, \href
  {https://ui.adsabs.harvard.edu/abs/2009ApJ...703...74N} {703, 74}

\bibitem[\protect\citeauthoryear{{Nelson} et~al.,}{{Nelson}
  et~al.}{2018}]{Nelson2018}
{Nelson} D.,  et~al., 2018, \mn@doi [\mnras] {10.1093/mnras/sty656}, \href
  {https://ui.adsabs.harvard.edu/abs/2018MNRAS.477..450N} {477, 450}

\bibitem[\protect\citeauthoryear{{Oppenheimer} \& {Dav{\'e}}}{{Oppenheimer} \&
  {Dav{\'e}}}{2009}]{oppenheimer2009}
{Oppenheimer} B.~D.,  {Dav{\'e}} R.,  2009, \mn@doi [\mnras]
  {10.1111/j.1365-2966.2009.14676.x}, \href
  {http://adsabs.harvard.edu/abs/2009MNRAS.395.1875O} {395, 1875}

\bibitem[\protect\citeauthoryear{{Oppenheimer} \& {Schaye}}{{Oppenheimer} \&
  {Schaye}}{2013a}]{oppenheimer2013a}
{Oppenheimer} B.~D.,  {Schaye} J.,  2013a, \mn@doi [\mnras]
  {10.1093/mnras/stt1043}, \href
  {http://adsabs.harvard.edu/abs/2013MNRAS.434.1043O} {434, 1043}

\bibitem[\protect\citeauthoryear{{Oppenheimer} \& {Schaye}}{{Oppenheimer} \&
  {Schaye}}{2013b}]{oppenheimer2013b}
{Oppenheimer} B.~D.,  {Schaye} J.,  2013b, \mn@doi [\mnras]
  {10.1093/mnras/stt1150}, \href
  {https://ui.adsabs.harvard.edu/abs/2013MNRAS.434.1063O} {434, 1063}

\bibitem[\protect\citeauthoryear{Oppenheimer, Dav{\'e}, Katz, Kollmeier  \&
  Weinberg}{Oppenheimer et~al.}{2012}]{oppenheimer2012}
Oppenheimer B.~D.,  Dav{\'e} R.,  Katz N.,  Kollmeier J.~A.,   Weinberg D.~H.,
  2012, Monthly Notices of the Royal Astronomical Society, 420, 829

\bibitem[\protect\citeauthoryear{{Oppenheimer} et~al.,}{{Oppenheimer}
  et~al.}{2016}]{oppenheimer2016}
{Oppenheimer} B.~D.,  et~al., 2016, \mn@doi [\mnras] {10.1093/mnras/stw1066},
  \href {http://adsabs.harvard.edu/abs/2016MNRAS.460.2157O} {460, 2157}

\bibitem[\protect\citeauthoryear{{Oppenheimer}, {Segers}, {Schaye}, {Richings}
  \& {Crain}}{{Oppenheimer} et~al.}{2018}]{oppenheimer2018}
{Oppenheimer} B.~D.,  {Segers} M.,  {Schaye} J.,  {Richings} A.~J.,   {Crain}
  R.~A.,  2018, \mn@doi [\mnras] {10.1093/mnras/stx2967}, \href
  {https://ui.adsabs.harvard.edu/abs/2018MNRAS.474.4740O} {474, 4740}

\bibitem[\protect\citeauthoryear{{Peeples} et~al.,}{{Peeples}
  et~al.}{2019}]{peeples2019}
{Peeples} M.~S.,  et~al., 2019, \mn@doi [\apj] {10.3847/1538-4357/ab0654},
  \href {https://ui.adsabs.harvard.edu/abs/2019ApJ...873..129P} {873, 129}

\bibitem[\protect\citeauthoryear{{Planck Collaboration} et~al.,}{{Planck
  Collaboration} et~al.}{2014}]{planck2014}
{Planck Collaboration} et~al., 2014, \mn@doi [\aap]
  {10.1051/0004-6361/201321591}, \href
  {http://adsabs.harvard.edu/abs/2014A%26A...571A..16P} {571, A16}

\bibitem[\protect\citeauthoryear{{Puchwein} \& {Springel}}{{Puchwein} \&
  {Springel}}{2013}]{Puchwein2013}
{Puchwein} E.,  {Springel} V.,  2013, \mn@doi [\mnras] {10.1093/mnras/sts243},
  \href {https://ui.adsabs.harvard.edu/abs/2013MNRAS.428.2966P} {428, 2966}

\bibitem[\protect\citeauthoryear{{Puchwein}, {Haardt}, {Haehnelt}  \&
  {Madau}}{{Puchwein} et~al.}{2019}]{puchwein2019}
{Puchwein} E.,  {Haardt} F.,  {Haehnelt} M.~G.,   {Madau} P.,  2019, \mn@doi
  [\mnras] {10.1093/mnras/stz222}, \href
  {https://ui.adsabs.harvard.edu/abs/2019MNRAS.485...47P} {485, 47}

\bibitem[\protect\citeauthoryear{{Rahmati}, {Pawlik}, {Rai{\v c}evi{\'c}}  \&
  {Schaye}}{{Rahmati} et~al.}{2013}]{rahmati2013}
{Rahmati} A.,  {Pawlik} A.~H.,  {Rai{\v c}evi{\'c}} M.,   {Schaye} J.,  2013,
  \mn@doi [\mnras] {10.1093/mnras/stt066}, \href
  {http://adsabs.harvard.edu/abs/2013MNRAS.430.2427R} {430, 2427}

\bibitem[\protect\citeauthoryear{{Rahmati}, {Schaye}, {Crain}, {Oppenheimer},
  {Schaller}  \& {Theuns}}{{Rahmati} et~al.}{2016}]{rahmati2016}
{Rahmati} A.,  {Schaye} J.,  {Crain} R.~A.,  {Oppenheimer} B.~D.,  {Schaller}
  M.,   {Theuns} T.,  2016, \mn@doi [\mnras] {10.1093/mnras/stw453}, \href
  {http://adsabs.harvard.edu/abs/2016MNRAS.459..310R} {459, 310}

\bibitem[\protect\citeauthoryear{{Samui}, {Subramanian}  \& {Srianand}}{{Samui}
  et~al.}{2008}]{Samui2008}
{Samui} S.,  {Subramanian} K.,   {Srianand} R.,  2008, \mn@doi [\mnras]
  {10.1111/j.1365-2966.2008.12932.x}, \href
  {https://ui.adsabs.harvard.edu/abs/2008MNRAS.385..783S} {385, 783}

\bibitem[\protect\citeauthoryear{{Samui}, {Subramanian}  \& {Srianand}}{{Samui}
  et~al.}{2010}]{Samui2010}
{Samui} S.,  {Subramanian} K.,   {Srianand} R.,  2010, \mn@doi [\mnras]
  {10.1111/j.1365-2966.2009.16099.x}, \href
  {https://ui.adsabs.harvard.edu/abs/2010MNRAS.402.2778S} {402, 2778}

\bibitem[\protect\citeauthoryear{{Scannapieco}, {Tissera}, {White}  \&
  {Springel}}{{Scannapieco} et~al.}{2005}]{Scannapieco2005}
{Scannapieco} C.,  {Tissera} P.~B.,  {White} S.~D.~M.,   {Springel} V.,  2005,
  \mn@doi [\mnras] {10.1111/j.1365-2966.2005.09574.x}, \href
  {https://ui.adsabs.harvard.edu/abs/2005MNRAS.364..552S} {364, 552}

\bibitem[\protect\citeauthoryear{{Schaye}, {Aguirre}, {Kim}, {Theuns}, {Rauch}
  \& {Sargent}}{{Schaye} et~al.}{2003}]{schaye2003}
{Schaye} J.,  {Aguirre} A.,  {Kim} T.-S.,  {Theuns} T.,  {Rauch} M.,
  {Sargent} W. L.~W.,  2003, \mn@doi [\apj] {10.1086/378044}, \href
  {https://ui.adsabs.harvard.edu/abs/2003ApJ...596..768S} {596, 768}

\bibitem[\protect\citeauthoryear{Shakura \& Sunyaev}{Shakura \&
  Sunyaev}{1973}]{shakura1973}
Shakura N.~I.,  Sunyaev R.~A.,  1973, Astronomy and Astrophysics, 24, 337

\bibitem[\protect\citeauthoryear{{Shull}, {Danforth}  \& {Tilton}}{{Shull}
  et~al.}{2014}]{shull2014}
{Shull} J.~M.,  {Danforth} C.~W.,   {Tilton} E.~M.,  2014, \mn@doi [\apj]
  {10.1088/0004-637X/796/1/49}, \href
  {http://adsabs.harvard.edu/abs/2014ApJ...796...49S} {796, 49}

\bibitem[\protect\citeauthoryear{{Shull}, {Moloney}, {Danforth}  \&
  {Tilton}}{{Shull} et~al.}{2015}]{shull2015}
{Shull} J.~M.,  {Moloney} J.,  {Danforth} C.~W.,   {Tilton} E.~M.,  2015,
  \mn@doi [\apj] {10.1088/0004-637X/811/1/3}, \href
  {http://adsabs.harvard.edu/abs/2015ApJ...811....3S} {811, 3}

\bibitem[\protect\citeauthoryear{{Simcoe}}{{Simcoe}}{2011}]{simcoe2011}
{Simcoe} R.~A.,  2011, \mn@doi [\apj] {10.1088/0004-637X/738/2/159}, \href
  {https://ui.adsabs.harvard.edu/abs/2011ApJ...738..159S} {738, 159}

\bibitem[\protect\citeauthoryear{{Simcoe}, {Sargent}  \& {Rauch}}{{Simcoe}
  et~al.}{2004}]{simcoe2004}
{Simcoe} R.~A.,  {Sargent} W. L.~W.,   {Rauch} M.,  2004, \mn@doi [\apj]
  {10.1086/382777}, \href
  {https://ui.adsabs.harvard.edu/abs/2004ApJ...606...92S} {606, 92}

\bibitem[\protect\citeauthoryear{{Songaila} \& {Cowie}}{{Songaila} \&
  {Cowie}}{1996}]{Songaila1996}
{Songaila} A.,  {Cowie} L.~L.,  1996, \mn@doi [\aj] {10.1086/118018}, \href
  {https://ui.adsabs.harvard.edu/abs/1996AJ....112..335S} {112, 335}

\bibitem[\protect\citeauthoryear{{Springel}}{{Springel}}{2005}]{springel2005}
{Springel} V.,  2005, \mn@doi [\mnras] {10.1111/j.1365-2966.2005.09655.x},
  \href {http://adsabs.harvard.edu/abs/2005MNRAS.364.1105S} {364, 1105}

\bibitem[\protect\citeauthoryear{{Springel} \& {Hernquist}}{{Springel} \&
  {Hernquist}}{2003}]{springel2003a}
{Springel} V.,  {Hernquist} L.,  2003, \mn@doi [\mnras]
  {10.1046/j.1365-8711.2003.06206.x}, \href
  {https://ui.adsabs.harvard.edu/abs/2003MNRAS.339..289S} {339, 289}

\bibitem[\protect\citeauthoryear{Springel, Di~Matteo  \& Hernquist}{Springel
  et~al.}{2005a}]{springel2005a}
Springel V.,  Di~Matteo T.,   Hernquist L.,  2005a, Monthly Notices of the
  Royal Astronomical Society, 361, 776

\bibitem[\protect\citeauthoryear{{Springel}, {Di Matteo}  \&
  {Hernquist}}{{Springel} et~al.}{2005b}]{springeldematteo2005}
{Springel} V.,  {Di Matteo} T.,   {Hernquist} L.,  2005b, \mn@doi [\mnras]
  {10.1111/j.1365-2966.2005.09238.x}, \href
  {https://ui.adsabs.harvard.edu/abs/2005MNRAS.361..776S} {361, 776}

\bibitem[\protect\citeauthoryear{{Stern} et~al.,}{{Stern}
  et~al.}{2021}]{stern2021}
{Stern} J.,  et~al., 2021, \mn@doi [\apj] {10.3847/1538-4357/abd776}, \href
  {https://ui.adsabs.harvard.edu/abs/2021ApJ...911...88S} {911, 88}

\bibitem[\protect\citeauthoryear{{Suresh}, {Bird}, {Vogelsberger}, {Genel},
  {Torrey}, {Sijacki}, {Springel}  \& {Hernquist}}{{Suresh}
  et~al.}{2015}]{surech2015}
{Suresh} J.,  {Bird} S.,  {Vogelsberger} M.,  {Genel} S.,  {Torrey} P.,
  {Sijacki} D.,  {Springel} V.,   {Hernquist} L.,  2015, \mn@doi [\mnras]
  {10.1093/mnras/stu2762}, \href
  {https://ui.adsabs.harvard.edu/abs/2015MNRAS.448..895S} {448, 895}

\bibitem[\protect\citeauthoryear{{Suresh}, {Rubin}, {Kannan}, {Werk},
  {Hernquist}  \& {Vogelsberger}}{{Suresh} et~al.}{2017}]{suresh2017}
{Suresh} J.,  {Rubin} K. H.~R.,  {Kannan} R.,  {Werk} J.~K.,  {Hernquist} L.,
  {Vogelsberger} M.,  2017, \mn@doi [\mnras] {10.1093/mnras/stw2499}, \href
  {https://ui.adsabs.harvard.edu/abs/2017MNRAS.465.2966S} {465, 2966}

\bibitem[\protect\citeauthoryear{{Sutherland} \& {Dopita}}{{Sutherland} \&
  {Dopita}}{1993}]{sutherland1993}
{Sutherland} R.~S.,  {Dopita} M.~A.,  1993, \mn@doi [\apjs] {10.1086/191823},
  \href {https://ui.adsabs.harvard.edu/abs/1993ApJS...88..253S} {88, 253}

\bibitem[\protect\citeauthoryear{Tepper-Garc{\'\i}a, Richter, Schaye, Booth,
  Vecchia, Theuns  \& Wiersma}{Tepper-Garc{\'\i}a et~al.}{2011}]{tepper2011}
Tepper-Garc{\'\i}a T.,  Richter P.,  Schaye J.,  Booth C.,  Vecchia C.~D.,
  Theuns T.,   Wiersma R.~P.,  2011, Monthly Notices of the Royal Astronomical
  Society, 413, 190

\bibitem[\protect\citeauthoryear{{Tepper-Garc{\'\i}a}, {Richter}  \&
  {Schaye}}{{Tepper-Garc{\'\i}a} et~al.}{2013}]{tepper2013}
{Tepper-Garc{\'\i}a} T.,  {Richter} P.,   {Schaye} J.,  2013, \mn@doi [\mnras]
  {10.1093/mnras/stt1712}, \href
  {https://ui.adsabs.harvard.edu/abs/2013MNRAS.436.2063T} {436, 2063}

\bibitem[\protect\citeauthoryear{{Tillman}, {Burkhart}, {Tonnesen}, {Bird},
  {Bryan}, {Angl{\'e}s-Alc{\'a}zar}, {Dav{\'e}}  \& {Genel}}{{Tillman}
  et~al.}{2022}]{Tillman2022}
{Tillman} M.~T.,  {Burkhart} B.,  {Tonnesen} S.,  {Bird} S.,  {Bryan} G.~L.,
  {Angl{\'e}s-Alc{\'a}zar} D.,  {Dav{\'e}} R.,   {Genel} S.,  2022, arXiv
  e-prints, \href {https://ui.adsabs.harvard.edu/abs/2022arXiv221002467T} {p.
  arXiv:2210.02467}

\bibitem[\protect\citeauthoryear{{Tripp}, {Sembach}, {Bowen}, {Savage},
  {Jenkins}, {Lehner}  \& {Richter}}{{Tripp} et~al.}{2008}]{tripp2008}
{Tripp} T.~M.,  {Sembach} K.~R.,  {Bowen} D.~V.,  {Savage} B.~D.,  {Jenkins}
  E.~B.,  {Lehner} N.,   {Richter} P.,  2008, \mn@doi [\apjs] {10.1086/587486},
  \href {http://adsabs.harvard.edu/abs/2008ApJS..177...39T} {177, 39}

\bibitem[\protect\citeauthoryear{{Tumlinson}, {Peeples}  \& {Werk}}{{Tumlinson}
  et~al.}{2017}]{Tumlinson2017}
{Tumlinson} J.,  {Peeples} M.~S.,   {Werk} J.~K.,  2017, \mn@doi [\araa]
  {10.1146/annurev-astro-091916-055240}, \href
  {https://ui.adsabs.harvard.edu/abs/2017ARA&A..55..389T} {55, 389}

\bibitem[\protect\citeauthoryear{{Upton Sanderbeck}, {McQuinn}, {D'Aloisio}  \&
  {Werk}}{{Upton Sanderbeck} et~al.}{2018}]{Upton2018}
{Upton Sanderbeck} P.~R.,  {McQuinn} M.,  {D'Aloisio} A.,   {Werk} J.~K.,
  2018, \mn@doi [\apj] {10.3847/1538-4357/aaeff2}, \href
  {https://ui.adsabs.harvard.edu/abs/2018ApJ...869..159U} {869, 159}

\bibitem[\protect\citeauthoryear{{Verner}, {Ferland}, {Korista}  \&
  {Yakovlev}}{{Verner} et~al.}{1996}]{verner1996}
{Verner} D.~A.,  {Ferland} G.~J.,  {Korista} K.~T.,   {Yakovlev} D.~G.,  1996,
  \mn@doi [\apj] {10.1086/177435}, \href
  {https://ui.adsabs.harvard.edu/abs/1996ApJ...465..487V} {465, 487}

\makeatother
\end{thebibliography}

\bsp	
\label{lastpage}

\end{document}